\documentclass[12pt,twoside, a4paper]{article}
\pdfoutput=1

\def\mc{\mathcal}
\def\ul{\underline}

\usepackage[dvips]{graphicx}
\usepackage{amssymb}
\usepackage{amssymb,amsmath}
\usepackage{graphicx}
\usepackage{dsfont}
\usepackage{caption}
\usepackage{subcaption}
\usepackage{mathtools}
\usepackage{verbatim}
\usepackage{graphicx}
\usepackage{multirow}
\usepackage[outline]{contour}
\usepackage{xcolor,colortbl}
\usepackage{pdflscape}
\input{epsf.sty}  
\contourlength{.1pt}
\graphicspath{ {./Figures/} }
\numberwithin{equation}{section}

\begin{document}
\begin{center}
\LARGE{\textbf{Supersymmetric domain walls in maximal 6D gauged supergravity I}}
\end{center}
\vspace{1 cm}
\begin{center}
\large{\textbf{Parinya Karndumri}$^a$ and \textbf{Patharadanai Nuchino}$^b$}
\end{center}
\begin{center}
String Theory and Supergravity Group, Department
of Physics, Faculty of Science, Chulalongkorn University, 254 Phayathai Road, Pathumwan, Bangkok 10330, Thailand
\end{center}
E-mail: $^a$parinya.ka@hotmail.com \\
E-mail: $^b$danai.nuchino@hotmail.com \vspace{1 cm}\\
\begin{abstract}
We find a large class of supersymmetric domain wall solutions from six-dimensional $N=(2,2)$ gauged supergravity with various gauge groups. In general, the embedding tensor lives in $\mathbf{144}_c$ representation of the global symmetry $SO(5,5)$. We explicitly construct the embedding tensors in $\mathbf{15}^{-1}$ and $\overline{\mathbf{40}}^{-1}$ representations of $GL(5)\sim \mathbb{R}^+\times SL(5)\subset SO(5,5)$ leading to $CSO(p,q,5-p-q)$ and $CSO(p,q,4-p-q)\ltimes\mathbb{R}^4_{\boldsymbol{s}}$ gauge groups, respectively. These gaugings can be obtained from $S^1$ reductions of seven-dimensional gauged supergravity with $CSO(p,q,5-p-q)$ and $CSO(p,q,4-p-q)$ gauge groups. As in seven dimensions, we find half-supersymmetric domain walls for purely magnetic or purely electric gaugings with the embedding tensors in $\mathbf{15}^{-1}$ or $\overline{\mathbf{40}}^{-1}$ representations, respectively. In addition, for dyonic gauge groups with the embedding tensors in both $\mathbf{15}^{-1}$ and $\overline{\mathbf{40}}^{-1}$ representations, the domain walls turn out to be $\frac{1}{4}$-supersymmetric as in the seven-dimensional analogue. By the DW/QFT duality, these solutions are dual to maximal and half-maximal super Yang-Mills theories in five dimensions. All of the solutions can be uplifted to seven dimensions and further embedded in type IIB or M-theories by the well-known consistent truncation of the seven-dimensional $N=4$ gauged supergravity.
\end{abstract}
\newpage
\section{Introduction}
Supersymmetric domain walls in gauged supergravities in various space-time dimensions have provided a useful tool for studying various aspects of the AdS/CFT correspondence since the original proposal in \cite{maldacena}, see also \cite{Gubser_AdS_CFT,Witten_AdS_CFT}. In particular, these solutions play an important role in the so-called DW/QFT correspondence \cite{DW_QFT1,DW_QFT2,DW_QFT3}, a generalization of the AdS/CFT correspondence to non-conformal field theories. They are also useful in studying some aspects of cosmology, see for example \cite{DW_cosmology1,DW_cosmology2,DW_cosmology3}. Due to their importance in many areas of applications, many domain wall solutions in gauged supergravities have been found in different space-time dimensions \cite{Eric_DW_10D,9D_DW_Cowdall,9D_DW,8D_DW1,8D_DW2,7D_DW,Pope_DW_massive,DW_algebraic_curve,Eric_DW_maximal_SUGRA,
Pope_symmetric_poten,DW_Hull,4D_DW,3DN4_DW,3DN10_DW,2D_DW1,2D_DW2}. A systematic classification of supersymmetric domain walls from maximal gauged supergravity in various space-time dimensions can also be found in \cite{Eric_SUSY_DW}. 

In this paper, we are interested in maximal $N=(2,2)$ six-dimensional gauged supergravity with $SO(5,5)$ global symmetry. Compared to other dimensions, supersymmetric solutions to this six-dimensional gauged supergravity have not been systematically studied since the original construction of the ungauged $N=(2,2)$ supergravity long ago in \cite{Tanii}. The first $N=(2,2)$ six-dimensional gauged supergravity with $SO(5)$ gauge group has been constructed in \cite{6D_SO(5)} by performing an $S^1$ reduction of the $SO(5)$ maximal gauged supergravity in seven dimensions \cite{7D_Pernici}. More recently, the most general gaugings have been constructed and classified in \cite{6D_Max_Gauging} using the embedding tensor formalism. From the results of \cite{6D_Max_Gauging}, there are two particularly interesting classes of gaugings under $GL(5)$ and $SO(4,4)$ subgroups of $SO(5,5)$. The former contains gaugings obtained from an $S^1$ reduction of seven-dimensional maximal gauged supergravity while the latter can be truncated to half-maximal $N=(1,1)$ gauged supergravity. 

We will consider only gaugings in the first class with the embedding tensor in $\mathbf{15}^{-1}$ and $\overline{\mathbf{40}}^{-1}$ representations of $GL(5)$. These gaugings have known seven-dimensional origins via an $S^1$ reduction and can also be embedded in string/M-theory using the truncations to maximal gauged supergravity in seven dimensions. The fact that there does not exist an $N=4$ superconformal symmetry in five dimensions \cite{Nahm_res} is in agreement with the recent classification of maximally supersymmetric $AdS$ vacua given in \cite{Max_SUSY_AdS}. This implies that there is no AdS$_6$/CFT$_5$ duality in the case of $32$ supercharges. Therefore, maximally supersymmetric vacuum solutions of the $N=(2,2)$ gauged supergravity are expected to be half-supersymmetric domain walls. In this work, we will systematically study this type of solutions and give a large number of them including $\frac{1}{4}$-supersymmetric solutions.

It has been shown recently that maximally supersymmetric Yang-Mills theory in five dimensions plays an important role in the dynamics of (conformal) field theories in both higher and lower dimensions via a number of dualities, see for example \cite{Douglas_5D_MSYM,5D_MSYM_1,5D_MSYM_2,5D_MSYM_3,5D_MSYM_4,5D_MSYM_5}. In particular, this theory could even be used to define the less known $N=(2,0)$ superconformal field theory in six dimensions compactified on $S^1$. The latter is well-known to describe the dynamics of strongly coupled theory on M5-branes. Accordingly, we expect that supersymmetric domain walls of the maximal gauged supergravity in six dimensions could be useful in studying various aspects of the maximal super Yang-Mills theory in five dimensions via the DW/QFT correspondence. A simple domain wall solution with $SO(5)$ symmetry has already been given in \cite{6D_SO(5)} for $SO(5)$ gauging, see \cite{Bobev_spherical_brane1} and \cite{Bobev_spherical_brane2} for the holographic interpretation of this solution. In this paper, we extend this study by including a large class of supersymmetric domain walls with different unbroken symmetries in $N=(2,2)$ gauged supergravity with various gauge groups.  

The paper is organized as follows. In section \ref{6DN=(2,2)gSUGRA}, the construction of six-dimensional maximal gauged supergravity in the embedding tensor formalism is reviewed. Supersymmetric domain wall solutions from gaugings in $\mathbf{15}^{-1}$, $\overline{\mathbf{40}}^{-1}$, and $(\mathbf{15}+\overline{\mathbf{40}})^{-1}$ representations are respectively given in sections \ref{pure15Sec}, \ref{pure40Sec}, and \ref{YandZsec}. Conclusions and discussions are given in section \ref{Discuss}. Branching rules for relevant $SO(5,5)$ representations under $GL(5)$ are given in appendix \ref{AppA}. The conventions on symplectic-Majorana-Weyl Spinors in six-dimensional space-time used throughout this work are collected in appendix \ref{SMWspinor}. Finally, consistent truncation ansatze for seven-dimensional $SO(5)$ gauged supergravity on $S^1$ giving rise to $SO(5)$ maximal gauged supergravity in six dimensions are reviewed in appendix \ref{TrunAnsz}.

\section{$N=(2,2)$ gauged supergravity in six dimensions}\label{6DN=(2,2)gSUGRA}
We begin by giving a brief review of six-dimensional $N=(2,2)$ gauged supergravity in the embedding tensor formalism constructed in \cite{6D_Max_Gauging}. We will mainly collect relevant formulae for constructing the embedding tensor and finding supersymmetric domain wall solutions. For more details, we refer the reader to the original construction in \cite{6D_Max_Gauging}.

As in other dimensions, $N=(2,2)$ maximal supersymmetry in six dimensions allows only a unique graviton supermultiplet with the following field content
\begin{equation}\label{6DSUGRAmultiplet}
\left(e^{\hat{\mu}}_\mu, B_{\mu\nu m}, A^{A}_\mu, {V_A}^{\alpha\dot{\alpha}}, \psi_{+\mu\alpha}, \psi_{-\mu\dot{\alpha}}, \chi_{+a\dot{\alpha}}, \chi_{-\dot{a}\alpha}\right).
\end{equation}
Most of the conventions are the same as in \cite{6D_Max_Gauging}. Curved and flat space-time indices are respectively denoted by $\mu,\nu,\ldots=0,1,\ldots,5$ and $\hat{\mu},\hat{\nu},\ldots=0,1,\ldots,5$. Lower and upper $m,n,\ldots=1,\ldots,5$ indices label fundamental and anti-fundamental representations of $GL(5)\subset SO(5,5)$, respectively. Indices $A,B,\ldots =1,\ldots,16$ describe Majorana-Weyl spinors of the $SO(5,5)$ duality symmetry. We also note that according to this convention, the electric two-form potentials $B_{\mu\nu m}$ transform as $\mathbf{5}$ under $GL(5)$ while the vector fields $A^{A}_\mu$ transform as $\mathbf{16}_c$ under $SO(5,5)$.

Fermionic fields, transforming under the local $SO(5)\times SO(5)$ symmetry, are symplectic-Majorana-Weyl (SMW) spinors, see appendix \ref{SMWspinor} for more detail on the convention. Indices $\alpha,\ldots=1,\ldots,4$ and $\dot{\alpha},\ldots=\dot{1},\ldots,\dot{4}$ are respectively two sets of $SO(5)$ spinor indices in $SO(5)\times SO(5)$. Similarly, vector indices of the two $SO(5)$ factors are denoted by $a,\ldots=1,\ldots,5$ and $\dot{a},\ldots=\dot{1},\ldots,\dot{5}$. We use $\pm$ to indicate space-time chiralities of the spinors. Under the local $SO(5)\times SO(5)$ symmetry, the two sets of gravitini $\psi_{+\mu\alpha}$ and $\psi_{-\mu\dot{\alpha}}$ transform as $(\mathbf{4},\mathbf{1})$ and $(\mathbf{1},\mathbf{4})$ while the spin-$\frac{1}{2}$ fields $\chi_{+a\dot{\alpha}}$ and $\chi_{-\dot{a}\alpha}$ transform as $(\mathbf{5},\mathbf{4})$ and $(\mathbf{4},\mathbf{5})$.

In ungauged supergravity, only the electric two-forms $B_{\mu\nu m}$ appear in the Lagrangian while the magnetic duals ${B_{\mu\nu}}^m$ transforming in $\overline{\mathbf{5}}$ representation of $GL(5)$ are introduced on-shell. The electric and magnetic two-forms are combined into a vector representation $\mathbf{10}$ of the full global symmetry group $SO(5,5)$ denoted by $B_{\mu\nu M}=(B_{\mu\nu m}, {B_{\mu\nu}}^m)$. Therefore, only the subgroup $GL(5)\subset SO(5,5)$ is a manifest off-shell symmetry of the theory. On the other hand, the full $SO(5,5)$ duality symmetry is the on-shell symmetry interchanging field equations and Bianchi identities of the two-form potentials. However, the most general gaugings of the ungauged supergravity can involve a symmetry that is not a subgroup of the off-shell $GL(5)$ symmetry. Moreover, the magnetic two-forms can also appear in the gauged Lagrangian via topological terms. 

In $N=(2,2)$ supergravity, there are $25$ scalar fields parametrizing the coset space $SO(5,5)/\left(SO(5)\times SO(5)\right)$. In chiral spinor representation, we can describe the coset manifold by a coset representative ${V_A}^{\alpha\dot{\beta}}$ transforming under the global $SO(5,5)$ and local $SO(5)\times SO(5)$ by left and right multiplications, respectively. The inverse elements ${(V^{-1})_{\alpha\dot{\beta}}}^A$ will be denoted by ${V^A}_{\alpha\dot{\beta}}$ satisfying the relations
\begin{equation}\label{VViProp}
{V_A}^{\alpha\dot{\beta}}{V^B}_{\alpha\dot{\beta}}=\delta^B_A\qquad \textrm{and}\qquad {V_A}^{\alpha\dot{\beta}}{V^A}_{\gamma\dot{\delta}}=\delta^{\alpha}_{\gamma}\delta^{\dot{\beta}}_{\dot{\delta}}\, .
\end{equation}
In vector representation, the coset representative is given by a $10\times10$ matrix ${\mathcal{V}_M}^{\underline{A}}=({\mathcal{V}_M}^{a}, {\mathcal{V}_M}^{\dot{a}})$ with $\underline{A}=(a,\dot{a})$. This is related to the coset representative in chiral spinor representation by the following relations
\begin{eqnarray}
{\mathcal{V}_M}^a&=&\frac{1}{16}V^{A\alpha\dot{\alpha}}(\Gamma_M)_{AB}{(\gamma^a)_{\alpha\dot{\alpha}}}^{\beta\dot{\beta}}{V^B}_{\beta\dot{\beta}},\label{VVrel1}\\{\mathcal{V}_M}^{\dot{a}}&=&-\frac{1}{16}V^{A\alpha\dot{\alpha}}(\Gamma_M)_{AB}{(\gamma^{\dot{a}})_{\alpha\dot{\alpha}}}^{\beta\dot{\beta}}{V^B}_{\beta\dot{\beta}}\, .\label{VVrel2}
\end{eqnarray}
In these equations, $(\Gamma_M)_{AB}$ and ${(\Gamma_{\ul{A}})_{\alpha\dot{\alpha}}}^{\beta\dot{\beta}}=({(\gamma_a)_{\alpha\dot{\alpha}}}^{\beta\dot{\beta}},{(\gamma_{\dot{a}})_{\alpha\dot{\alpha}}}^{\beta\dot{\beta}})$ are respectively $SO(5,5)$ gamma matrices in non-diagonal $\eta_{MN}$ and diagonal $\eta_{\ul{A}\ul{B}}$ bases, see appendix \ref{SpinorBranc} for more detail.  
\\
\indent The inverse will be denoted by $\mathcal{V}^{M\underline{A}}$ satisfying the following relations 
\begin{equation}
\mathcal{V}^{Ma}{\mathcal{V}_M}^b=\delta^{ab},\qquad\mathcal{V}^{M\dot{a}}{\mathcal{V}_M}^{\dot{b}}=\delta^{\dot{a}\dot{b}},\qquad\mathcal{V}^{Ma}{\mathcal{V}_M}^{\dot{a}}=0
\end{equation}
and
\begin{equation}
{\mathcal{V}_M}^a\mathcal{V}^{Na}-{\mathcal{V}_M}^{\dot{a}}\mathcal{V}^{N\dot{a}}=\delta^N_M\, .\label{V_Vi_vec}
\end{equation}
In these equations, we have explicitly raised the $SO(5)\times SO(5)$ vector index $\underline{A}=(a,\dot{a})$ resulting in a minus sign in equation \eqref{V_Vi_vec}.

The most general gaugings of six-dimensional $N=(2,2)$ supergravity can be efficiently described by using the embedding tensor ${\Theta_A}^{MN}$. This tensor introduces the minimal coupling of various fields via the covariant derivative
\begin{equation}\label{gauge_covariant_derivative}
D_\mu=\partial_\mu-gA^A_\mu\ {\Theta_A}^{MN}\boldsymbol{t}_{MN}
\end{equation}
where $g$ is a gauge coupling constant. The embedding tensor identifies generators $X_A={\Theta_A}^{MN}\boldsymbol{t}_{MN}$ of the gauge group $G_0\subset SO(5,5)$ with particular linear combinations of the $SO(5,5)$ generators $\boldsymbol{t}_{MN}$. Supersymmetry requires the embedding tensor to transform as $\mathbf{144}_c$ representation of $SO(5,5)$. Accordingly, ${\Theta_A}^{MN}$ can be parametrized in term of a vector-spinor $\theta^{AM}$ of $SO(5,5)$ as 
\begin{equation}
{\Theta_A}^{MN}\ =\ -\theta^{B[M}(\Gamma^{N]})_{BA}\ \equiv \ \left(\Gamma^{[M}\theta^{N]}\right)_A
\end{equation}
with $\theta^{AM}$ subject to the constraint
\begin{equation}\label{MainLC}
(\Gamma_M)_{AB}\,\theta^{BM}\ =\ 0\, .
\end{equation}
\indent With the $SO(5,5)$ generators in vector and spinor representations given by
\begin{equation}
{(\boldsymbol{t}_{MN})_P}^Q=4\eta_{P[M}\delta^Q_{N]}\qquad\text{and}\qquad {(\boldsymbol{t}_{MN})_A}^B\ =\ {(\Gamma_{MN})_A}^B
\end{equation}
in which $\eta_{MN}$ is the off-diagonal $SO(5,5)$ invariant tensor given in \eqref{off-diag-eta}, the corresponding gauge generators take the forms
\begin{equation}\label{DefGaugeGen}
{(X_A)_M}^N=2\left(\Gamma_{M}\theta^{N}\right)_A+2\left(\Gamma^{N}\theta_{M}\right)_A\qquad\text{and}\qquad {(X_A)_B}^C= \left(\Gamma^{M}\theta^{N}\right)_A{(\Gamma_{MN})_B}^C\, .
\end{equation}
\indent For consistency, the gauge generators must form a closed subalgebra of $SO(5,5)$, so the embedding tensor needs to satisfy the quadratic constraint
\begin{equation}
\left[X_A,X_B\right]\ = \ -{(X_A)_B}^C\,X_C\, .
\end{equation} 
In terms of $\theta^{AM}$, the quadratic constraint reduces to the following two conditions
\begin{equation}\label{QC}
\theta^{AM}\theta^{BN}\eta_{MN}\ =\ 0,\qquad \theta^{AM}\theta^{B[N}(\Gamma^{P]})_{AB}=0\, .
\end{equation} 
It follows that any $\theta^{AM}\in\mathbf{144}_c$ satisfying this quadratic constraint defines a consistent gauging of the theory.
\\
\indent To identify possible gaugings, we first decompose $\theta^{AM}$ under a given subgroup of $SO(5, 5)$. As pointed out before, the $GL(5)$ subgroup of $SO(5,5)$ is of particular interest since this is the symmetry of the ungauged Lagrangian. As given in \cite{6D_Max_Gauging}, $\theta^{AM}\in\mathbf{144}_c$ decomposes under $GL(5)\subset SO(5,5)$ as
\begin{equation}\label{mainthetaDec}
\mathbf{144}_c\ \rightarrow\ \overline{\mathbf{5}}^{+3}\,\oplus\,\mathbf{5}^{+7}\,\oplus\,\mathbf{10}^{-1}\,\oplus\,\mathbf{15}^{-1}\,\oplus\,\mathbf{24}^{-5}\,\oplus\,\overline{\mathbf{40}}^{-1}\,\oplus\,\overline{\mathbf{45}}^{+3}\, .
\end{equation}
The explicit form of all the seven irreducible components can be found in appendix \ref{Apptheta}. In this case, determining consistent gaugings is to find the irreducible components satisfying the quadratic constraint \eqref{QC}.

By decomposing the $SO(5,5)$ vector index under $GL(5)$, we can write $\theta^{AM}=(\theta^{Am},\theta^{A}_m)$ with $\theta^{Am}$ and $\theta^{A}_m$ containing the following irreducible components
\begin{eqnarray}
\theta^{Am}&=&\overline{\mathbf{5}}^{+3}\,\oplus\,\mathbf{10}^{-1}\,\oplus\,\mathbf{24}^{-5}\,\oplus\,\overline{\mathbf{40}}^{-1},\label{splitthetaDec1}\\ \theta^{A}_m&=&\overline{\mathbf{5}}^{+3}\,\oplus\,\mathbf{5}^{+7}\,\oplus\,\mathbf{10}^{-1}\,\oplus\,\mathbf{15}^{-1}\,\oplus\,\overline{\mathbf{45}}^{+3}\, .\label{splitthetaDec2}
\end{eqnarray}
It is easily seen that the first equation in \eqref{QC} is automatically satisfied for purely electric or purely magnetic gaugings that involve only $\theta^{Am}$ or $\theta^{A}_m$ components. We note that as pointed out in \cite{6D_Max_Gauging}, gaugings triggered by $\theta^{Am}$ are electric in the sense that only electric two-forms participate in the resulting gauged theory while magnetic gaugings triggered by $\theta^A_m$ involve magnetic two-forms together with additional three-form tensor fields. Comparing \eqref{splitthetaDec1} and \eqref{splitthetaDec2} to \eqref{mainthetaDec}, we immediately see that gaugings in $\mathbf{24}^{-5}\,\oplus\,\overline{\mathbf{40}}^{-1}$ and $\mathbf{5}^{+7}\,\oplus\,\mathbf{15}^{-1}\,\oplus\,\overline{\mathbf{45}}^{+3}$ representations are respectively purely electric and purely magnetic whereas those in $\overline{\mathbf{5}}^{+3}\,\oplus\,\mathbf{10}^{-1}$ representation correspond to dyonic gaugings involving both electric and magnetic two-forms. Other dyonic gaugings can also arise from combinations of various electric and magnetic components leading to many possible gauge groups.

Apart from the minimal coupling implemented by the covariant derivative \eqref{gauge_covariant_derivative}, gaugings also lead to hierarchies of non-abelian vector and tensor fields of various ranks. However, since we are only interested in domain wall solutions which only involve the metric and scalar fields, we will, from now on, set all vector and tensor fields to zero. It is straightforward to verify that this is indeed a consistent truncation. With only the metric and scalars non-vanishing, the bosonic Lagrangian of the maximal $N=(2,2)$ gauged supergravity takes the form
\begin{equation}
e^{-1}\mathcal{L}=\frac{1}{4}R-\frac{1}{16}P_{\mu}^{a\dot{a}}P^\mu_{a\dot{a}}-\mathbf{V},\label{bosonic_L}
\end{equation}
and supersymmetry transformations of fermionic fields are given by
\begin{eqnarray}
\delta\psi_{+\mu\alpha}&=& D_\mu\epsilon_{+\alpha}+\frac{g}{4}\hat{\gamma}_\mu {T_\alpha}^{\dot{\beta}}\epsilon_{-\dot{\beta}},\label{1stSUSY}\\
\delta\psi_{-\mu\dot{\alpha}}&=& D_\mu\epsilon_{-\dot{\alpha}}-\frac{g}{4}\hat{\gamma}_\mu {T^{\beta}}_{\dot{\alpha}}\epsilon_{+\beta},\label{2ndSUSY}\\
\delta\chi_{+a\dot{\alpha}}&=&\frac{1}{4}P^\mu_{a\dot{a}}\hat{\gamma}_\mu{(\gamma^{\dot{a}})_{\dot{\alpha}}}^{\dot{\beta}}\epsilon_{-\dot{\beta}}+2g{(T_{a})^\beta}_{\dot{\alpha}}\epsilon_{+\beta}-\frac{g}{2}{T^{\alpha}}_{\dot{\alpha}}{(\gamma_a)_\alpha}^\beta\epsilon_{+\beta},\label{3rdSUSY}\\
\delta\chi_{-\dot{a}\alpha}&=&\frac{1}{4}P^\mu_{a\dot{a}}\hat{\gamma}_\mu{(\gamma^a)_\alpha}^\beta\epsilon_{+\beta}+2g{(T_{\dot{a}})_{\alpha}}^{\dot{\beta}}\epsilon_{-\dot{\beta}}+\frac{g}{2}{T_{\alpha}}^{\dot{\alpha}}{(\gamma_{\dot{a}})_{\dot{\alpha}}}^{\dot{\beta}}\epsilon_{-\dot{\beta}}.\label{4thtSUSY}
\end{eqnarray}
\indent The covariant derivatives of supersymmetry parameters, $\epsilon_{+\alpha}$ and $\epsilon_{-\dot{\alpha}}$, are defined by
\begin{eqnarray}
D_\mu\epsilon_{+\alpha}&=& \partial_\mu\epsilon_{+\alpha}+\frac{1}{4}{\omega_\mu}^{\nu\rho}\hat{\gamma}_{\nu\rho}\epsilon_{+\alpha}+\frac{1}{4}Q_\mu^{ab}{(\gamma_{ab})_\alpha}^\beta\epsilon_{+\beta},\label{CoDivEp+}\\
D_\mu\epsilon_{-\dot{\alpha}}&=& \partial_\mu\epsilon_{-\dot{\alpha}}+\frac{1}{4}{\omega_\mu}^{\nu\rho}\hat{\gamma}_{\nu\rho}\epsilon_{-\dot{\alpha}}+\frac{1}{4}Q_\mu^{\dot{a}\dot{b}}{(\gamma_{\dot{a}\dot{b}})_{\dot{\alpha}}}^{\dot{\beta}}\epsilon_{-\dot{\beta}}\label{CoDivEp-}
\end{eqnarray}
with $\hat{\gamma}_\mu=e_\mu^{\hat{\mu}}\hat{\gamma}_{\hat{\mu}}$. Matrices $\hat{\gamma}_{\hat{\mu}}$ are space-time gamma matrices, see the convention in appendix \ref{SMWspinor}. For simplicity, we will suppress all space-time spinor indices.

The scalar vielbein $P_{\mu}^{a\dot{a}}$ and $SO(5)\times SO(5)$ composite connections, $Q_\mu^{ab}$ and $Q_\mu^{\dot{a}\dot{b}}$, are given by
\begin{eqnarray}
P_{\mu}^{a\dot{a}}&=&\frac{1}{4}{(\gamma^a)}^{\alpha\beta}{(\gamma^{\dot{a}})}^{\dot{\alpha}\dot{\beta}}{V^A}_{\alpha\dot{\alpha}}\partial_\mu V_{A\beta\dot{\beta}},\label{PDef}\\
Q_{\mu}^{ab}&=&\frac{1}{8}{(\gamma^{ab})}^{\alpha\beta}\Omega^{\dot{\alpha}\dot{\beta}} {V^A}_{\alpha\dot{\alpha}}\partial_\mu V_{A\beta\dot{\beta}},\label{QuDef}\\
Q_{\mu}^{\dot{a}\dot{b}}&=&\frac{1}{8}\Omega^{\alpha\beta}{(\gamma^{\dot{a}\dot{b}})}^{\dot{\alpha}\dot{\beta}}{V^A}_{\alpha\dot{\alpha}}\partial_\mu V_{A\beta\dot{\beta}}\label{QdDef}
\end{eqnarray}
in which $\Omega^{\alpha\beta}$ and $\Omega^{\dot{\alpha}\dot{\beta}}$ are the two $USp(4)$ symplectic forms whose explicit forms can be found in \eqref{USp(4)Omegas}. These definitions can be derived from the following relation
\begin{equation}
{V^A}_{\alpha\dot{\alpha}}\partial_\mu V_{A\beta\dot{\beta}}=\frac{1}{4}P_{\mu}^{a\dot{a}}(\gamma_a)_{\alpha\beta}(\gamma_{\dot{a}})_{\dot{\alpha}\dot{\beta}}+\frac{1}{4}Q_{\mu}^{ab}(\gamma_{ab})_{\alpha\beta}\Omega_{\dot{\alpha}\dot{\beta}}+\frac{1}{4}Q_{\mu}^{\dot{a}\dot{b}}\Omega_{\alpha\beta}(\gamma_{\dot{a}\dot{b}})_{\dot{\alpha}\dot{\beta}}.
\end{equation}
\indent The scalar potential is given by
\begin{eqnarray}\label{scalarPot}
\mathbf{V}&=&\frac{g^2}{2}\theta^{AM}\theta^{BN}{\mathcal{V}_M}^a{\mathcal{V}_N}^b\left[{V_A}^{\alpha\dot{\alpha}}{(\gamma_a)_\alpha}^\beta{(\gamma_b)_\beta}^\gamma V_{B\gamma\dot{\alpha}}\right]\nonumber\\
&=&-\frac{g^2}{2}\left[T^{\alpha\dot{\alpha}}T_{\alpha\dot{\alpha}}-2(T^a)^{\alpha\dot{\alpha}}(T_a)_{\alpha\dot{\alpha}}\right]\label{scalarPot}
\end{eqnarray}
where we have introduced the T-tensors defined by
\begin{equation}\label{TTenDef}
(T^a)^{\alpha\dot{\alpha}}={\mathcal{V}_M}^a\theta^{AM}{V_A}^{\alpha\dot{\alpha}},\qquad (T^{\dot{a}})^{\alpha\dot{\alpha}}=-{\mathcal{V}_M}^{\dot{a}}\theta^{AM}{V_A}^{\alpha\dot{\alpha}}
\end{equation}
with
\begin{equation}
T^{\alpha\dot{\alpha}}\equiv (T^a)^{\beta\dot{\alpha}}{(\gamma_a)_\beta}^\alpha=-(T^{\dot{a}})^{\alpha\dot{\beta}}{(\gamma_{\dot{a}})_{\dot{\beta}}}^{\dot{\alpha}}.
\end{equation}
\section{Supersymmetric domain walls from gaugings in $\mathbf{15}^{-1}$ representation}\label{pure15Sec}
In this section, we consider gauge groups arising from the embedding tensor in $\mathbf{15}^{-1}$ representation. These are purely magnetic gaugings with the corresponding embedding tensor given by
\begin{equation}
\theta^{A}_m\ = \ \mathbb{T}^{An}Y_{nm}\, .
\end{equation}
The matrix $\mathbb{T}^{An}$ is the inverse of the transformation matrix $\mathbb{T}_{An}$ given in \eqref{TranMatTDef1} and $Y_{mn}$ is a symmetric $5\times 5$ matrix.

As previously mentioned, for $\theta^{Am}=0$, the embedding tensor $\theta^{AM}=(\,0\,,\,\mathbb{T}^{An}Y_{nm}\,)$ automatically satisfies the quadratic constraint \eqref{QC}. Therefore, every symmetric tensor $Y_{mn}$ defines a viable gauging in $\mathbf{15}^{-1}$ representation. As in \cite{7D_Max_Gauging}, we can use $SL(5)\subset GL(5)$ symmetry to bring $Y_{mn}$ to the form
\begin{equation}\label{diagYmn}
Y_{mn}\ =\ \text{diag}(\underbrace{1,..,1}_p,\underbrace{-1,..,-1}_q,\underbrace{0,..,0}_r)
\end{equation}
where $p+q+r=5$. 

Under $GL(5)$, the gauge generators transforming as a spinor $\mathbf{16}_s$ of $SO(5,5)$ decompose as follows
\begin{equation}\label{GaugeGenSplit}
X_A\ =\ \mathbb{T}_{Am}X^m+\mathbb{T}_{A}^{mn}X_{mn}+\mathbb{T}_{A\ast}X_\ast\, .
\end{equation}
For the embedding tensor in $\mathbf{15}^{-1}$ representation, the only non-vanishing gauge generators are given by 
\begin{equation}
X_{mn}=2Y_{p[m}{\boldsymbol{t}^p}_{n]}
\end{equation}
with ${\boldsymbol{t}^m}_{n}$ being $GL(5)$ generators. In vector representation, the explicit form of $X_{mn}$ is given by
\begin{equation}
{(X_{mn})_P}^Q\ =\ -4(\delta^Q_{[m}Y_{n]p}\delta^p_P+\eta_{P[m}Y_{n]q}\eta^{qQ}).
\end{equation}
These generators satisfy the commutation relations
\begin{equation}
[X_{mn},X_{pq}]={(X_{mn})_{pq}}^{rs}X_{rs}
\end{equation}
in which ${(X_{mn})_{pq}}^{rs}=2{(X_{mn})_{[p}}^{[r}\delta_{q]}^{s]}$. Therefore, the corresponding gauge group is determined to be
\begin{equation}
G_0\ =\ CSO(p,q,r)\ =\ SO(p,q) \ltimes \mathbb{R}^{(p+q)\cdot r}.
\end{equation}
These gaugings arise from an $S^1$ reduction of seven-dimensional maximal gauged supergravity with the same gauge groups. In the case of $SO(5)$ gauge group ($p=5$ and $q=r=0$), the complete reduction ansatz has already been constructed in \cite{6D_SO(5)}.

\subsection{Supersymmetric domain walls}\label{super_DW}
In order to find supersymmetric domain wall solutions, we take the space-time metric to be the standard domain wall ansatz
\begin{equation}\label{DWmetric}
ds_6^2=e^{2A(r)}\eta_{\bar{\mu} \bar{\nu}}dx^{\bar{\mu}} dx^{\bar{\nu}}+dr^2
\end{equation}
where $\bar{\mu},\bar{\nu}=0,1,\ldots,4$, and $A(r)$ is a warped factor depending only on the radial coordinate $r$. To parametrize the coset representative of $SO(5,5)/(SO(5)\times SO(5))$, we first identify the corresponding non-compact generators of $SO(5,5)$ in the basis with diagonal $SO(5,5)$ metric $\eta_{\ul{A}\ul{B}}$. These are given by
\begin{equation}
\hat{\boldsymbol{t}}_{a\dot{b}}\,=\,{\mathbb{M}_{a}}^M{\mathbb{M}_{\dot{b}}}^N\boldsymbol{t}_{MN}
\end{equation}
where ${\mathbb{M}_{\underline{A}}}^M=({\mathbb{M}_{a}}^M,{\mathbb{M}_{\dot{a}}}^M)$ is the inverse of the transformation matrix $\mathbb{M}$ given in \eqref{offDiagTrans}. 
\\
\indent We then split these generators into two parts that are symmetric and antisymmetric in $a$ and $\dot{b}$ indices as follows
\begin{equation}\label{noncomsep}
\hat{\boldsymbol{t}}_{a\dot{b}}\,=\,\hat{\boldsymbol{t}}^+_{a\dot{b}}+\hat{\boldsymbol{t}}^-_{a\dot{b}}
\end{equation}
with
\begin{equation}
\hat{\boldsymbol{t}}^+_{a\dot{b}}\,=\,\frac{1}{2}\left(\hat{\boldsymbol{t}}_{a\dot{b}}+\hat{\boldsymbol{t}}_{b\dot{a}}\right)\qquad\text{ and }\qquad \hat{\boldsymbol{t}}^-_{a\dot{b}}\,=\,\frac{1}{2}\left(\hat{\boldsymbol{t}}_{a\dot{b}}-\hat{\boldsymbol{t}}_{b\dot{a}}\right).
\end{equation}
It is now straightforward to check that symmetric generators $\hat{\boldsymbol{t}}^+_{a\dot{b}}$ are given by $\frac{1}{2}\left({\boldsymbol{t}^m}_n+{\boldsymbol{t}^n}_m\right)$ which are non-compact generators of $GL(5)$. Accordingly, the scalars corresponding to these generators parametrize the submanifold $GL(5)/SO(5)$. The antisymmetric generators $\hat{\boldsymbol{t}}^-_{a\dot{b}}$ correspond to the shift generators $\boldsymbol{s}_{mn}$. Therefore, the $25$ non-compact generators decompose into
\begin{equation}\label{15repscalarDEC}
\underbrace{25}_{\hat{\boldsymbol{t}}_{a\dot{b}}}\ \rightarrow\ \underbrace{1+14}_{\hat{\boldsymbol{t}}^+_{a\dot{b}}}\,+\underbrace{10}_{\boldsymbol{s}_{mn}}.
\end{equation}
\indent We can also separate the trace part of $\hat{\boldsymbol{t}}^+_{a\dot{b}}$, corresponding to the dilaton scalar field $\varphi$ in $GL(5)/SO(5)\sim \mathbb{R}^+\times SL(5)/SO(5)$ scalar coset. This generator is the $\mathbb{R}^+\sim SO(1,1)$ generator defined in \eqref{DefDOp}. In terms of $\hat{\boldsymbol{t}}^+_{a\dot{b}}$, this is given by
\begin{equation}\label{DefdilatonOp}
\boldsymbol{d}\,=\,\hat{\boldsymbol{t}}^+_{1\dot{1}}+\hat{\boldsymbol{t}}^+_{2\dot{2}}+\hat{\boldsymbol{t}}^+_{3\dot{3}}+\hat{\boldsymbol{t}}^+_{4\dot{4}}+\hat{\boldsymbol{t}}^+_{5\dot{5}}.
\end{equation}
The remaining generators can be identified as the fourteen non-compact generators corresponding to scalar fields $\{\phi_1,...,\phi_{14}\}$ in the $SL(5)/SO(5)$ coset. These generators are given by the symmetric traceless part
\begin{equation}
\tilde{\boldsymbol{t}}_{a\dot{b}}\,=\,\hat{\boldsymbol{t}}^+_{a\dot{b}}-\frac{1}{5}\boldsymbol{d}\,\delta_{a\dot{b}}
\end{equation}
satisfying $\delta^{a\dot{b}}\tilde{\boldsymbol{t}}_{a\dot{b}}=0$. 
\\
\indent The other ten scalars denoted by $\{\varsigma_1,...,\varsigma_{10}\}$ correspond to the shift generators $\boldsymbol{s}_{mn}$. These will be called the axions or shift scalars in this work. The decomposition in equation \eqref{15repscalarDEC} is in agreement with that in \cite{6D_SO(5)} in which the consistent circle reduction of seven-dimensional $SO(5)$ gauged supergravity giving rise to $SO(5)$ gauged theory in six dimensions is performed. From a higher-dimensional perspective, the fourteen scalars are the seven-dimensional scalars parameterizing the $SL(5)/SO(5)$ coset in seven dimensions while the dilaton and shift scalars descend from the reduction of seven-dimensional metric and vector fields, respectively, see appendix \ref{TrunAnsz} for more detail. 
\\
\indent By this decomposition of the scalar fields, we can rewrite the kinetic terms of the scalars in \eqref{bosonic_L} and obtain the following bosonic Lagrangian
\begin{equation}
e^{-1}\mathcal{L}=\frac{1}{4}R-G_{IJ}\partial_\mu\Phi^I\partial^\mu\Phi^J-\mathbf{V}\label{Ex_bosonic_L}
\end{equation}
in which $G_{IJ}$ is a symmetric matrix depending on scalar fields denoted by $\Phi^I=\{\varphi,\phi_1,\ldots,\phi_{14},\varsigma_1,\ldots,\varsigma_{10}\}$ with $I,J=1,\ldots,25$.
\\
\indent In order to find supersymmetric solutions, we consider first-order BPS equations derived from the supersymmetry transformations of fermionic fields in the background with vanishing fermionic fields. In this section, we only discuss a general structure of the procedure leaving a more detailed analysis and explicit results in subsequent sections. We begin with the variations of the gravitini which are given by
\begin{eqnarray} 
\delta \psi_{+\bar{\mu}\alpha}&:&\qquad A'\hat{\gamma}_{r}\epsilon_{+\alpha}+\frac{1}{2}gT_{\alpha\dot{\alpha}}\epsilon_-^{\dot{\alpha}}=0,\label{eq1_GBPS}\\
\delta \psi_{-\bar{\mu}\dot{\alpha}}&:&\qquad A'\hat{\gamma}_{r}\epsilon_{-\dot{\alpha}}-\frac{1}{2}gT_{\dot{\alpha}\alpha}\epsilon_+^{\alpha}=0\, .\label{eq2_GBPS}
\end{eqnarray}
In these equations, we have used the notation $A'=\frac{dA}{dr}$. We will use a prime to denote an $r$-derivative throughout the paper.
\\
\indent Multiply the first equation by $A'\hat{\gamma}_{r}$ and use the second equation or vice-versa, we find the following consistency conditions
\begin{eqnarray}
{A'}^2{\delta_\alpha}^\beta &=&\frac{1}{4}g^2T_{\alpha\dot{\alpha}}\Omega^{\dot{\alpha}\dot{\beta}}T_{\gamma\dot{\beta}}\Omega^{\beta\gamma}=\mc{W}^2{\delta_\alpha}^\beta,\\
{A'}^2{\delta_{\dot{\alpha}}}^{\dot{\beta}} &=&\frac{1}{4}g^2T_{\dot{\alpha}\alpha}\Omega^{\alpha\beta}T_{\beta\dot{\gamma}}\Omega^{\dot{\beta}\dot{\gamma}}
=\mc{W}^2{\delta_{\dot{\alpha}}}^{\dot{\beta}}
\end{eqnarray}
in which we have introduced the ``superpotential'' $\mc{W}$. We then obtain the BPS equations for the warped factor
\begin{equation}
A'=\pm\mathcal{W}\, .\label{Ap_eq}
\end{equation}
Using this result in equations \eqref{eq1_GBPS} and \eqref{eq2_GBPS} leads to the following projectors on the Killing spinors
\begin{equation}
\hat{\gamma}_{r}\epsilon_{+\alpha}=P_{\alpha\dot{\alpha}}\epsilon^{\dot{\alpha}}_-\qquad \textrm{and}\qquad \hat{\gamma}_{r}\epsilon_{-\dot{\alpha}}=P_{\dot{\alpha}\alpha}\epsilon^{\alpha}_+
\end{equation}
with 
\begin{equation}
P_{\alpha\dot{\alpha}}=-\frac{1}{2}g\frac{T_{\alpha\dot{\alpha}}}{A'}\qquad \textrm{and}\qquad P_{\dot{\alpha}\alpha}=\frac{1}{2}g\frac{T_{\dot{\alpha}\alpha}}{A'}
\end{equation}
satisfying ${P_\alpha}^{\dot{\alpha}}{P_{\dot{\alpha}}}^\beta={\delta_\alpha}^\beta$ and ${P_{\dot{\alpha}}}^\alpha{P_\alpha}^{\dot{\beta}}={\delta_{\dot{\alpha}}}^{\dot{\beta}}$. The conditions $\delta \psi_{+r\alpha}=0$ and $\delta\psi_{-r\dot{\alpha}}=0$ determine the Killing spinors as functions of the radial coordinate $r$ as usual.
\\
\indent Using these projectors in $\delta\chi_{+a\dot{\alpha}}=0$ and $\delta\chi_{-\dot{a}\alpha}=0$ equations, we eventually obtain the BPS equations for scalars. These equations are of the form 
\begin{equation}\label{BPSGenEq}
{\Phi^I}'=\mp 2G^{IJ}\frac{\partial\mathcal{W}}{\partial\Phi^J}
\end{equation}
in which $G^{IJ}$ is the inverse of the scalar metric $G_{IJ}$ defined in \eqref{Ex_bosonic_L}.
\\
\indent In addition, the scalar potential can also be written in term of $\mc{W}$ as
\begin{equation}\label{PopularSP}
\mathbf{V}=2G^{IJ}\frac{\partial\mathcal{W}}{\partial\Phi^I}\frac{\partial\mathcal{W}}{\partial\Phi^J}-5\mathcal{W}^2\, .
\end{equation}
It is well-known that the BPS equations of the form \eqref{Ap_eq} and \eqref{BPSGenEq} satisfy the second-order field equations derived from the bosonic Lagrangian \eqref{Ex_bosonic_L} with the scalar potential given by \eqref{PopularSP}, see \cite{SPotWay1, SPotWay2, SPotWay3, SPotWay4, SPotWay5, SPotWay6} for more detail.  
\\
\indent As in other dimensions, we will follow the approach introduced in \cite{New_Extrema} to explicitly find supersymmetric domain wall solutions involving only a subset of the $25$ scalars that is invariant under a particular subgroup $H_0\subset G_0$ to make the analysis more traceable. 

\subsection{$SO(5)$ symmetric domain walls}
We first consider supersymmetric domain walls with the maximal unbroken symmetry $SO(5)\subset CSO(p,q,5-p-q)$. The only gauge group containing $SO(5)$ as a subgroup is $SO(5)$ with $Y_{mn}=\delta_{mn}$. In this case, only the dilaton $\varphi$ corresponding to the non-compact generator \eqref{DefdilatonOp} is invariant under $SO(5)$. Thus, the coset representative can be written as
\begin{equation}\label{YSO(5)coset}
V=e^{\varphi\boldsymbol{d}}\, .
\end{equation}
We recall that this coset representative is a $16\times 16$ matrix with an index structure ${V_A}^B$. To compute the T-tensor, we need to write the $SO(5)\times SO(5)$ index as a pair of $SO(5)$ spinor indices resulting in the coset representative of the form ${V_A}^{\alpha\dot{\alpha}}$. To achieve this, we use the transformation matrices $\boldsymbol{p}$ introduced in \eqref{thepmatrix} so that ${V_A}^{\alpha\dot{\alpha}}$ and its inverse ${V^A}_{\alpha\dot{\alpha}}$ are given by
\begin{equation}
{V_A}^{\alpha\dot{\alpha}}={V_A}^B{\boldsymbol{p}_B}^{\alpha\dot{\alpha}}\qquad\text{and}
\qquad{V^A}_{\alpha\dot{\alpha}}={(V^{-1})_B}^A{\boldsymbol{p}^B}_{\alpha\dot{\alpha}}\, .
\end{equation}
With all these, it is now straightforward to find the T-tensor 
\begin{equation}\label{15repSO(5)T-tenor}
T^{\alpha\dot{\beta}}=\frac{5}{2\sqrt{2}}e^{\varphi}\, \Omega^{\alpha\beta}\delta^{\dot{\beta}}_\beta=\frac{2}{g}\mathcal{W}\, \Omega^{\alpha\beta}\delta^{\dot{\beta}}_\beta
\end{equation}
from which the superpotential is given by
\begin{equation}\label{YSO(5)SPot}
\mathcal{W}=\frac{5g}{4\sqrt{2}}e^{\varphi}.
\end{equation}
The resulting scalar potential reads
\begin{equation}\label{YSO(4)Pot}
\mathbf{V}=-\frac{15g^2}{4}e^{2\varphi}
\end{equation}
which does not admit any stationary points.
\\
\indent The general analysis given above leads to the BPS equation for the warped factor
\begin{equation}
A'=\frac{5g}{4\sqrt{2}}e^{\varphi}
\end{equation}
and the following projector
\begin{equation}\label{pureYProj}
\hat{\gamma}_r\epsilon_\pm=\epsilon_\mp\, .
\end{equation}
For definiteness, we have chosen a particular sign choice in the $A'$ equation and the $\hat{\gamma}_{r}$ projector. The condition $\delta \psi_{\pm r}=0$ gives the standard solution for the Killing spinors 
\begin{equation}
\epsilon_\pm=e^{\frac{A(r)}{2}}\epsilon_{\pm}^{0}\label{DW_Killing_spinor}	
\end{equation}	
with the constant spinors $\epsilon_{\pm}^{0}$ satisfying $\hat{\gamma}_r\epsilon_{\pm}^{0}=\epsilon_{\mp}^{0}$. Accordingly, the solution is half-supersymmetric. 
\\
\indent The BPS equation for the dilaton can be found from the condition $\delta\chi_\pm=0$ with the projector \eqref{pureYProj}. This results in a simple equation
\begin{equation}\label{YSO(5)BPS}
\varphi'=-\frac{g}{4\sqrt{2}}e^{\varphi}\, .
\end{equation}
All of these equations can be readily solved to obtain the solution
\begin{equation}\label{YSO(5)DW}
A=5\ln\left(\frac{gr}{4\sqrt{2}}-C\right)\qquad \textrm{and}\qquad  \varphi=-\ln\left(\frac{gr}{4\sqrt{2}}-C\right).
\end{equation}
The integration constant $C$ can be removed by shifting the radial coordinate $r$. We have also neglected an additive integration constant for $A$ since it can be absorbed by rescaling the coordinates $x^{\bar\mu}$. This is the $SO(5)$ domain wall originally found in \cite{6D_SO(5)}. In order to recover the same form of the solution, we redefine the radial coordinate as $r\rightarrow \frac{4\sqrt{2}}{g}\left[C+(3\sqrt{2}gr+C)^{-\frac{1}{24}}\right]$ and set $\varphi=\frac{1}{2\sqrt{10}}\sigma$.

\subsection{$SO(4)$ symmetric domain walls}\label{YSO(4)}
We now look for more complicated solutions with $SO(4)$ symmetry. The gauge groups that contain $SO(4)$ as a subgroup are $SO(5)$, $SO(4,1)$, and $CSO(4,0,1)$. To incorporate all of these gauge groups within a single framework, we write the embedding tensor in the form
\begin{equation}\label{YSO(4)YMN}
Y_{mn}=\text{diag}(1,1,1,1,\kappa)
\end{equation}
with $\kappa=1,0,-1$ corresponding to $SO(5)$, $CSO(4,0,1)$, and $SO(4,1)$ gauge groups, respectively. 

There are two $SO(4)$ singlet scalars. The first one is the dilaton corresponding to the non-compact generator \eqref{DefdilatonOp}, and the other one comes from the $SL(5)/SO(5)$ coset corresponding to the non-compact generator
\begin{equation}\label{YSO(4)Ys}
\mathcal{Y}=\hat{\boldsymbol{t}}^+_{1\dot{1}}+\hat{\boldsymbol{t}}^+_{2\dot{2}}+\hat{\boldsymbol{t}}^+_{3\dot{3}}+\hat{\boldsymbol{t}}^+_{4\dot{4}}-4\,\hat{\boldsymbol{t}}^+_{5\dot{5}}.
\end{equation}
Using the coset representative 
\begin{equation}
V=e^{\varphi\boldsymbol{d}+\phi\mathcal{Y}}\,,
\end{equation}
we find that the T-tensor is given by 
\begin{equation}\label{15repSO(4)T-tenor}
T^{\alpha\dot{\beta}}=\frac{1}{2\sqrt{2}}e^{\varphi-4\phi}(4+\kappa e^{20\phi})\, \Omega^{\alpha\beta}\delta^{\dot{\beta}}_\beta=\frac{2}{g}\mathcal{W}\, \Omega^{\alpha\beta}\delta^{\dot{\beta}}_\beta\, .
\end{equation}
This leads to the superpotential and the scalar potential of the form
\begin{eqnarray}
\mathcal{W}&=&\frac{g}{4\sqrt{2}}e^{\varphi-4\phi}(4+\kappa e^{20\phi}),\label{YSO(4)SPot}\\
\mathbf{V}&=&-\frac{g^2}{4}e^{2\varphi-8\phi}\left(8+8\kappa e^{20\phi}-\kappa^2e^{40\phi}\right).\label{YSO(4)Pot}
\end{eqnarray}
\indent Using the projector \eqref{pureYProj}, we find the BPS equations
\begin{eqnarray}
A'&=&\frac{g}{4\sqrt{2}}e^{\varphi-4\phi}(4+\kappa e^{20\phi}),\label{SO(4)BPS1}\\
\varphi'&=&-\frac{g}{20\sqrt{2}}e^{\varphi-4\phi}(4+\kappa e^{20\phi}),\\
\phi'&=&\frac{g}{5\sqrt{2}}e^{\varphi-4\phi}(1-\kappa e^{20\phi}).\label{SO(4)BPS3}
\end{eqnarray}
The resulting solutions for the dilation $\varphi$ and the warped factor $A$ as functions of $\phi$ are given by
\begin{eqnarray}
\varphi&=&-\phi+C+\frac{1}{16}\ln\left(1-\kappa e^{20\phi}\right),\\
A&=&-5\varphi\,=\,5\phi-5C-\frac{5}{16}\ln\left(1-\kappa e^{20\phi}\right).
\end{eqnarray}        
To obtain the solution for $\phi$, we change $r$ to a new radial coordinate $\rho$ defined by $\frac{d\rho}{dr}=e^{\varphi+6\phi}$. The solution for $\phi$ is then given by
\begin{equation}
e^{10\phi}=\frac{1}{\sqrt{\kappa}}\tanh\left[\sqrt{\kappa}(\sqrt{2}g\rho+C_1)\right]
\end{equation}        
for an integration constant $C_1$. It is useful to note that for $\kappa=-1$, the solution for $\phi$ can be written as
\begin{equation}
e^{10\phi}=\tan\left[\sqrt{2}g\rho+C_1\right].
\end{equation}        
For $\kappa=0$, the solution is simply given by 
\begin{equation}
e^{10\phi}=\sqrt{2}g\rho+C_1\, .
\end{equation}        

\subsection{$SO(3)\times SO(2)$ symmetric domain walls}
We now consider $SO(3)\times SO(2)$ residual symmetry, which is possible only for $SO(5)$ and $SO(3,2)$ gauge groups. In this case, we write the embedding tensor as
\begin{equation}
Y_{mn}=\textrm{diag}(1,1,1,\kappa,\kappa)
 \end{equation}
with $\kappa=1$ and $\kappa=-1$ corresponding to $SO(5)$ and $SO(3,2)$, respectively. The $SO(3)\times SO(2)$ symmetry is generated by $X_{ij}$, $i,j=1,2,3$, and $X_{45}$. There are three singlet scalars corresponding to the dilaton and the following non-compact generators 
\begin{equation}\label{YSO32noncom}
\mathcal{Y}_1=2\,\hat{\boldsymbol{t}}^+_{1\dot{1}}+2\,\hat{\boldsymbol{t}}^+_{2\dot{2}}+2\,\hat{\boldsymbol{t}}^+_{3\dot{3}}-3\,\hat{\boldsymbol{t}}^+_{4\dot{4}}-3\,\hat{\boldsymbol{t}}^+_{5\dot{5}},\qquad
\mathcal{Y}_2=\boldsymbol{s}_{45}\, .
\end{equation}
With the coset representative
\begin{equation}
V=e^{\varphi\boldsymbol{d}+\phi\mathcal{Y}_1+\varsigma\mathcal{Y}_2},
\end{equation}
we find the scalar potential
\begin{equation}\label{YSO3xSO2Pot}
\mathbf{V}=-\frac{3g^2}{4}e^{2(\varphi-8\phi)}(1+4\kappa e^{20\phi}).
\end{equation}
The superpotential reads
\begin{equation}\label{YSO3xSO2fullSPot}
\mathcal{W}=\frac{g}{4\sqrt{2}}e^{\varphi-8\phi}\sqrt{(3+2\kappa e^{20\phi})^2+8\kappa^2\varsigma^2e^{40\phi}}
\end{equation}
which can be found from the T-tensor given by
\begin{equation}\label{15repSO(3)xSO(20T-tenorAll}
T^{\alpha\dot{\beta}}=\frac{1}{2\sqrt{2}}e^{\varphi-8\phi}(3+2\kappa e^{20\phi})\, \Omega^{\alpha\beta}\delta^{\dot{\beta}}_\beta-\sqrt{2}\kappa\varsigma e^{\varphi+12\phi}\delta^{\alpha\dot{\beta}}.
\end{equation}
\indent In this case, it turns out that consistency of the supersymmetry conditions from $\delta\chi_{\pm}$ requires $\varsigma=0$. Therefore, in order to find a consistent set of BPS equations, we need to truncate the axion out. With $\varsigma=0$, the superpotential is given by
\begin{equation}
\mathcal{W}=\frac{g}{4\sqrt{2}}e^{\varphi-8\phi}(3+2\kappa e^{20\phi}).
\end{equation}
With the projector \eqref{pureYProj}, we find the following BPS equations
\begin{eqnarray}
A'&=&\frac{g}{4\sqrt{2}}e^{\varphi-8\phi}(3+2\kappa e^{20\phi}),\\
\varphi'&=&-\frac{g}{20\sqrt{2}}e^{\varphi-8\phi}(3+2\kappa e^{20\phi}),\\
\phi'&=&\frac{g}{5\sqrt{2}}e^{\varphi-8\phi}(1-\kappa e^{20\phi}).
\end{eqnarray}
It can be verified that all these equations satisfy the corresponding field equations as expected.         

With a new radial coordinate $\rho$ given by $\frac{d\rho}{dr}=e^{\varphi+2\phi}$, we obtain the domain wall solution
\begin{eqnarray}
\varphi&=&-\frac{3\phi}{4}+C+\frac{1}{16}\ln\left(1-\kappa e^{20\phi}\right),\\
A&=&-5\varphi\,=\,\frac{15\phi}{4}-5C-\frac{5}{16}\ln\left(1-\kappa e^{20\phi}\right),\\
e^{10\phi}&=&\frac{1}{\sqrt{\kappa}}\tanh\left[\sqrt{\kappa}(\sqrt{2}g\rho+C_1)\right].
\end{eqnarray}

\subsection{$SO(3)$ symmetric domain walls}
We now move to domain wall solutions with $SO(3)$ symmetry. Many gauge groups contain $SO(3)$ as a subgroup with the embedding tensor parameterized by
\begin{equation}
Y_{mn}=\text{diag}(1,1,1,\kappa,\lambda)
\end{equation}
for $\kappa,\lambda=0,\pm1$. With this embedding tensor, the $SO(3)$ symmetry is generated by $X_{mn}$, $m,n=1,2,3$. In addition to the dilaton, there are four singlet scalars corresponding to the following non-compact generators
\begin{eqnarray}
\mathcal{Y}_1&=&2\,\hat{\boldsymbol{t}}^+_{1\dot{1}}+2\,\hat{\boldsymbol{t}}^+_{2\dot{2}}+2\,\hat{\boldsymbol{t}}^+_{3\dot{3}}-3\,\hat{\boldsymbol{t}}^+_{4\dot{4}}-3\,\hat{\boldsymbol{t}}^+_{5\dot{5}},\qquad\mathcal{Y}_2\,=\,\hat{\boldsymbol{t}}^+_{4\dot{5}},\nonumber\\\mathcal{Y}_3&=&\hat{\boldsymbol{t}}^+_{4\dot{4}}-\hat{\boldsymbol{t}}^+_{5\dot{5}},\qquad\quad \mathcal{Y}_4\,=\,\boldsymbol{s}_{45}\, .\label{YSO(3)Ys}
\end{eqnarray}
With the only exception for $\kappa=\lambda=0$ corresponding to $CSO(3, 0, 2)$ gauge group, we need to truncate out the scalar corresponding to $\boldsymbol{s}_{45}$ generator in order to find a consistent set of BPS equations as in the previous case. For the moment, we will set this shift scalar to zero and consider the particular case of $\kappa=\lambda=0$ afterward.

For vanishing shift scalars, the coset representative is given by
\begin{equation}
V=e^{\varphi\boldsymbol{d}+\phi_1\mathcal{Y}_1+\phi_2\mathcal{Y}_2+\phi_3\mathcal{Y}_3}
\end{equation}
giving rise to the superpotential and the scalar potential of the form
\begin{eqnarray}
\mathcal{W}&=&\frac{ge^{\varphi-8\phi_1}}{4\sqrt{2}}\left[3+e^{20\phi_1}\left((\kappa+\lambda)\cosh{2\phi_2}\cosh{4\phi_3}-(\kappa-\lambda)\sinh{4\phi_3}\right)\right],\qquad\label{15genSO(3)SPot}\\
\mathbf{V}&=&-\frac{g^2e^{2(\varphi-8\phi_1)}}{4}\left[3+6e^{20\phi_1}\left((\kappa+\lambda)\cosh{2\phi_2}\cosh{4\phi_3}-(\kappa-\lambda)\sinh{4\phi_3}\right)\phantom{\frac{1}{2}}\right.\nonumber\\
&&\left.+\frac{e^{40\phi_1}}{4}\left(\kappa^2+10\kappa\lambda+\lambda^2-(3\kappa^2-2\kappa\lambda+3\lambda^2)\cosh{8\phi_3}\phantom{\frac{1}{2}}\right.\right.\nonumber\\
&&\left.\left.\phantom{\frac{1}{2}}-2(\kappa+\lambda)^2\cosh{4\phi_2}\cosh^2{4\phi_3}+4(\kappa^2-\lambda^2)\cosh{2\phi_2}\sinh{8\phi_3}\right)\right].\label{15genSO(3)Pot}
\end{eqnarray}
We also note the matrix $G^{IJ}$ in this case
\begin{equation}
G^{IJ}=\frac{1}{60}\text{diag}(6,1,60\text{sech}^2{4\phi_3},15)
\end{equation}
for $\Phi_I=\{\varphi,\phi_1,\phi_2,\phi_3\}$ with $I,J=1,2,3,4$. 

In this case, the Killing spinors are different from the ansatz given in \eqref{DW_Killing_spinor} due to the non-vanishing composite connections $Q_r^{45}$ and $Q_r^{\dot{4}\dot{5}}$ appearing in $\delta \psi_{\pm r}=0$ conditions. In more detail, there are additional terms involving ${(\gamma_{45})_\alpha}^\beta\epsilon_{+\beta}$ and ${(\gamma_{\dot{4}\dot{5}})_{\dot{\alpha}}}^{\dot{\beta}}\epsilon_{-\dot{\beta}}$ in the covariant derivative of the supersymmetry parameters, see equations \eqref{CoDivEp+} and \eqref{CoDivEp-}. According to this, we modify the ansatz for the Killing spinors to
\begin{equation}
\epsilon_+=e^{\frac{A(r)}{2}+B(r)\gamma_{45}}\epsilon_{+}^{0}\qquad\text{ and }\qquad\epsilon_-=e^{\frac{A(r)}{2}+B(r)\gamma_{\dot{4}\dot{5}}}\epsilon_{-}^{0}\label{SO(3)DW_Killing_spinor}	
\end{equation}
where $B(r)$ is an $r$-dependent function, and $\epsilon_{\pm}^{0}$ are constant symplectic-Majorana-Weyl spinors satisfying $\hat{\gamma}_r\epsilon_{\pm}^{0}=\epsilon_{\mp}^{0}$. 

Using this ansatz for the Killing spinors satisfying the projector \eqref{pureYProj}, we find the following set of BPS equations from the supersymmetry transformations of fermions
\begin{eqnarray}
A'&=&\frac{ge^{\varphi-8\phi_1}}{4\sqrt{2}}\left[3+e^{20\phi_1}\left((\kappa+\lambda)\cosh{2\phi_2}\cosh{4\phi_3}-(\kappa-\lambda)\sinh{4\phi_3}\right)\right],\label{GEN15SO(2)BPSeq1}\\
\varphi'&=&-\frac{ge^{\varphi-8\phi_1}}{20\sqrt{2}}\left[3+e^{20\phi_1}\left((\kappa+\lambda)\cosh{2\phi_2}\cosh{4\phi_3}-(\kappa-\lambda)\sinh{4\phi_3}\right)\right],\qquad\ \, \\
\phi'_1&=&\frac{ge^{\varphi-8\phi_1}}{10\sqrt{2}}\left[2-e^{20\phi_1}\left((\kappa+\lambda)\cosh{2\phi_2}\cosh{4\phi_3}-(\kappa-\lambda)\sinh{4\phi_3}\right)\right],\\
\phi'_2&=&-\frac{g}{\sqrt{2}}e^{\varphi+12\phi_1}(\kappa+\lambda)\sinh{2\phi_2}\,\text{sech }{4\phi_3},\\
\phi'_3&=&\frac{ge^{\varphi+12\phi_1}}{2\sqrt{2}}\left((\kappa-\lambda)\cosh{4\phi_3}-(\kappa+\lambda)\cosh{2\phi_2}\sinh{4\phi_3}\right)\label{GEN15SO(2)BPSeq5}
\end{eqnarray}        
together with 
\begin{equation}
B'=-\frac{g}{2\sqrt{2}}e^{\varphi+12\phi_1}(\kappa+\lambda)\sinh{2\phi_2}\tanh{4\phi_3}.
\end{equation}
To find explicit solutions, we will separately discuss various possible values of $\kappa$ and $\lambda$.

\subsubsection{Domain walls in $CSO(4,0,1)$ and $CSO(3,1,1)$ gauge groups}
For $\lambda=0$ and $\kappa\neq 0$, the gauge groups are given by $CSO(4,0,1)$ and $CSO(3,1,1)$ for $\kappa=1$ and $\kappa=-1$, respectively. Using a new radial coordinate $\rho$ defined by $\frac{d\rho}{dr}=e^{\varphi+12\phi_1}$, a domain wall solution to the BPS equations can be obtained 
\begin{eqnarray}
\phi_2&=&\frac{1}{4}\ln\left[\frac{g^2\rho^2(1+2C_3)^2+2(1+C_3)^2}{g^2\rho^2(1+2C_3)^2+2C_3^2}\right],\label{SO3_DW_CSO41_1}\\
\phi_3&=&\frac{1}{8}\ln\left[\frac{e^{2\phi_2}-C_3e^{4\phi_2}+C_3+1}{e^{2\phi_2}+C_3e^{4\phi_2}-C_3-1}\right],\label{SO3_DW_CSO41_2}\\
\phi_1&=&\frac{1}{20}\ln\left[\frac{\kappa\left(2+C_1(e^{4\phi_2}-1)\right)}{\sqrt{(1-e^{4\phi_2})\left(C_3^2e^{4\phi_2}-(C_3+1)^2\right)}}\right],\label{SO3_DW_CSO41_3}\\
\varphi&=&\frac{\phi_1}{2}+C-\frac{1}{16}\ln\left[\frac{C_1(e^{4\phi_2}-1)+2}{e^{4\phi_2}-1}\right],\label{SO3_DW_CSO41_4}\\
A&=&-5\varphi\,=\,-\frac{5\phi_1}{2}-5C+\frac{5}{16}\ln\left[\frac{C_1(e^{4\phi_2}-1)+2}{e^{4\phi_2}-1}\right]\label{SO3_DW_CSO41_5}
\end{eqnarray}
together with
\begin{equation}
B=\frac{1}{4}\sin^{-1}\left[C_3\sqrt{\frac{e^{4\phi_2}-1}{2C_3+1}}\,\right]+\frac{1}{4}\tan^{-1}\left[\sqrt{\frac{(1-e^{4\phi_2})(C_3+1)^2}{C_3^2e^{4\phi_2}-(C_3+1)^2}}\,\right].
\end{equation}
We have chosen integration constants for $\phi_2$ and $B$ to be zero for simplicity.

\subsubsection{Domain walls in $SO(4,1)$ gauge group}
In $SO(4,1)$ gauge group with $\kappa=-\lambda=1$, the BPS equations give $\phi_2'=B'=0$. Accordingly, we can set $B=0$ and $\phi_2=0$. We can readily verify that this is a consistent truncation. Taking $\phi_2=0$ and redefining the radial coordinate $r$ to $\rho$ as given in the $CSO(4,0,1)$ and $CSO(3,1,1)$ gauge groups, we obtain a domain wall solution 
\begin{eqnarray}
\phi_3&=&\frac{1}{2}\tanh^{-1}\left[\tan\left[\sqrt{2}g\rho+C_3\right]\right],\\
\phi_1&=&\frac{1}{20}\ln\left[e^{4\phi_3}(C_1+1)+C_1e^{-4\phi_3}\right],\\
\varphi&=&\frac{\phi_1}{2}+C+\frac{1}{16}\ln\left[\frac{e^{8\phi_3}+1}{2\left(C_1+e^{8\phi_3}(C_1+1)\right)}\right],\\
A&=&-5\varphi\,=\,-\frac{5\phi_1}{2}-5C-\frac{5}{16}\ln\left[\frac{e^{8\phi_3}+1}{2\left(C_1+e^{8\phi_3}(C_1+1)\right)}\right].
\end{eqnarray}

\subsubsection{Domain walls in $SO(5)$ and $CSO(3,2)$ gauge groups}
For $\kappa=\lambda=\pm 1$ corresponding to $SO(5)$ and $SO(3,2)$ gauge groups. we find the following domain wall solution
\begin{eqnarray}
\phi_2&=&\frac{1}{4}\ln \left[\frac{1-2e^{2\sqrt{2}g\kappa\rho}+e^{4\sqrt{2}g\kappa\rho}+e^{4\sqrt{2}g\kappa\rho+2C_3}}
{1+2e^{2\sqrt{2}g\kappa\rho}+e^{4\sqrt{2}g\kappa\rho}+e^{4\sqrt{2}g\kappa\rho+2C_3}}\right],\\
\phi_3&=&\frac{1}{8}\ln\left[\frac{2e^{2\phi_2}+e^{4\phi_2+C_3}-e^{C_3}}{2e^{2\phi_2}-e^{4\phi_2+C_3}+e^{C_3}}\right],\\
\phi_1&=&\frac{\phi_2}{10}+\frac{1}{20}\ln\left[\frac{\kappa e^{-2\phi_2}\left(C_1(e^{4\phi_2}-1)-2\right)}{\sqrt{4e^{4\phi_2}-e^{2C_3}(e^{4\phi_2}-1)^2}}\right],\\
\varphi&=&\frac{\phi_1}{2}+C-\frac{1}{16}\ln\left[e^{4\phi_2}-1\right]+\frac{1}{16}\ln[C_1(e^{4\phi_2}-1)-2],\\
A&=&-5\varphi\,=\,-\frac{5\phi_1}{2}-5C+\frac{5}{16}\ln\left[e^{4\phi_2}-1\right]+\frac{1}{16}\ln[C_1(e^{4\phi_2}-1)-2]\qquad
\end{eqnarray}
in terms of the new radial coordinate $\rho$ defined previously. The function $B(r)$ appearing in the Killing spinors is given in term of $\phi_2$ as
\begin{equation}
B=-\frac{1}{8}\tan^{-1}\left[\frac{1-e^{4\phi_2}+2e^{-2C_3}}{\sqrt{4e^{4\phi_2-2C_3}-(e^{4\phi_2}-1)^2}}\right]-\frac{1}{8}\tan^{-1}\left[\frac{e^{4\phi_2}(1+2e^{-2C_3})-1}{\sqrt{4e^{4\phi_2-2C_3}-(e^{4\phi_2}-1)^2}}\right]
\end{equation}
in which the integration constant has been set to zero.

\subsubsection{Domain walls in $CSO(3,0,2)$ gauge group}\label{GENCSO302Soln}
In the case of $CSO(3,0,2)$ gauge group with $\kappa=\lambda=0$, supersymmetry allows a non-vanishing axion corresponding to $\mathcal{Y}_4$ generator. We write the coset representative as
\begin{equation}
V=e^{\varphi\boldsymbol{d}+\phi_1\mathcal{Y}_1+\phi_2\mathcal{Y}_2+\phi_3\mathcal{Y}_3+\varsigma\mathcal{Y}_4}
\end{equation}
and find a simple scalar potential
\begin{equation}\label{SPSO(3)POT}
\mathbf{V}=-\frac{3g^2}{4}e^{2(\varphi-8\phi_1)}\, .
\end{equation}
We also note that this potential does not depend on $\varsigma$ and can be obtained from \eqref{15genSO(3)Pot} by setting $\kappa=\lambda=0$. This potential can also be written in the form \eqref{PopularSP} using the superpotential
\begin{equation}
\mathcal{W}=\frac{3g}{4\sqrt{2}}e^{\varphi-8\phi_1}
\end{equation}
and the symmetric matrix
\begin{equation}
G^{IJ}=\begin{pmatrix}	\frac{1}{10} & 0 & 0 & 0 & \frac{2\varsigma}{5} \\ 0 & \frac{1}{60} & 0 & 0 & -\frac{\varsigma}{5} \\  0 & 0 & \text{sech}^2{4\phi_3} & 0 & 0 \\ 0 & 0 & 0 & \frac{1}{4} & 0 \\ \frac{2\varsigma}{5} & -\frac{\varsigma}{5} & 0 & 0 & 1+4\varsigma^2 \end{pmatrix}
\end{equation}
for $\Phi^I=\{\varphi,\phi_1,\phi_2,\phi_3,\varsigma\}$, $I,J=1,2,3,4,5$. 

With all these and the usual ansatz for the Killing spinors \eqref{DW_Killing_spinor} together with the projector \eqref{pureYProj}, we find the BPS equations 
\begin{eqnarray}
A'&=&\frac{3g}{4\sqrt{2}}e^{\varphi-8\phi_1},\quad
\varphi'=-\frac{3g}{20\sqrt{2}}e^{\varphi-8\phi_1},\quad
\phi'_1=\frac{g}{5\sqrt{2}}e^{\varphi-8\phi_1},\nonumber\\
\phi'_2&=&\phi'_3\ =\ 0,\qquad\ \,
\varsigma'=-\frac{3g}{\sqrt{2}}e^{\varphi-8\phi_1}\varsigma\, .
\end{eqnarray}
Except for an additional equation for $\varsigma$, these are the BPS equations obtained from \eqref{GEN15SO(2)BPSeq1} to \eqref{GEN15SO(2)BPSeq5} by setting $\kappa=\lambda=0$. Furthermore, $\phi_2$ and $\phi_3$ can be consistently truncated out since the scalar potential \eqref{SPSO(3)POT} is independent of $\phi_2$ and $\phi_3$. 
\\
\indent With all these, we find a domain wall solution
\begin{eqnarray}
\phi_1&=&\frac{1}{4}\ln\left[\frac{2}{5}(\sqrt{2}g\rho+C_1)\right],\label{GENCSO302Soln1}\\
\varphi&=&C-\frac{1}{8}\ln\left[\frac{2}{5}(\sqrt{2}g\rho+C_1)\right],\\
\varsigma&=&C_4e^{-15\phi_1}=\frac{C_4}{\left(\frac{2}{5}(\sqrt{2}g\rho+C_1)\right)^{\frac{15}{4}}},\\
A&=&-5\varphi\,=\,-5C+\frac{5}{8}\ln\left[\frac{2}{5}(\sqrt{2}g\rho+C_1)\right]\label{GENCSO302Soln3}
\end{eqnarray}
in which $\rho$ is a new radial coordinate defined by $\frac{d\rho}{dr}=e^{\varphi-4\phi_1}$. It should also be noted that the axion $\varsigma$ can also be truncated out. 
 
\subsection{$SO(2)\times SO(2)$ symmetric domain walls}\label{YSO(2)xSO(2)Sec}
As a final example of domain wall solutions in $\mathbf{15}^{-1}$ representation, we consider an $SO(2)\times SO(2)$ unbroken symmetry. In this case, the embedding tensor for all possible gauge groups takes the form
\begin{equation}\label{SO3xSO2Y}
Y_{mn}=\textrm{diag}(1,1,\kappa,\kappa,\lambda)
\end{equation} 
for $\lambda=0,\pm 1$ and $\kappa=\pm 1$. These gauge groups are $SO(5)$ ($\kappa=\lambda=1$), $SO(4,1)$ ($\kappa=-\lambda=1$), $SO(3,2)$ ($\kappa=-\lambda=-1$), $CSO(4,0,1)$ ($\kappa=1,\lambda=0$), and $CSO(2,2,1)$ ($\kappa=-1,\lambda=0$).

There are five scalars invariant under $SO(2)\times SO(2)$ generated by $X_{12}$ and $X_{34}$. As usual, one of these is the dilaton and the other four are associated with the following non-compact generators
\begin{equation}
\overline{\mathcal{Y}}_1=\hat{\boldsymbol{t}}^+_{1\dot{1}}+\hat{\boldsymbol{t}}^+_{2\dot{2}}-2\,\hat{\boldsymbol{t}}^+_{5\dot{5}},\qquad\overline{\mathcal{Y}}_2=\hat{\boldsymbol{t}}^+_{3\dot{3}}+\hat{\boldsymbol{t}}^+_{4\dot{4}}-2\,\hat{\boldsymbol{t}}^+_{5\dot{5}},\qquad\overline{\mathcal{Y}}_3=\boldsymbol{s}_{12},\qquad\overline{\mathcal{Y}}_4=\boldsymbol{s}_{34}\, .
\end{equation}
As in many previous cases, we need to truncate out the axions corresponding to the shift generators $\boldsymbol{s}_{12}$ and $\boldsymbol{s}_{34}$ in order to find a consistent set of BPS equations that are compatible with the field equations. We then take the coset representative of the form
\begin{equation}\label{YSO(2)SO(2)coset}
V=e^{\varphi\boldsymbol{d}+\phi_1\overline{\mathcal{Y}}_1+\phi_2\overline{\mathcal{Y}}_2}\, .
\end{equation}
The resulting scalar potential reads
\begin{equation}\label{YSO(2)xSO(2)PotSub}
\mathbf{V}=-\frac{g^2}{4}e^{2\left(\varphi-2(\phi_1+\phi_2)\right)}\left[4\kappa(2+\lambda e^{12\phi_1+8\phi_2})+\lambda e^{8\phi_1+12\phi_2}(4-\lambda e^{12\phi_1+8\phi_2})\right]
\end{equation}
which can be written in terms of the superpotential 
\begin{equation}
\mathcal{W}=\frac{g}{4\sqrt{2}}e^{\varphi}(2e^{-4\phi_1}+2\kappa e^{-4\phi_2}+\lambda e^{8(\phi_1+\phi_2)})
\end{equation}
using
\begin{equation}
G^{IJ}=\begin{pmatrix}	\frac{1}{10} & 0 & 0\\ 0 & \frac{3}{20} & -\frac{1}{10} \\  0 & -\frac{1}{10} & \frac{3}{20} \end{pmatrix}
\end{equation}
for $\Phi^I=\{\varphi,\phi_1,\phi_2\}$, $I,J=1,2,3$. 

Using the projector \eqref{pureYProj} together with the Killing spinors \eqref{DW_Killing_spinor}, we find the following BPS equations 
\begin{eqnarray}
A'&=&\frac{g}{4\sqrt{2}}e^{\varphi}(2e^{-4\phi_1}+2\kappa e^{-4\phi_2}+\lambda e^{8(\phi_1+\phi_2)}),\label{SO(2)xSO(2)BPS1}\\
\varphi'&=&-\frac{g}{20\sqrt{2}}e^{\varphi}(2e^{-4\phi_1}+2\kappa e^{-4\phi_2}+\lambda e^{8(\phi_1+\phi_2)}),\\
\phi'_1&=&\frac{g}{5\sqrt{2}}e^{\varphi}(3e^{-4\phi_1}-2\kappa e^{-4\phi_2}-\lambda e^{8(\phi_1+\phi_2)}),\\
\phi'_2&=&\frac{g}{5\sqrt{2}}e^{\varphi}(3\kappa e^{-4\phi_2}-2e^{-4\phi_1}-\lambda e^{8(\phi_1+\phi_2)}).\label{SO(2)xSO(2)BPS4}
\end{eqnarray}        
Solving these BPS equations gives a domain wall solution 
\begin{eqnarray}\label{YSO(2)xSO(2)DW}
\phi_2&=&-\frac{3\sqrt{2}g\rho}{10}+\frac{3}{20}\ln\left[\kappa e^{2\sqrt{2}g\rho}+C_2\right]-\frac{1}{20}\ln\left[C_1e^{-2\sqrt{2}g\rho}+\lambda\right],\\
\phi_1&=&-\frac{2\phi_2}{3}-\frac{1}{12}\ln\left[C_1e^{-2\sqrt{2}g\rho}+\lambda\right],\\
\varphi&=&-\frac{\phi_2}{6}-\frac{3\sqrt{2}g\rho}{24}+C-\frac{1}{48}\ln\left[C_1e^{-2\sqrt{2}g\rho}+\lambda\right],\\
A&=&-5\varphi\,=\,\frac{5\phi_2}{6}+\frac{15\sqrt{2}g\rho}{24}-5C+\frac{5}{48}\ln\left[C_1e^{-2\sqrt{2}g\rho}+\lambda\right]
\end{eqnarray}
in which $\rho$ is the new radial coordinate defined by the relation $\frac{d\rho}{dr}=e^{\varphi-4\phi_1}$.

For domain walls preserving smaller residual symmetries such as $SO(2)_{\text{diag}}\subset SO(2)\times SO(2)$ and $SO(2)$, there are many more scalars, and the analysis is much more involved without any possibility for complete analytic solutions. We will not consider these cases in this work.

\section{Domain walls from gaugings in $\overline{\mathbf{40}}^{-1}$ representation}\label{pure40Sec}
In this section, we consider gaugings in which the irreducible part of the embedding tensor transforms in $\overline{\mathbf{40}}^{-1}$ representation. These gauged theories are obtained from a consistent circle reduction of the maximal seven-dimensional $CSO(p,q,4-p-q)$ gauged supergravity constructed in \cite{7D_Max_Gauging}. 

In six dimensions, gaugings in $\overline{\mathbf{40}}^{-1}$ representation are purely electric and triggered by
\begin{equation}
\theta^{Am}\ = \ \mathbb{T}^A_{np}U^{np,m}
\end{equation}
where $U^{mn,p}=U^{[mn],p}$ satisfying $U^{[mn,p]}=0$. With $\theta^{AM}=(\,\mathbb{T}^A_{np}U^{np,m}\,,\,0\,)$, the second condition from the quadratic constraint \eqref{QC} reduces to
\begin{equation}\label{40QC}
U^{mn,r}U^{pq,s}\varepsilon_{mnpqt}\ = \ 0\, .
\end{equation}
This condition can be solved by setting
\begin{equation}
U^{mn,p}\ =\ v^{[m}w^{n]p}
\end{equation}
in which $v^m$ is a $GL(5)$ vector and $w^{mn}$ is a symmetric tensor, $w^{mn}=w^{(mn)}$. 

To classify possible gauge groups, we follow \cite{7D_Max_Gauging} by using the $SL(5)$ symmetry to further fix $v^m=\delta^m_5$ and split the index $m=(i,5)$, $i=1,..,4$. For simplicity, we also restrict to cases with $w^{i5}=w^{55}=0$. The remaining $SL(4)$ residual symmetry can be used to diagonalize the $4\times4$ block corresponding to $w^{ij}$ as
\begin{equation}
w^{ij}\ =\ \text{diag}(\underbrace{1,..,1}_p,\underbrace{-1,..,-1}_q,\underbrace{0,..,0}_r)
\end{equation}
with $p+q+r=4$. From the decomposition in \eqref{GaugeGenSplit}, we find that in this case, only $X_{ij}$ and $X^i$ gauge generators are non-vanishing. The generators $X_{ij}$ are given in terms of the $GL(5)$ generators while $X^i$ only involve the shift generators. Explicitly, these generators are given by 
\begin{equation}\label{GenCSO(p,q,4-p-q)X}
X_{ij}=\frac{1}{\sqrt{2}}\varepsilon_{ijkm}w^{kl}{\boldsymbol{t}^m}_l\qquad\textrm{and}\qquad X^i=w^{ij}\boldsymbol{s}_{5j}\, .
\end{equation}
\indent It is now straightforward to show that the gauge generators satisfy the following commutation relations
\begin{equation}
[X^i,X^j]=0,\qquad [X_{ij},X^k]={(X_{ij})_l}^kX^l,\qquad [X_{ij},X_{kl}]={(X_{ij})_{kl}}^{mn}X_{mn}
\end{equation}
in which ${(X_{ij})_{kl}}^{mn}=2{(X_{ij})_{[k}}^{[m}\delta_{l]}^{n]}$. This implies that the corresponding gauge group is of the form
\begin{equation}\label{Zgaugegroup}
G_0=CSO(p,q,4-p-q)\ltimes \mathbb{R}^{4}_{\boldsymbol{s}}=SO(p,q) \ltimes \left(\mathbb{R}^{(p+q)(4-p-q)}\times\mathbb{R}^{4}_{\boldsymbol{s}}\right).
\end{equation}
The $CSO(p,q,4-p-q)$ factor and the four-dimensional translation group from the shift symmetries $\mathbb{R}^{4}_{\boldsymbol{s}}$ are respectively generated by $X_{ij}$ and $X^i$. 

We should note here that the corresponding gauge group in seven dimensions is just $CSO(p,q,4-p-q)$. After an $S^1$ reduction, this gauge group is accompanied by a translation group $\mathbb{R}^{4}_{\boldsymbol{s}}$. As pointed out in \cite{6D_Max_Gauging}, the complete off-shell symmetry group of the maximal six-dimensional gauged supergravity is $GL(5)\ltimes\mathbf{10}^{-4}$, with $\mathbf{10}^{-4}$ being shift symmetries of scalar fields. The gauge group given in \eqref{Zgaugegroup} is embedded in $GL(5)\ltimes\mathbf{10}^{-4}$ as $CSO(p,q,4-p-q)\subset GL(5)$ and $\mathbb{R}^{4}_{\boldsymbol{s}}\subset \mathbf{10}^{-4}$. We also note that in vector representation of $SO(5,5)$, the gauge generators are given by 
\begin{equation}\label{CSO(p,q,4-p-q)X}
{(X^i)_P}^Q=4\eta_{P[j}\delta^Q_{k]}\delta^j_5w^{ki},\qquad
{(X_{ij})_P}^Q=\sqrt{2}\varepsilon_{ijkm}w^{kl}(\delta_P^m\delta^Q_l-\eta^{mP}\eta_{lP}).
\end{equation}
\indent By splitting the $SO(5)\times SO(5)$ vector indices as $a=(i,5)$ and $\dot{a}=(\dot{i},\dot{5})$, we find the following decomposition for non-compact generators of $SL(5)\subset GL(5)\subset SO(5,5)$ under $SL(5)\rightarrow SL(4)\times SO(1,1)$ 
\begin{equation}
\tilde{\boldsymbol{t}}_{a\dot{b}}\rightarrow\left(\tilde{\boldsymbol{t}}_{i\dot{j}},\,\tilde{\boldsymbol{t}}_{i\dot{5}},\,\tilde{\boldsymbol{t}}_{5\dot{5}}\right).
\end{equation}
Since the $SL(5)$ generators $\tilde{\boldsymbol{t}}_{a\dot{b}}$ are traceless, the generator $\tilde{\boldsymbol{t}}_{5\dot{5}}$ is related to the trace part of $\tilde{\boldsymbol{t}}_{i\dot{j}}$ according to $\tilde{\boldsymbol{t}}_{1\dot{1}}+\tilde{\boldsymbol{t}}_{2\dot{2}}+\tilde{\boldsymbol{t}}_{3\dot{3}}+\tilde{\boldsymbol{t}}_{4\dot{4}}=-\tilde{\boldsymbol{t}}_{5\dot{5}}$. It is then convenience to define new non-compact generators $\overline{\boldsymbol{t}}_{i\dot{j}}$ as
\begin{equation}
\overline{\boldsymbol{t}}_{i\dot{j}}=\tilde{\boldsymbol{t}}_{i\dot{j}}+\frac{1}{4}\tilde{\boldsymbol{t}}_{5\dot{5}}\delta_{i\dot{j}}
\end{equation}
which are symmetric traceless. The nine scalar fields corresponding to these generators then parametrize an $SL(4)/SO(4)$ coset. The other four scalars associated with $\tilde{\boldsymbol{t}}_{i\dot{5}}=\hat{\boldsymbol{t}}^+_{i\dot{5}}$ are nilpotent scalars and will be denoted by $b_i$ as in seven dimensions. In addition, there are also ten axions corresponding to the antisymmetric shift generators as in the previous section. 

As in the previous section, we will systematically find supersymmetric domain walls invariant under some residual symmetries of the $CSO(p,q,4-p-q)$ factor in the gauge group.

\subsection{$SO(4)$ symmetric domain walls}\label{ZSO(4)}
We first consider domain walls with the largest possible unbroken symmetry namely $SO(4)$. The only gauge group containing $SO(4)$ as a subgroup is $SO(4)\ltimes \mathbb{R}^{4}_{\boldsymbol{s}}$ with the embedding tensor parametrized by $w^{ij}=\delta^{ij}$. The $SO(4)$ symmetry is generated by $X_{ij}$, $i,j=1,2,3,4$, generators.

There are two $SO(4)$ singlet scalars given by the dilaton $\varphi$ and another dilatonic scalar corresponding to the $SO(1,1)$ factor in $SL(4)\times SO(1,1)\subset SL(5)$. The latter is given by the non-compact generator 
\begin{equation}\label{40repDil}
\widetilde{\mathcal{Y}}_0=\hat{\boldsymbol{t}}^+_{1\dot{1}}+\hat{\boldsymbol{t}}^+_{2\dot{2}}+\hat{\boldsymbol{t}}^+_{3\dot{3}}+\hat{\boldsymbol{t}}^+_{4\dot{4}}-4\,\hat{\boldsymbol{t}}^+_{5\dot{5}}
\end{equation}
and will be denoted by $\phi_0$.

The coset representative can be written as 
\begin{equation}
V=e^{\varphi\boldsymbol{d}+\phi_0\widetilde{\mathcal{Y}}_0}
\end{equation}
leading to the T-tensor given by
\begin{equation}\label{40repSO(4)T-tenor}
T^{\alpha\dot{\beta}}=e^{\varphi-4\phi_0}\, (\gamma^5)^{\alpha\beta}\delta^{\dot{\beta}}_\beta=\frac{2}{g}\mathcal{W}\, (\gamma^5)^{\alpha\beta}\delta^{\dot{\beta}}_\beta\,
\end{equation}
with the superpotential 
\begin{equation}
\mathcal{W}=\frac{g}{2}e^{\varphi-4\phi_0}\, .
\end{equation}
The appearance of $\gamma^5$ rather than other $SO(5)$ gamma matrices is due to the specific choice of $v^m=\delta^m_5$ for the tensor $U^{mn,p}$. The scalar potential can also be directly computed and is given by
\begin{equation}\label{ZSO(4)Pot}
\mathbf{V}=-g^2e^{2\varphi-8\phi_0}\, .
\end{equation}
\indent The Killing spinors are given by the same ansatz as in \eqref{DW_Killing_spinor} but in this case subject to the following projector
\begin{equation}\label{pureZProj}
\hat{\gamma}_r\epsilon_\pm=\gamma^5\epsilon_\mp.
\end{equation}
because of the appearance of $\gamma^5$ in the T-tensor. With this new projector, it is now straightforward to derive the following BPS equations
\begin{equation}\label{ZSO(4)BPS}
A'=\frac{g}{2}e^{\varphi-4\phi_0},\qquad \varphi'=-\frac{g}{10}e^{\varphi-4\phi_0},\qquad \phi'_0= \frac{g}{10}e^{\varphi-4\phi_0}\, .
\end{equation}
These equations are solved by the solution
\begin{eqnarray}
\varphi&=&4C-\frac{1}{5}\ln\left[\frac{gr}{2}+C_0\right],\\
\phi_0&=&C+\frac{1}{5}\ln\left[\frac{gr}{2}+C_0\right],\\
A&=&-5\varphi\ =\ \ln\left[\frac{gr}{2}+C_0\right]-20C\, .
\end{eqnarray}

\subsection{$SO(3)$ symmetric domain walls}
We now look for more complicated solutions with $SO(3)$ symmetry. Gauge groups with an $SO(3)$ subgroup are $SO(4)\ltimes \mathbb{R}^{4}_{\boldsymbol{s}}$, $SO(3,1)\ltimes \mathbb{R}^{4}_{\boldsymbol{s}}$, and $CSO(3,0,1)\ltimes \mathbb{R}^{4}_{\boldsymbol{s}}$ which are collectively described by the symmetric tensor
\begin{equation}
w^{ij}=\textrm{diag}(1,1,1,\kappa)
\end{equation}
for $\kappa=1,-1,0$, respectively.

The residual symmetry $SO(3)$ is generated by the generators $X_{\hat{i}4}$ with $\hat{i}=1,2,3$. Apart from the two dilatons, there are three additional $SO(3)$ singlet scalars, one from the $SL(4)/SO(4)$ coset and the other two from symmetric and antisymmetric axions denoted by $b$ and $\varsigma$. These three singlets correspond to the following $SO(5,5)$ non-compact generators 
\begin{equation}
\widetilde{\mathcal{Y}}_1=\hat{\boldsymbol{t}}^+_{1\dot{1}}+\hat{\boldsymbol{t}}^+_{2\dot{2}}+\hat{\boldsymbol{t}}^+_{3\dot{3}}-3\,\hat{\boldsymbol{t}}^+_{4\dot{4}},\qquad \widetilde{\mathcal{Y}}_2=\hat{\boldsymbol{t}}^+_{4\dot{5}},\qquad \widetilde{\mathcal{Y}}_3=\boldsymbol{s}_{45}\, .
\end{equation}
\indent Using the coset representative of the form
\begin{equation}\label{ZSO(3)coset}
V=e^{\varphi\boldsymbol{d}+\phi_0\widetilde{\mathcal{Y}}_0+\phi\widetilde{\mathcal{Y}}_1+b\widetilde{\mathcal{Y}}_2+\varsigma\widetilde{\mathcal{Y}}_3},
\end{equation}
we find the scalar potential and the T-tensor given by 
\begin{equation}\label{ZSO(3)Pot}
\mathbf{V}=-\frac{g^2}{16}e^{2(\varphi-4(\phi_0+3\phi))}\left(6\kappa e^{16\phi}+(9e^{32\phi}+\kappa^2)\cosh{2b}\right)
\end{equation}
and 
\begin{equation}\label{40repSO(3)T-tenor}
T^{\alpha\dot{\beta}}=\frac{1}{4}e^{\varphi-4(\phi_0+3\phi)}\left[(3e^{16\phi}+\kappa)\cosh{b}\, (\gamma^5)^{\alpha\beta}\delta^{\dot{\beta}}_\beta+(3e^{16\phi}-\kappa)\sinh{b}\, (\gamma^4)^{\alpha\beta}\delta^{\dot{\beta}}_\beta\right].
\end{equation}
It turns out that consistency of the BPS equations from $\delta\chi_\pm$ conditions requires vanishing symmetric axion $b$ unless $\kappa=0$ corresponding to $CSO(3,0,1)\ltimes \mathbb{R}^{4}_{\boldsymbol{s}}$ gauge group.

\subsubsection{Domain walls without the symmetric axion}\label{40SO(3)nobSec}
With $b=0$, the scalar potential and superpotential read
\begin{eqnarray}
\mathbf{V}&=&-\frac{g^2}{8}e^{2(\varphi-4(\phi_0+3\phi))}(3e^{32\phi}+6\kappa e^{16\phi}-\kappa^2),\\
\mathcal{W}&=&\frac{g}{8}e^{\varphi-4(\phi_0+3\phi)}(3e^{16\phi}+\kappa).
\end{eqnarray}
\indent Imposing the projector \eqref{pureZProj} on the Killing spinors \eqref{DW_Killing_spinor}, we can derive the following set of BPS equations
\begin{eqnarray}\label{genZSO(3)BPS}
A'&=&\frac{g}{8}e^{\varphi-4(\phi_0+3\phi)}(3e^{16\phi}+\kappa),\\
\varphi'&=&-\frac{g}{40}e^{\varphi-4(\phi_0+3\phi)}(3e^{16\phi}+\kappa),\\
\phi'_0&=&\frac{g}{40}e^{\varphi-4(\phi_0+3\phi)}(3e^{16\phi}+\kappa),\\
\phi'&=&-\frac{g}{8}e^{\varphi-4(\phi_0+3\phi)}(3e^{16\phi}-\kappa),\\
\varsigma'&=&-ge^{\varphi-4(\phi_0+3\phi)}\varsigma\, .
\end{eqnarray}
From these equations, we can find the solutions for $A$, $\varphi$, and $\phi_0$ as functions of $\phi$ of the form
\begin{eqnarray}
\phi_0&=&\frac{\phi}{5}+C_0-\frac{1}{20}\ln(e^{16\phi}-\kappa),\\
\varphi&=&-\frac{\phi}{5}+C-C_0+\frac{1}{20}\ln(e^{16\phi}-\kappa),\\
A&=&-5\varphi\,=\,\phi-5C+5C_0-\frac{1}{4}\ln(e^{16\phi}-\kappa).
\end{eqnarray}
With the new radial coordinate $\rho$ defined by $\frac{d\rho}{dr}=e^{\varphi-4(\phi_0+\phi)}$, the solutions for $\phi$ and $\varsigma$ are given by
\begin{equation}
e^{8\phi}=\sqrt{\kappa}\tanh\left[\sqrt{\kappa}(g\rho+C_1)\right]\qquad \textrm{and}\qquad \varsigma=C_2\,\text{csch}\left[\sqrt{\kappa}(g\rho+C_1)\right].
\end{equation}
In particular, for $\kappa=-1$ and $\kappa=0$, we find respectively
\begin{equation}
e^{8\phi}=-\tan\left[\sqrt{\kappa}(g\rho+C_1)\right],\qquad \varsigma=C_2\,\text{csc}\left(g\rho+C_1\right)
\end{equation}
and
\begin{eqnarray}
e^{8\phi}=\frac{1}{(g\rho+C_1)},\qquad \varsigma=C_2\, .
\end{eqnarray}

\subsubsection{Domain walls with the symmetric axion}
For $\kappa=0$ corresponding to $CSO(3,0,1)\ltimes \mathbb{R}^{4}_{\boldsymbol{s}}$ gauge group, it is possible to find solutions with the symmetric axion $b$ non-vanishing. With $\kappa=0$, the scalar potential and the T-tensor are given by 
\begin{equation}
\mathbf{V}=-\frac{3g^2}{8}e^{2(\varphi-4\phi_0+4\phi)}\cosh{2b}
\end{equation}
and
\begin{equation}\label{40repSO(3)T-tenorwithb}
T^{\alpha\dot{\beta}}=\frac{3}{4}e^{\varphi-4(\phi_0-\phi)}\left[\cosh{b}\, (\gamma^5)^{\alpha\beta}\delta^{\dot{\beta}}_\beta+\sinh{b}\, (\gamma^4)^{\alpha\beta}\delta^{\dot{\beta}}_\beta\right].
\end{equation}
By the general procedure given in section \ref{super_DW}, we find the superpotential and $\hat{\gamma}_r$ projectors on the Killing spinors
\begin{equation}
\mathcal{W}=\frac{3g}{8}e^{\varphi-4\phi_0+4\phi}\sqrt{\cosh{2b}}
\end{equation}
and
\begin{eqnarray}\label{GenDWProj}
\hat{\gamma}_r\epsilon_{+\alpha}&=&\frac{\Omega_{\alpha\beta}}{\sqrt{\cosh{2b}}}\left[\cosh{b}\, (\gamma^5)^{\beta\gamma}\delta^{\dot{\alpha}}_\gamma+\sinh{b}\, (\gamma^4)^{\beta\gamma}\delta^{\dot{\alpha}}_\gamma\right]\epsilon_{-\dot{\alpha}},\\
\hat{\gamma}_r\epsilon_{-\dot{\alpha}}&=&-\frac{\Omega_{\dot{\alpha}\dot{\beta}}}{\sqrt{\cosh{2b}}}\left[\cosh{b}\, (\gamma^5)^{\alpha\beta}\delta^{\dot{\beta}}_\beta+\sinh{b}\, (\gamma^4)^{\alpha\beta}\delta^{\dot{\beta}}_\beta\right]\epsilon_{+\alpha}\, .
\end{eqnarray}
It should be noted that these projectors are not independent. Therefore, the resulting solutions will preserve half of the supersymmetry. Moreover, we can easily see that these projectors reduce to that given in \eqref{pureZProj} for $b=0$. 
\\
\indent With all these, we find the following set of BPS equations
\begin{eqnarray}
A'&=&\frac{3g}{8}e^{\varphi-4\phi_0+4\phi}\sqrt{\cosh{2b}},\qquad\quad\ \ 
B'\ =\ -\frac{3ge^{\varphi-4\phi_0+4\phi}\tanh{2b}}{8\sqrt{\cosh{2b}}},\\
\varphi'&=&-\frac{3g}{40}e^{\varphi-4\phi_0+4\phi}\sqrt{\cosh{2b}},\qquad\ \ \, 
\phi'_0\ =\ -\frac{3ge^{\varphi-4\phi_0+4\phi}(\cosh{4b}-9)}{320\cosh^{3/2}{2b}},\qquad\\
\phi'&=&-\frac{ge^{\varphi-4\phi_0+4\phi}(\cosh{4b}+7)}{64\cosh^{3/2}{2b}},\qquad
b'\ =\ -\frac{3ge^{\varphi-4\phi_0+4\phi}\sinh{2b}}{4\sqrt{\cosh{2b}}}
\end{eqnarray}
together with $\varsigma'=0$. Since the scalar potential does not depend on $\varsigma$, we can consistently truncate $\varsigma$ out by setting $\varsigma=0$. The domain wall solution to the above BPS equations is then given by
\begin{eqnarray}
\sqrt{\sinh{2b}}(\frac{3g\rho}{4}+C_b)&=&\,_2F_1(\frac{1}{4},\frac{1}{4},\frac{5}{4},-\frac{1}{\sinh^2{2b}}),\\
\varphi&=&C-\frac{b}{10}+\frac{1}{20}\ln(1-e^{4b}),\\
\phi_0&=&C_0-\frac{b}{40}-\frac{1}{20}\ln(1-e^{4b})+\frac{1}{16}\ln(1+e^{4b}),\\
\phi&=&C_1-\frac{b}{24}+\frac{1}{12}\ln(1-e^{4b})-\frac{1}{16}\ln(1+e^{4b}),\\
A&=&-5\varphi\,=\,-5C+\frac{b}{2}-\frac{1}{4}\ln(1-e^{4b}),\\
B&=&\frac{1}{2}\tan^{-1}\left(e^{2b}\right)+C_B
\end{eqnarray}
in which the new radial coordinate $\rho$ is defined by $\frac{d\rho}{dr}=e^{\varphi-4\phi_0+4\phi}$, and $_2F_1$ is the hypergeometric function. 

\subsection{$SO(2)\times SO(2)$ symmetric domain walls}\label{ZSO(2)xSO(2)}
Domain walls preserving $SO(2)\times SO(2)$ symmetry can be found in $SO(4)\ltimes \mathbb{R}^{4}_{\boldsymbol{s}}$ and $SO(2,2)\ltimes \mathbb{R}^{4}_{\boldsymbol{s}}$ gauge groups described by the embedding tensor with
\begin{equation}
w^{ij}=\textrm{diag}(1,1,\kappa,\kappa), \qquad \kappa=1,-1\, .
\end{equation}
\indent In addition to the two dilatons, there are three $SO(2)\times SO(2)$ singlet scalars corresponding to the following $SO(5,5)$ non-compact generators
\begin{equation}
\widehat{\mathcal{Y}}_1=\hat{\boldsymbol{t}}^+_{1\dot{1}}+\hat{\boldsymbol{t}}^+_{2\dot{2}}-\hat{\boldsymbol{t}}^+_{3\dot{3}}-\hat{\boldsymbol{t}}^+_{4\dot{4}},\qquad
\widehat{\mathcal{Y}}_2=\boldsymbol{s}_{12},\qquad
\widehat{\mathcal{Y}}_3=\boldsymbol{s}_{34}\, .
\end{equation}
In this case, a consistent set of BPS equations can be found only when the scalars corresponding to $\widehat{\mathcal{Y}}_2$ and $\widehat{\mathcal{Y}}_3$ generators vanish.

With the coset representative
\begin{equation}
V=e^{\varphi\boldsymbol{d}+\phi_0\widetilde{\mathcal{Y}}_0+\phi\widehat{\mathcal{Y}}_1},
\end{equation} 
the scalar potential and superpotential are given by
\begin{equation}\label{ZSO(4)Pot}
\mathbf{V}=-g^2\kappa e^{2\varphi-8\phi_0}\qquad \textrm{and}\qquad \mathcal{W}=\frac{g}{4}e^{\varphi-4(\phi_0+\phi)}(e^{8\phi}+\kappa).
\end{equation}
\indent With all these and the usual Killing spinors \eqref{DW_Killing_spinor} subject to the projector \eqref{pureZProj}, the resulting BPS equations read
\begin{eqnarray}
A'&=&\frac{g}{4}e^{\varphi-4(\phi_0+\phi)}(e^{8\phi}+\kappa),\\
\varphi'&=&-\frac{g}{20}e^{\varphi-4(\phi_0+\phi)}(e^{8\phi}+\kappa),\\
\phi'_0&=&\frac{g}{20}e^{\varphi-4(\phi_0+\phi)}(e^{8\phi}+\kappa),\\
\phi'&=&-\frac{g}{4}e^{\varphi-4(\phi_0+\phi)}(e^{8\phi}-\kappa).
\end{eqnarray}
Using a new radial coordinate $\rho$ defined by $\frac{d\rho}{dr}=e^{\varphi-2\phi_0}$, we find a domain wall solution
\begin{eqnarray}
\phi_0&=&\frac{\phi}{5}+C_0-\frac{1}{20}\ln\left(e^{8\phi}-\kappa\right),\\
\varphi&=&-\frac{\phi}{5}+C-C_0+\frac{1}{20}\ln\left(e^{8\phi}-\kappa\right),\\
A&=&-5\varphi\,=\,\phi-5C+5C_0-\frac{1}{4}\ln\left(e^{8\phi}-\kappa\right),\\
e^{4\phi}&=&\sqrt{\kappa}\tanh\left[\sqrt{\kappa}(g\rho+C_1)\right].
\end{eqnarray}

\subsection{$SO(2)$ symmetric domain walls}
As a final example in this case, we consider $SO(2)$ symmetric domain walls. There are many gauge groups admitting an $SO(2)$ subgroup. They are collectively characterized by the following component of the embedding tensor
\begin{equation}\label{ZSO(2)w}
w^{ij}=\text{diag}(1,1,\kappa,\lambda).
\end{equation}
\indent Together with the two dilatons, there are additional nine $SO(2)$ singlet scalars. Three of them are in the $SL(4)/SO(4)$ coset corresponding to non-compact generators
\begin{equation}
\overline{\mathcal{Y}}_1=\hat{\boldsymbol{t}}^+_{1\dot{1}}+\hat{\boldsymbol{t}}^+_{2\dot{2}}-\hat{\boldsymbol{t}}^+_{3\dot{3}}-\hat{\boldsymbol{t}}^+_{4\dot{4}},\qquad
\overline{\mathcal{Y}}_2=\hat{\boldsymbol{t}}^+_{3\dot{4}},\qquad
\overline{\mathcal{Y}}_3=\hat{\boldsymbol{t}}^+_{3\dot{3}}-\hat{\boldsymbol{t}}^+_{4\dot{4}}\, .
\end{equation}
The remaining ones consist of two nilpotent scalars associated with 
\begin{equation}
\overline{\mathcal{Y}}_4=\hat{\boldsymbol{t}}^+_{3\dot{5}},\qquad
\overline{\mathcal{Y}}_5=\hat{\boldsymbol{t}}^+_{4\dot{5}}
\end{equation}
and four shift scalars corresponding to
\begin{equation}
\overline{\mathcal{Y}}_6=\boldsymbol{s}_{12},\qquad
\overline{\mathcal{Y}}_7=\boldsymbol{s}_{35},\qquad
\overline{\mathcal{Y}}_8=\boldsymbol{s}_{45},\qquad
\overline{\mathcal{Y}}_9=\boldsymbol{s}_{34}\, .
\end{equation}
However, dealing with all eleven scalars turns out to be highly complicated, so we perform a subtruncation by setting the shift scalar corresponding to $\boldsymbol{s}_{12}$ and the two nilpotent scalars to zero. It is straightforward to verify that this is a consistent truncation and still gives interesting solutions. We now end up with eight singlet scalars with the coset representative
\begin{equation}\label{40SO(2)FullCoset}
V=e^{\varphi\boldsymbol{d}+\phi_0\widetilde{\mathcal{Y}}_0+\phi_1\overline{\mathcal{Y}}_1
+\phi_2\overline{\mathcal{Y}}_2+\phi_3\overline{\mathcal{Y}}_3+\varsigma_1\overline{\mathcal{Y}}_7+\varsigma_2\overline{\mathcal{Y}}_8+\varsigma_3\overline{\mathcal{Y}}_9}\, .
\end{equation} 
Consistency of the resulting BPS equations requires vanishing of the shift scalar $\varsigma_3$ unless $\kappa=\lambda=0$ corresponding to $CSO(2,0,2)\ltimes \mathbb{R}^4_{\boldsymbol{s}}$ gauge group. In what follows, we will for the moment set $\varsigma_3=0$ and separately consider the $CSO(2,0,2)\ltimes \mathbb{R}^4_{\boldsymbol{s}}$ gauge group with $\varsigma_3\neq0$.

With $\varsigma_3=0$, we can compute the scalar potential and the superpotential of the form
\begin{eqnarray}
\mathbf{V}&=&-\frac{g^2}{32}e^{2\left(\varphi-4(\phi_0+\phi_1)\right)}\left[\kappa^2+10\kappa\lambda+\lambda^2-2(\kappa+\lambda)^2\cosh{4\phi_2}\cosh^2{4\phi_3}\phantom{\frac{1}{\sqrt{2}}}\right.\nonumber\\&&\left.\phantom{+\frac{1}{\sqrt{2}}}-(3\kappa^2-2\kappa\lambda+3\lambda^2)\cosh{8\phi_3}+16(\kappa-\lambda)e^{8\phi_1}\sinh{4\phi_3}\right.\nonumber\\&&\left.\phantom{+\frac{1}{\sqrt{2}}}+4(\kappa+\lambda)\cosh{2\phi_2}\left(4e^{8\phi_1}\cosh{4\phi_3}-(\kappa-\lambda)\sinh{8\phi_3}\right)\right],\label{ZSO(2)Pot}
\end{eqnarray}
\begin{eqnarray}
\mathcal{W}&=&\frac{g}{8}e^{\varphi-4(\phi_0+\phi_1)}\left[2e^{8\phi_1}+(\kappa+\lambda)\cosh{2\phi_2}\cosh{4\phi_3}+(\kappa-\lambda)\sinh{4\phi_3}\right].\nonumber\\
\end{eqnarray}
This scalar potential can be written in term of the superpotential according to \eqref{PopularSP} using
\begin{equation}
G^{IJ}=\begin{pmatrix} G^{ij}	& G^{iy} \\ G^{xj}	& G^{xy}  \end{pmatrix}
\end{equation}
where
\begin{equation}
G^{ij}=\frac{1}{40}\text{diag}(4,1,5,40\text{sech}^2{4\phi_3},10),\qquad 
G^{xj}= \begin{pmatrix} \frac{2\varsigma_1}{5} & -\frac{3\varsigma_1}{20} & -\frac{\varsigma_1}{4} & \frac{4\varsigma_2e^{12\phi_3}}{(1+e^{8\phi_3})^2} & \frac{\varsigma_1}{2} \\ \frac{2\varsigma_2}{5} & -\frac{3\varsigma_2}{20} & -\frac{\varsigma_2}{4} & \frac{4\varsigma_1e^{12\phi_3}}{(1+e^{8\phi_3})^2} & \frac{\varsigma_2}{2} \end{pmatrix},
\end{equation}
and
\begin{equation}
G^{xy}=\begin{pmatrix}  1+4\varsigma_1^2+\varsigma_2^2e^{8\phi_3}\,\text{sech}^2{4\phi_3} & \frac{2\varsigma_1\varsigma_2(1+4e^{8\phi_3}+e^{16\phi_3})}{(1+e^{8\phi_3})^2} \\ \frac{2\varsigma_1\varsigma_2(1+4e^{8\phi_3}+e^{16\phi_3})}{(1+e^{8\phi_3})^2}& 1+4\varsigma_2^2+\varsigma_1^2e^{8\phi_3}\,\text{sech}^2{4\phi_3} \end{pmatrix}.
\end{equation}
Here, we have denoted $\Phi^I=\{\varphi,\phi_0,\phi_1,\phi_2,\phi_3,\varsigma_1,\varsigma_2\}=\{\Phi^i,\Phi^x\}$ for $i,j=1,2,\ldots,5$ and $x,y=6,7$. Note also that the scalar potential for $CSO(2,0,2)\ltimes \mathbb{R}^{4}_{\boldsymbol{s}}$ gauge group with $\kappa=\lambda=0$ vanishes identically leading to a family of Minkowski vacua. 

Imposing the projector \eqref{pureZProj} on the Killing spinors of the form
\begin{equation}
\epsilon_+=e^{\frac{A(r)}{2}+B(r)\gamma_{34}}\epsilon_{+}^{0}\qquad\text{ and }\qquad\epsilon_-=e^{\frac{A(r)}{2}+B(r)\gamma_{\dot{3}\dot{4}}}\epsilon_{-}^{0},	
\end{equation}
we obtain the following set of BPS equations 
\begin{eqnarray}
A'&\hspace{-0.1cm}=&\hspace{-0.1cm}\frac{g}{8}e^{\varphi-4(\phi_0+\phi_1)}\left[2e^{8\phi_1}+(\kappa+\lambda)\cosh{2\phi_2}\cosh{4\phi_3}+(\kappa-\lambda)\sinh{4\phi_3}\right],\label{40SO(2)GenBPS1}\\
\varphi'&\hspace{-0.1cm}=&\hspace{-0.2cm}-\frac{g}{40}e^{\varphi-4(\phi_0+\phi_1)}\left[2e^{8\phi_1}+(\kappa+\lambda)\cosh{2\phi_2}\cosh{4\phi_3}+(\kappa-\lambda)\sinh{4\phi_3}\right],\qquad\ \label{40SO(2)GenBPS2}\\
\phi'_0&\hspace{-0.1cm}=&\hspace{-0.1cm}\frac{g}{40}e^{\varphi-4(\phi_0+\phi_1)}\left[2e^{8\phi_1}+(\kappa+\lambda)\cosh{2\phi_2}\cosh{4\phi_3}+(\kappa-\lambda)\sinh{4\phi_3}\right],\label{40SO(2)GenBPS3}\\
\phi_1'&\hspace{-0.1cm}=&\hspace{-0.2cm}-\frac{g}{8}e^{\varphi-4(\phi_0+\phi_1)}\left[2e^{8\phi_1}-(\kappa+\lambda)\cosh{2\phi_2}\cosh{4\phi_3}-(\kappa-\lambda)\sinh{4\phi_3}\right],\label{40SO(2)GenBPS4}\\
\phi_2'&\hspace{-0.1cm}=&\hspace{-0.1cm}-\frac{g}{2}e^{\phi_0-4(\phi_0+\phi_1)}(\kappa+\lambda)\sinh{2\phi_2}\,\text{sech}\,{4\phi_3},\label{40SO(2)GenBPS5}\\
\phi_3'&\hspace{-0.1cm}=&\hspace{-0.1cm}-\frac{g}{4}e^{\phi_0-4(\phi_0+\phi_1)}\left((\kappa+\lambda)\cosh{2\phi_2}\sinh{3\phi_3}+(\kappa-\lambda)\cosh{4\phi_3}\right)\label{40SO(2)GenBPS6}
\end{eqnarray}
together with 
\begin{eqnarray}
B'&=&-\frac{g}{4}e^{\phi_0-4(\phi_0+\phi_1)}(\kappa+\lambda)\sinh{2\phi_2}\tanh{4\phi_3},\label{40SO(2)GenBPS7}\\
\varsigma'_1&\hspace{-0.1cm}=&\hspace{-0.2cm}-\frac{ge^{\varphi+4\phi_3}}{2e^{4(\phi_0+\phi_1)}}\left[\varsigma_1\left(\kappa-\lambda+(\kappa+\lambda)\cosh{2\phi_2}\right)+\varsigma_2(\kappa+\lambda)\sinh{2\phi_2}\,\text{sech}\,{4\phi_3}\right],\nonumber\\ \label{40SO(2)GenBPS8}\\
\varsigma'_2&\hspace{-0.1cm}=&\hspace{-0.2cm}-\frac{ge^{\varphi-4\phi_3}}{2e^{4(\phi_0+\phi_1)}}\left[\varsigma_1(\kappa+\lambda)\sinh{2\phi_2}\,\text{sech}\,{4\phi_3}-\varsigma_2\left(\kappa-\lambda-(\kappa+\lambda)\cosh{2\phi_2}\right)\right].\nonumber\\\label{40SO(2)GenBPS9}
\end{eqnarray}
We are unable to completely solve these equations for arbitrary values of the parameters $\kappa$ and $\lambda$. However, the solutions can be separately found for each value of $\kappa$ and $\lambda$.

\subsubsection{Domain walls in $SO(3,1)\ltimes\mathbb{R}^4_{\boldsymbol{s}}$ gauge group}
In this case, we set $\kappa=-\lambda=1$, and the BPS equations give $B'=\phi_2'=0$. We can again truncate $\phi_2$ out and set the constant $B=0$. As a result, we find a domain wall solution
\begin{eqnarray}
\phi_1&=&\frac{1}{2}\phi_3-\frac{1}{8}\ln\left[1+C_1(1+e^{8\phi_3})\right],\\
\phi_0&=&C_0+\frac{1}{10}\phi_3-\frac{1}{20}\ln (1+e^{8\phi_3})+\frac{1}{40}\ln\left[1+C_1(1+e^{8\phi_3})\right],\\
\varphi&=&C-C_0-\frac{1}{10}\phi_3+\frac{1}{20}\ln (1+e^{8\phi_3})-\frac{1}{40}\ln\left[1+C_1(1+e^{8\phi_3})\right],\\
A&=&\hspace{-0.2cm}-5\varphi\, = \, 5(C_0-C)+\frac{1}{2}\phi_3-\frac{1}{4}\ln (1+e^{8\phi_3})+\frac{1}{8}\ln\left[1+C_1(1+e^{8\phi_3})\right],\qquad\quad\\
\phi_3&=&\frac{1}{4}\ln\tan(C_3-g\rho),\\
\varsigma_1&=&C_4\sec(C_3-g\rho),\\
\varsigma_2&=&C_5\csc(C_3-g\rho)
\end{eqnarray}
with $\rho$ defined by $\frac{d\rho}{dr}=e^{\varphi-4(\phi_0+\phi_1)}$.

\subsubsection{Domain walls in $CSO(3,0,1)\ltimes\mathbb{R}^4_{\boldsymbol{s}}$ and $CSO(2,1,1)\ltimes\mathbb{R}^4_{\boldsymbol{s}}$ gauge groups}
For $\lambda=0$ and $\kappa=\pm 1$ corresponding to $CSO(3,0,1)\ltimes\mathbb{R}^4_{\boldsymbol{s}}$ and $CSO(2,1,1)\ltimes\mathbb{R}^4_{\boldsymbol{s}}$ gauge groups, the domain wall solution is given by
\begin{eqnarray}
\phi_1&=&\frac{1}{16}\ln\left[(e^{4\phi_2}-1)(1+2e^{C_3}+e^{2C_3}-e^{2C_3+4\phi_2})\right]-\frac{1}{8}\ln(4-2e^{4\phi_2}),\qquad\, \\
\phi_2&=&\frac{1}{4}\ln\left[\frac{4(1+e^{C_3})^2+(1+2e^{C_3})^2g^2\rho^2}{4e^{2C_3}+(1+2e^{C_3})^2g^2\rho^2}\right],\\
\phi_3&=&\frac{1}{8}\ln\left[\frac{(e^{2\phi_2}-1)(1+e^{C_3}+e^{C_3+2\phi_2})}{1+e^{C_3}+e^{2\phi_2}-e^{C_3+4\phi_2}}\right],\\
\phi_0&=&C_0-\frac{1}{5}\phi_1+\frac{1}{40}\ln(1-e^{4\phi_2})-\frac{1}{40}\ln\left[1+2e^{C_3}+e^{2C_3}-e^{2C_3+4\phi_2}\right],\quad\\
\varphi&=&C-\phi_0,\\
A&=&-5\varphi
\end{eqnarray}
together with
\begin{equation}
B=C_B+\frac{1}{4}\sin^{-1}\left[e^{C_3}\sqrt{\frac{e^{4\phi_2}-1}{1+2e^{C_3}}}\right]+\frac{1}{4}\tan^{-1}\left[\sqrt{\frac{(e^{4\phi_2}-1)(1+e^{C_3})^2}{1+2e^{C_3}+e^{2C_3}-e^{2C_3+4\phi_2}}}\right].
\end{equation}
In this solution, we have defined the coordinate $\rho$ by $\frac{d\rho}{dr}=e^{-2\phi_0-2\phi_1}$ and set the integration constant for $\phi_2$ solution to be $C_2=\frac{1}{16(1+2e^{C_3})^2}$ in order to simplify the expression for the solution. We also note that the two gauge groups have exactly the same domain wall solution since the parameter $\kappa$ does not appear anywhere in the solution. In more detail, $\kappa^2$ appears in $\phi_2$ solution as $g^2\kappa^2\rho^2$, but this term is simply given by $g^2\rho^2$ for $\kappa=\pm1$.  
\\
\indent For the remaining scalars $\varsigma_1$ and $\varsigma_2$, we are not able to analytically find their solutions. We can instead perform a numerical analysis to find these solutions, but we will not pursue any further along this direction. In any case, these scalars can be consistently truncated out since they do not appear in the scalar potential. 

\subsubsection{Domain walls in $SO(4)\ltimes\mathbb{R}^4_{\boldsymbol{s}}$ and $SO(2,2)\ltimes\mathbb{R}^4_{\boldsymbol{s}}$ gauge groups}
In this case, we set $\kappa=\lambda=\pm 1$ corresponding to $SO(4)\ltimes\mathbb{R}^4_{\boldsymbol{s}}$ and $SO(2,2)\ltimes\mathbb{R}^4_{\boldsymbol{s}}$ gauge groups. As in the previous case, the resulting BPS equations are very complicated to find explicit solutions. Therefore, we will set $\varsigma_1=\varsigma_2=0$ and find the domain wall solution for the remaining fields as follows
\begin{eqnarray}
\phi_1&\hspace{-0.2cm}=&\hspace{-0.2cm}\frac{1}{16}\ln\left[e^{4\phi_2}-e^{2C_3}(e^{4\phi_2}-1)^2\right]-\frac{1}{8}\ln(2-e^{4\phi_2}),\\
\phi_2&\hspace{-0.2cm}=&\hspace{-0.2cm}\frac{1}{4}\ln\left[\frac{1-2e^{2g\kappa \rho}+e^{4g\kappa \rho}+4e^{2C_3}}{1+2e^{2g\kappa \rho}+e^{4g\kappa \rho}+4e^{2C_3}}\right],\\
\phi_3&\hspace{-0.2cm}=&\hspace{-0.2cm}\frac{1}{8}\ln\left[\frac{e^{2\phi_2}-e^{C_3}+e^{C_3+4\phi_2}}{e^{2\phi_2}+e^{C_3}-e^{C_3+4\phi_2}}\right],\\
\phi_0&\hspace{-0.2cm}=&\hspace{-0.2cm}C_0-\frac{\phi_1}{5}-\frac{1}{20}\ln(e^{4\phi_2}-1)+\frac{1}{4}\ln\left[e^{4\phi_2}-e^{2C_3}+e^{4\phi_2+2C_3}(2-e^{4\phi_2})\right],\qquad\ \ \\
\varphi&\hspace{-0.2cm}=&\hspace{-0.2cm}C-\phi_0,\\
B&=&C_B-\frac{1}{8}\tan^{-1}\left[\frac{e^{-C_3}(e^{4\phi_2}-2e^{2C_3}+2e^{2C_3+4\phi_2})}{2\sqrt{e^{4\phi_2}-e^{2C_3}+2e^{2C_3+4\phi_2}-e^{2C_3+8\phi_2}}}\right]\nonumber\\&&-\frac{1}{8}\tan^{-1}\left[\frac{e^{-C_3}(1+2e^{2C_3}+2e^{2C_3+4\phi_2})}{2\sqrt{e^{4\phi_2}(1+2e^{2C_3})-2e^{2C_3}-e^{2C_3+8\phi_2}}}\right],\\
A&\hspace{-0.2cm}=&\hspace{-0.2cm}-5\varphi
\end{eqnarray}
with $\frac{d\rho}{dr}=e^{\varphi-4(\phi_0+\phi_1)}$. 
\subsubsection{Domain walls in $CSO(2,0,2)\ltimes\mathbb{R}^4_{\boldsymbol{s}}$ gauge group}
Finally, we consider the case of $\kappa=\lambda=0$ corresponding to $CSO(2,0,2)\ltimes\mathbb{R}^4_{\boldsymbol{s}}$ gauge group. Using the coset representative \eqref{40SO(2)FullCoset}, we find the T-tensor given by
\begin{equation}
T^{\alpha\dot{\beta}}=\frac{1}{2}e^{\varphi-4(\phi_0-\phi_1)}\left[\, (\gamma^5)^{\alpha\beta}\delta^{\dot{\beta}}_\beta+2\varsigma_3\, (\gamma^{12})^{\alpha\beta}\delta^{\dot{\beta}}_\beta\right].
\end{equation}
By the general procedure given in section \ref{super_DW}, we find the superpotential
\begin{equation}\label{40SO(2)SpeSPot}
\mathcal{W}=\frac{g}{4}e^{\varphi-4(\phi_0-\phi_1)}\sqrt{\varsigma_3^2+1}
\end{equation}
and the following projectors   
\begin{eqnarray}
\hat{\gamma}_r\epsilon_{+\alpha}&=&\Omega_{\alpha\beta}\frac{\left[\, (\gamma^5)^{\beta\gamma}\delta^{\dot{\beta}}_\gamma+2\varsigma_3\, (\gamma^{12})^{\beta\gamma}\delta^{\dot{\beta}}_\gamma\right]}{\sqrt{\varsigma_3^2+1}}\epsilon_{-\dot{\beta}},\\
\hat{\gamma}_r\epsilon_{-\dot{\alpha}}&=&-\Omega_{\dot{\alpha}\dot{\beta}}\frac{\left[\, (\gamma^5)^{\alpha\gamma}\delta^{\dot{\beta}}_\gamma+2\varsigma_3\, (\gamma^{12})^{\alpha\gamma}\delta_{\dot{\gamma}}^\beta\right]}{\sqrt{\varsigma_3^2+1}}\epsilon_{+\alpha}\, .
\end{eqnarray}
As expected for half-supersymmetric solutions, these projectors are not independent. In addition, for $\varsigma_3=0$, they reduce to a simpler projector given in \eqref{pureZProj}. At this point, it is useful to note that for this gauge group, the scalar potential vanishes as previously mentioned, so there exists a six-dimensional Minkowski vacuum for this gauge group. However, the superpotential \eqref{40SO(2)SpeSPot} does not have any stationary points, so this Minkowski vacuum is not supersymmetric. 

With the following ansatz for the Killing spinors
\begin{equation}
\epsilon_+=e^{\frac{A(r)}{2}+B(r)\gamma_{34}}\epsilon_{+}^{0}\quad\text{ and }\quad\epsilon_-=e^{\frac{A(r)}{2}-B(r)\gamma_{\dot{3}\dot{4}}}\epsilon_{-}^{0},	
\end{equation}
we obtain the BPS equations
\begin{eqnarray}
A'&=&\frac{g}{4}e^{\varphi-4(\phi_0-\phi_1)}\sqrt{\varsigma_3^2+1},\qquad
\varphi'\ =\ -\frac{ge^{\varphi-4(\phi_0-\phi_1)}(1+20\varsigma_3^2)}{20\sqrt{\varsigma_3^2+1}},\nonumber\\
\phi'_0&=&\frac{ge^{\varphi-4(\phi_0-\phi_1)}}{20\sqrt{\varsigma_3^2+1}},\qquad
\phi_1'\ =\ -\frac{ge^{\varphi-4(\phi_0-\phi_1)}}{4\sqrt{\varsigma_3^2+1}},\qquad
\phi_2'\ =\ \phi_3'\ =\ 0,\nonumber\\
\varsigma_1'&=&-\frac{4ge^{\varphi-4(\phi_0-\phi_1)}\varsigma_3^2\varsigma_1}{\sqrt{\varsigma_3^2+1}},\qquad
\varsigma_2'\ =\ -\frac{4ge^{\varphi-4(\phi_0-\phi_1)}\varsigma_3^2\varsigma_2}{\sqrt{\varsigma_3^2+1}},\nonumber\\
\varsigma_3'&=&-ge^{\varphi-4(\phi_0-\phi_1)}\varsigma_3\sqrt{\varsigma_3^2+1},\qquad
B'\ =\ \frac{ge^{\varphi-4(\phi_0-\phi_1)}}{20\sqrt{\varsigma_3^2+1}}\, .
\end{eqnarray}
With a new radial coordinate $\rho$ defined by $\frac{d\rho}{dr}=e^{\varphi-4\phi_0}$, the corresponding solution is given by
\begin{eqnarray}
A&=&-\frac{1}{4}\ln\varsigma_3,\qquad\quad\ \ B\ = \ C_B-\frac{1}{2}\tan^{-1}{2\varsigma_3},\\
\varphi&=&C+\frac{1}{20}\ln\varsigma_3+\frac{1}{10}\ln(1+4\varsigma_3^2),\\
\phi_0&=&C_0+\frac{1}{20}\ln\varsigma_3-\frac{1}{40}\ln(1+4\varsigma_3^2),\\
\phi_1&=&C_1+\frac{1}{4}\ln\varsigma_3-\frac{1}{8}\ln(1+4\varsigma_3^2),\\
\varsigma_1&=&C_4\sqrt{1+4\varsigma_3^2},\qquad
\varsigma_2\ =\ C_5\sqrt{1+4\varsigma_3^2},\\
\varsigma_3&=&\frac{1}{g\rho e^{4C_1}+C_6}\, .
\end{eqnarray}

\section{Domain walls from gaugings in $(\mathbf{15} +\overline{\mathbf{40}})^{-1}$ representation}\label{YandZsec}
We now consider gaugings with non-vanishing components of the embedding tensor in both $\mathbf{15}^{-1}$ and $\overline{\mathbf{40}}^{-1}$ representations. These gaugings are dyonic with the embedding tensor containing both electric and magnetic parts. The full embedding tensor is given by $\theta^{AM}=(\theta^{Am}, \theta^{A}_m)$ with
\begin{equation}\label{15+40theta}
\theta^{Am}\ = \ \mathbb{T}^A_{np}U^{np,m}\qquad \textrm{and}\qquad \theta^{A}_m\ = \ \mathbb{T}^{An}Y_{nm}
\end{equation}
for $Y_{mn}=Y_{(mn)}$ and $U^{mn,p}=U^{[mn],p}$ satisfying $U^{[mn,p]}=0$. 

However, for dyonic gaugings, the first condition in the quadratic constraint \eqref{QC} is not automatically satisfied. For the embedding tensor given in \eqref{15+40theta}, we find that this constraint imposes the following condition
\begin{equation}\label{YZQC}
U^{np,m}Y_{qm}=0\, .
\end{equation}
To solve this condition, we follow \cite{7D_Max_Gauging} and split the $GL(5)$ index as $m=(i,x)$. By choosing a suitable basis, we can take $Y_{mn}$ to be
\begin{equation}\label{YZdiagY}
Y_{ij}=\text{diag}(+1,...,+1,-1,...,-1)\qquad \text{and}\qquad Y_{xy}=0\, .
\end{equation}
The constraint \eqref{YZQC} then implies that only the components $U^{xy,z}$ and $U^{ix,y}=U^{i(x,y)}$ are non-vanishing. As a result, the embedding tensor is parametrized by the following tensors
\begin{equation}\label{YZtensor}
Y_{ij},\qquad U^{i(x,y)},\qquad U^{xy,z}\, .
\end{equation}
We now consider different possible gauge groups with $\text{rank}Y=0,1,...,5$. There are two trivial cases for $\text{rank}Y=5$ with $U^{mn,p}=0$ and $\text{rank}Y=0$ with all $Y_{mn}=0$. These correspond respectively to gaugings in $\mathbf{15}^{-1}$ and $\overline{\mathbf{40}}^{-1}$ representations and have already been considered in the previous two sections.  

For $\text{rank}Y=4$, only $U^{i5,5}$ can be non-vanishing, but another condition from the quadratic constraint \eqref{QC} requires $U^{i5,5}=0$. Accordingly, the corresponding gauge groups are given by $CSO(4,0,1)$, $CSO(3,1,1)$ and $CSO(2,2,1)$ which again have been considered in section \ref{pure15Sec}.

In the following, we will study supersymmetric domain walls in the two non-trivial cases with $\text{rank}Y=3$ and $\text{rank}Y=2$. Gaugings in these cases are expected to arise from a circle reduction of seven-dimensional maximal gauged supergravity with the embedding tensor in both $\mathbf{15}$ and $\overline{\mathbf{40}}$ representations of $SL(5)$. Similar to the seven-dimensional solutions given in \cite{7D_DW}, we will find that in these gaugings, the domain walls are $\frac{1}{4}$-BPS preserving eight supercharges. For the case of $\text{rank}Y=1$, the second condition from the quadratic constraint \eqref{QC} is much more complicated to find a non-trivial solution for $U^{i(x,y)}$ and $U^{xy,z}$. We refrain from discussing this case here.
\subsection{$\frac{1}{4}$-BPS domain walls for $\text{rank}Y=3$}
We first consider the case of $\text{rank}Y=3$ with $i,j=1,2,3$. The second condition from the quadratic constraint \eqref{QC} becomes
\begin{equation}\label{mixt=3QuadCon}
\varepsilon_{ijk}U^{jx,z}\varepsilon_{zw}U^{kw,y}=\frac{1}{\sqrt{2}}Y_{ij}U^{jx,y}
\end{equation}
which can be solved by $U^{ix,y}$ of the form 
\begin{equation}
U^{ix,y}=-\frac{1}{2\sqrt{2}}\varepsilon^{xz}{(\Sigma^i)_z}^y
\end{equation}
where ${(\Sigma^i)_x}^y$ are $2\times 2$ matrices. In terms of these $\Sigma^i$, the quadratic constraint \eqref{mixt=3QuadCon} can be rewritten as
\begin{equation}\label{Impmixt=3QuadCon}
[\Sigma^i,\Sigma^j]=2\varepsilon^{ijk}Y_{kl}\Sigma^l.
\end{equation}
As pointed out in \cite{7D_Max_Gauging}, a real, non-vanishing solution for $U^{ix,y}$ is possible only for 
\begin{equation}\label{t=3Y}
Y_{ij}=\textrm{diag}(1,1,-1)
\end{equation} 
with the explicit form of $\Sigma^i$ given in terms of Pauli matrices as
\begin{equation}
\Sigma^1=\sigma_1,\qquad \Sigma^2=\sigma_3,\qquad \Sigma^3=i\sigma_2\, .
\end{equation}
The constraint \eqref{Impmixt=3QuadCon} is then the Lie algebra of a non-compact group $SO(2,1)$. It should also be noted that the tensor $U^{xy,z}$ is not constrained by this condition, so it can be parametrized by an arbitrary two-component vector $u^x$ as 
\begin{equation}
U^{xy,z}=\varepsilon^{xy}u^z\, .
\end{equation}
\indent We now consider the corresponding gauge algebra spanned by the following gauge generators
\begin{eqnarray}
X^x&=& -\frac{1}{2\sqrt{2}}\varepsilon^{yz}{(\Sigma^i)_z}^x\boldsymbol{s}_{iy}+\varepsilon^{yz}u^x\boldsymbol{s}_{yz},\label{t=3gaugeGen1}\\
X_{ij}&=&2Y_{k[i}{\boldsymbol{t}^k}_{j]}+2\sqrt{2}\varepsilon_{ijk}u^x{\boldsymbol{t}^k}_{x}-\frac{1}{2}\varepsilon_{ijk}{(\Sigma^k)_z}^x{\boldsymbol{t}^z}_{x},\\
X_{ix}&=& Y_{ik}{\boldsymbol{t}^k}_{x}+\frac{1}{2}\varepsilon_{ijk}{(\Sigma^j)_x}^z{\boldsymbol{t}^k}_{z}.\label{t=3gaugeGen3}
\end{eqnarray}
To determine the form of the corresponding gauge group, we explicitly evaluate these generators in vector representation and find the following commutation relations
\begin{eqnarray}
\left[X^x,X^y\right]&=&0,\qquad\, \left[X_{ij},X^x\right]={(X_{ij})_y}^xX^y,\qquad \left[X_{ix},X^y\right]=0,\\
\left[X_{ix},X_{jy}\right]&=&0,\qquad \left[X_{ij},X_{kx}\right]=-2{(X_{ij})_{kx}}^{ly}X_{ly},\\
\left[X_{ij},X_{kl}\right]&=&-{(X_{ij})_{kl}}^{pq}X_{pq}-2{(X_{ij})_{kl}}^{px}X_{px}.\label{t=3lastcommute}
\end{eqnarray}
Redefining the $X_{ij}$ generators as
\begin{equation}
\widetilde{X}_{ij}=X_{ij}-\frac{2\sqrt{2}}{3}\varepsilon_{ijk}\eta^{kl}X_{lx}u^x
\end{equation}
with $\eta^{ij}=\text{diag}(+1,+1,-1)$, we find that $\widetilde{X}_{ij}$ generate an $SO(2,1)$ subgroup with the Lie algebra
\begin{equation}
\left[\widetilde{X}_{ij},\widetilde{X}_{kl}\right]=-{(\widetilde{X}_{ij})_{kl}}^{pq}\widetilde{X}_{pq}\, .
\end{equation}
The remaining generators $X_{ix}$ and $X_x$, which transform non-trivially under $SO(2,1)$, generate two translation groups. Note also that there are only four independent $X_{ix}$ generators. With all these, the resulting gauge group is then given by
\begin{equation}
G_0=SO(2,1)\ltimes\left(\mathbb{R}^4\times\mathbb{R}^2_{\boldsymbol{s}}\right)
\end{equation}
in which $\mathbb{R}^2_{\boldsymbol{s}}$ is the translation group from the shift symmetries generated by $X^x$. As also pointed out in \cite{7D_Max_Gauging}, we see that the vector $u^x$ does not change the gauge algebra, so we can set $u^x=0$ for simplicity.

We now look for supersymmetric domain wall solutions invariant under $SO(2)\subset SO(2,1)$ generated by $X_{12}$. There are five $SO(2)$ singlet scalars corresponding to the non-compact generators 
\begin{eqnarray}
\mathbf{Y}_{\boldsymbol{d}}&=&\hat{\boldsymbol{t}}^+_{1\dot{1}}+\hat{\boldsymbol{t}}^+_{2\dot{2}}+\hat{\boldsymbol{t}}^+_{3\dot{3}}+\hat{\boldsymbol{t}}^+_{4\dot{4}}+\hat{\boldsymbol{t}}^+_{5\dot{5}},\\
\mathbf{Y}_1&=&2\,\hat{\boldsymbol{t}}^+_{1\dot{1}}+2\,\hat{\boldsymbol{t}}^+_{2\dot{2}}+2\,\hat{\boldsymbol{t}}^+_{3\dot{3}}-3\,\hat{\boldsymbol{t}}^+_{4\dot{4}}-3\,\hat{\boldsymbol{t}}^+_{5\dot{5}},\\
\mathbf{Y}_2&=&\hat{\boldsymbol{t}}^+_{1\dot{1}}+\hat{\boldsymbol{t}}^+_{2\dot{2}}-2\,\hat{\boldsymbol{t}}^+_{3\dot{3}}, \\
\mathbf{Y}_3&=&\boldsymbol{s}_{12},\\
\mathbf{Y}_4&=&\boldsymbol{s}_{45}.
\end{eqnarray}
Using the coset representative of the form
\begin{equation}\label{t=3coset}
V=e^{\varphi\mathbf{Y}_{\boldsymbol{d}}+\phi_1\mathbf{Y}_1+\phi_2\mathbf{Y}_2+\varsigma_1\mathbf{Y}_3+\varsigma_2\mathbf{Y}_4},
\end{equation}
we find the scalar potential
\begin{equation}\label{t=3Pot}
\mathbf{V}=-\frac{g^2}{4}e^{2(\varphi-8\phi_1+2\phi_2)}(e^{12\phi_2}+6)\, .
\end{equation}
Consistency of the BPS equations from $\delta \chi_\pm$ conditions requires $\varsigma_1=0$. After truncating out $\varsigma_1$, we find the T-tensor 
\begin{equation}
T^{\alpha\dot{\beta}}=\frac{2}{g}e^{\varphi-8\phi_1-4\phi_2}\left[\mathcal{W}_1(\delta^{\alpha}_1\delta^{\dot{\beta}}_3-\delta^{\alpha}_3\delta^{\dot{\beta}}_1)+\mathcal{W}_2(\delta^{\alpha}_2\delta^{\dot{\beta}}_4-\delta^{\alpha}_4\delta^{\dot{\beta}}_2)\right]
\end{equation}
with
\begin{equation}
\mathcal{W}_1=\frac{g}{4\sqrt{2}}e^{\varphi-8\phi_1-4\phi_2}(3-e^{12\phi_2}),\qquad
\mathcal{W}_2=\frac{g}{4\sqrt{2}}e^{\varphi-8\phi_1-4\phi_2}(1-e^{12\phi_2}).
\end{equation}
It turns out that only $\mc{W}_1$ gives rise to the superpotential in term of which the scalar potential can be written. 
\\
\indent With the superpotential given by $\mc{W}_1$, the unbroken supersymmetry corresponds to $\epsilon^1_\pm$ and $\epsilon^3_\pm$. Therefore, we set $\epsilon^2_\pm=\epsilon^4_\pm=0$ in the following analysis. Alternatively, we can implement this by imposing an additional projector of the form 
\begin{equation}
\gamma^3\epsilon_{\mp}=\epsilon_{\mp}\, .
\end{equation}
\indent By the same procedure as in the previous cases together with the projector \eqref{pureYProj}, we obtain the BPS equations, with $\varsigma_2=\varsigma$,
\begin{eqnarray}\label{YZt=3BPS}
A'&=&\frac{g}{4\sqrt{2}}e^{\varphi-8\phi_1-4\phi_2}(3-e^{12\phi_2}),\\
\varphi'&=&-\frac{g}{20\sqrt{2}}e^{\varphi-8\phi_1-4\phi_2}(3-e^{12\phi_2}),\\
\phi_1'&=&\frac{g}{15\sqrt{2}}e^{\varphi-8\phi_1-4\phi_2}(3-e^{12\phi_2}),\\
\phi_2'&=&\frac{g}{6\sqrt{2}}e^{\varphi-8\phi_1-4\phi_2}(3+e^{12\phi_2}),\\
\varsigma&=&-\frac{g}{\sqrt{2}}e^{\varphi-8\phi_1-4\phi_2}(3-e^{12\phi_2})\varsigma\, .
\end{eqnarray}
Introducing a new radial coordinate $\rho$ via $\frac{d\rho}{dr}=e^{\varphi-8\phi_1+2\phi_2}$, we find a domain wall solution 
\begin{eqnarray}
e^{6\phi_2}&=&\sqrt{\frac{3}{2}}\tan(\sqrt{3}g\rho+C_2),\\
\phi_1&=&C_1+\frac{2}{5}\phi_2-\frac{1}{20}\ln(3+2e^{12\phi_2}),\\
\varsigma&=&C_3e^{-6\phi_2}(3+2e^{12\phi_2})^{\frac{3}{4}},\\
\varphi&=&C-\frac{3}{4}C_1-\frac{3}{10}\phi_2-\frac{3}{80}\ln(3+2e^{12\phi_2}),\\
A&=&-5\varphi\ =\ -5C+\frac{15}{4}C_1+\frac{3}{2}\phi_2+\frac{3}{16}\ln(3+2e^{12\phi_2}).
\end{eqnarray}

\subsection{$\frac{1}{4}$-BPS domain walls for $\text{rank}Y=2$}
In this case, $i,j=1,2$, we have $Y_{ij}=\text{diag}(1,\pm1)$. The second condition from the quadratic constraint \eqref{QC} allows only the components $U^{xy,z}$, $x,y,\ldots=3,4,5$, which can be parametrized by a $3\times 3$ traceless matrix ${u_x}^y$ as
\begin{equation}\label{t=2Z}
U^{xy,z}=\frac{1}{2\sqrt{2}}\varepsilon^{xyt}{u_t}^z
\end{equation}
with ${u_x}^x=0$.
The non-vanishing gauge generators read
\begin{eqnarray}
X^x&=&\frac{1}{2\sqrt{2}}\epsilon^{yz,t}{u_t}^x\boldsymbol{s}_{yz},\\
 X_{12}&=&2Y_{k[1}{\boldsymbol{t}^k}_{2]}+\frac{1}{2}{u_x}^y{\boldsymbol{t}^x}_{y},\\
 X_{ix}&=&Y_{ij}{\boldsymbol{t}^j}_{x}-\frac{1}{2}\varepsilon_{ij}{u_x}^y{\boldsymbol{t}^j}_{y}
\end{eqnarray}
with the commutation relations given by
\begin{eqnarray}
\left[X^x,X^y\right]&=&0,\qquad \left[X^x,X_{iy}\right]=0,\qquad \left[X_{ix},X_{jy}\right]=0,\label{t=2gaugeCom1}\\
\left[X_{12},X^x\right]&=&{(X_{12})_y}^xX^y,\qquad \left[X_{12},X_{ix}\right]=-2{(X_{12})_{ix}}^{jy}X_{jy}.\label{t=2gaugeCom2}
\end{eqnarray}
$X^x$ and $X_{ix}$ commute with each other and separately generate two translation groups $\mathbb{R}^3_{\boldsymbol{s}}$ and $\mathbb{R}^6$ which transform non-trivially under $X_{12}$. The single $X_{12}$ generator in turn leads to a compact $SO(2)$ or a non-compact $SO(1,1)$ group for $Y_{ij}=\text{diag}(1,1)$ or $Y_{ij}=\text{diag}(1,-1)$, respectively. The corresponding gauge groups are then given by $SO(2)\ltimes\left(\mathbb{R}^6\times\mathbb{R}^3_{\boldsymbol{s}}\right)$ or $SO(1,1)\ltimes\left(\mathbb{R}^6\times\mathbb{R}^3_{\boldsymbol{s}}\right)$.

\subsubsection{Domain walls in $SO(2)\ltimes\left(\mathbb{R}^6\times\mathbb{R}^2_{\boldsymbol{s}}\right)$ gauge group}
To find solutions with a non-trivial residual symmetry, we will consider $SO(2)\ltimes\left(\mathbb{R}^6\times\mathbb{R}^2_{\boldsymbol{s}}\right)$ gauge group with $Y_{ij}=\delta_{ij}$. In vector representation, the $X_{12}$ generator is given by
\begin{equation}
{(X_{12})_m}^n=\begin{pmatrix} 	2i{(\sigma_2)_i}^j & 0_{2\times6} \\
						0_{2\times6} & {u_x}^y \end{pmatrix}.
\end{equation}
Accordingly, we choose the matrix ${u_x}^y$ to be
\begin{equation}\label{t=2u}
{u_x}^y=\begin{pmatrix} 0 & 0 & 0\\ 0 & 0 & -\lambda \\ 0 & \lambda & 0 \end{pmatrix}
\end{equation}
with $\lambda\in\mathbb{R}$. The $SO(2)$ subgroup is then embedded diagonally with only $X^{4}$ and $X^5$ non-vanishing. Thus, the corresponding gauge group, in this case, is given by $SO(2)\ltimes\left(\mathbb{R}^6\times\mathbb{R}^2_{\boldsymbol{s}}\right)$.

There are five $SO(2)$ singlets corresponding to the following non-compact generators commuting with $X_{12}$
\begin{eqnarray}
\overline{\mathbf{Y}}_{\boldsymbol{d}}&=&\hat{\boldsymbol{t}}^+_{1\dot{1}}+\hat{\boldsymbol{t}}^+_{2\dot{2}}+\hat{\boldsymbol{t}}^+_{3\dot{3}}+\hat{\boldsymbol{t}}^+_{4\dot{4}}+\hat{\boldsymbol{t}}^+_{5\dot{5}},\label{t=2Gensinglet1}\\
\overline{\mathbf{Y}}_1&=&3\,\hat{\boldsymbol{t}}^+_{1\dot{1}}+3\,\hat{\boldsymbol{t}}^+_{2\dot{2}}-2\,\hat{\boldsymbol{t}}^+_{3\dot{3}}-2\,\hat{\boldsymbol{t}}^+_{4\dot{4}}-2\,\hat{\boldsymbol{t}}^+_{5\dot{5}},\label{t=2Gensinglet2}\\
\overline{\mathbf{Y}}_2&=&-2\,\hat{\boldsymbol{t}}^+_{3\dot{3}}+\hat{\boldsymbol{t}}^+_{4\dot{4}}+\,\hat{\boldsymbol{t}}^+_{5\dot{5}}, \label{t=2Gensinglet3}\\
\overline{\mathbf{Y}}_3&=&\boldsymbol{s}_{12},\label{t=2Gensinglet4}\\
\overline{\mathbf{Y}}_4&=&\boldsymbol{s}_{45}\, .\label{t=2Gensinglet5}
\end{eqnarray}
With the coset representative
\begin{equation}\label{t=2cosetGen}
V=e^{\varphi\overline{\mathbf{Y}}_{\boldsymbol{d}}+\phi_1\overline{\mathbf{Y}}_1+\phi_2\overline{\mathbf{Y}}_2
+\varsigma_1\overline{\mathbf{Y}}_3+\varsigma_2\overline{\mathbf{Y}}_4},
\end{equation}
it turns out that the scalar potential vanishes identically. On the other hand, the T-tensor is given by
\begin{equation}\label{t=2T-tensorGen}
T^{\alpha\dot{\beta}}=\frac{e^{\varphi-12\phi_1}}{2\sqrt{2}}\left[\lambda\, (\gamma^3)^{\alpha\beta}+2\, \Omega^{\alpha\beta}+2\varsigma_1\left[\lambda\, (\gamma^{45})^{\alpha\beta}-2\, (\gamma^{12})^{\alpha\beta}\right]\right]\delta^{\dot{\beta}}_\beta
\end{equation}
or explicitly
\begin{equation}
T^{\alpha\dot{\beta}}=\frac{e^{\varphi-12\phi_1}}{2\sqrt{2}}\begin{pmatrix} 2(\lambda+2)\varsigma_1 & 0 & (\lambda+2) & 0 \\ 0 & 2(\lambda-2)\varsigma_1 & 0 & -(\lambda-2) \\ -(\lambda+2) & 0 & 2(\lambda+2)\varsigma_1 & 0 \\ 0 & (\lambda-2)& 0 & 2(\lambda-2)\varsigma_1\end{pmatrix}.
\end{equation}
This leads to two superpotentials
\begin{eqnarray}
\mathcal{W}_1&=&\frac{g}{4\sqrt{2}}e^{\varphi-12\phi_1}(\lambda+2)\sqrt{1+4\varsigma_1^2},\\
\mathcal{W}_2&=&\frac{g}{4\sqrt{2}}e^{\varphi-12\phi_1}(\lambda-2)\sqrt{1+4\varsigma_1^2}\, .
\end{eqnarray}
Unlike the previous $\text{rank}Y=3$ case, both of these give a valid superpotential in term of which the scalar potential can be written. As in the previous case, half of the supersymmetry is broken by choosing any one of these two possibilities which again corresponds to imposing an additional $\gamma^3$ projector of the form 
\begin{equation}\label{t=2Gam3Proj}
\gamma^3\epsilon_{\pm}=\epsilon_{\pm}\qquad \text{ or }\qquad \gamma^3\epsilon_{\pm}=-\epsilon_{\pm}
\end{equation}
for $\mc{W}=\mc{W}_1$ or $\mc{W}=\mc{W}_2$, respectively. Together with the usual $\hat{\gamma}_r$ projectors 
\begin{equation}\label{t=3GenDWProj}
\hat{\gamma}_r\epsilon_{+\alpha}=\Omega_{\alpha\beta}\frac{T^{\beta\dot{\beta}}}{A'}\epsilon_{-\dot{\beta}},\qquad
\hat{\gamma}_r\epsilon_{-\dot{\alpha}}=-\Omega_{\dot{\alpha}\dot{\beta}}\frac{T^{\alpha\dot{\beta}}}{A'}\epsilon_{+\alpha},
\end{equation}
the resulting solutions will preserve only eight supercharges or $\frac{1}{4}$ of the original supersymmetry.

With the following ansatz for the Killing spinors 
\begin{equation}\label{t=2Killing}
\epsilon_+=e^{\frac{A(r)}{2}+B(r)\gamma_{12}}\epsilon_{+}^{0}\qquad\text{ and }\qquad\epsilon_-=e^{\frac{A(r)}{2}-B(r)\gamma_{\dot{1}\dot{2}}}\epsilon_{-}^{0},
\end{equation}
for $\epsilon^0_\pm$ satisfying the projectors \eqref{t=2Gam3Proj} and \eqref{t=3GenDWProj}, we obtain the following BPS equations
\begin{eqnarray}
A'&=&\frac{g}{4\sqrt{2}}e^{\varphi-12\phi_1}(\lambda\pm2)\sqrt{1+4\varsigma_1^2},\qquad
B'\ =\ \frac{g\varsigma_1e^{\varphi-12\phi_1}(\lambda\pm2)}{\sqrt{2+8\varsigma_1^2}},\\
\varphi'&=&-\frac{ge^{\varphi-12\phi_1}(\lambda\pm2)(1+20\varsigma_1^2)}{20\sqrt{2+8\varsigma_1^2}},\qquad\ 
\phi'_1\ =\ \frac{ge^{\varphi-12\phi_1}(\lambda\pm2)}{10\sqrt{2+8\varsigma_1^2}},\\
\phi_2'&=&0,\qquad\varsigma'_1=\frac{g}{\sqrt{2}}\varsigma_1e^{\varphi-12\phi_1}(\lambda\pm2)\sqrt{1+4\varsigma_1^2},\\
\varsigma'_2&=& \frac{g}{\sqrt{2}}\varsigma_2e^{\varphi-12\phi_1}(\lambda\pm2)\sqrt{1+4\varsigma_1^2}\, .
\end{eqnarray}
The choices of plus or minus signs in these equations are correlated with the plus or minus signs of the two projectors given in \eqref{t=2Gam3Proj}.

We can consistently set $\phi_2=0$ and find a domain wall solution 
\begin{eqnarray}
A&=&-\frac{1}{4}\ln\varsigma_1,\quad B\ = \ C_B-\frac{1}{2}\tan^{-1}{2\varsigma_1},\label{t=2GenDW1}\\
\varphi&=&C+\frac{1}{20}\ln\varsigma_1+\frac{1}{10}\ln(1+4\varsigma_1^2),\\
\phi_1&=&C_1+\frac{1}{10}\ln\varsigma_1-\frac{1}{20}\ln(1+4\varsigma_1^2),\\
\varsigma_1&=&\frac{1}{2}\tan\left[\sqrt{2}e^{-10C_1}(\lambda\pm2)g\rho+C_3\right],\\
\varsigma_2&=&C_4\varsigma_1\label{t=2GenDW5}
\end{eqnarray}
where $\rho$ is the new radial coordinate defined by $\frac{d\rho}{dr}=e^{\varphi-2\phi_1}$.

\subsubsection{Domain walls in $CSO(2,0,2)\ltimes\mathbb{R}^2_{\boldsymbol{s}}$ gauge group}
From the previous result, there are special values of $\lambda=\pm2$ at which the $SO(2)\ltimes\left(\mathbb{R}^6\times\mathbb{R}^2_{\boldsymbol{s}}\right)$ gauge group reduces to $SO(2)\ltimes\left(\mathbb{R}^4\times\mathbb{R}^2_{\boldsymbol{s}}\right)\sim CSO(2,0,2)\ltimes\mathbb{R}^2_{\boldsymbol{s}}$. The two choices are equivalent, so we will choose $\lambda=2$ for definiteness.

In this case, there are nine scalars invariant under the residual $SO(2)$ symmetry generated by $X_{12}$. They are given by the five scalars associated with the non-compact generators given in \eqref{t=2Gensinglet1} to \eqref{t=2Gensinglet5} together with additional two symmetric and two shift scalars respectively corresponding to 
\begin{eqnarray}
\overline{\mathbf{Y}}_6&=&\hat{\boldsymbol{t}}^+_{1\dot{4}}+\hat{\boldsymbol{t}}^+_{2\dot{5}},\qquad\,
\overline{\mathbf{Y}}_7\ =\ \hat{\boldsymbol{t}}^+_{1\dot{5}}-\hat{\boldsymbol{t}}^+_{2\dot{4}},\\
\overline{\mathbf{Y}}_8&=&\boldsymbol{s}_{14}+\boldsymbol{s}_{25},\qquad
\overline{\mathbf{Y}}_9\ =\ \boldsymbol{s}_{15}-\boldsymbol{s}_{24}\, .
\end{eqnarray}
However, with this large number of scalar fields, the analysis is highly complicated. To make things more manageable, we will further truncate the nine scalars to the previous five singlets together with each of the two sets of axionic scalars separately.

Turning on two shift scalars, denoted by $\varsigma_3$ and $\varsigma_4$, corresponding to $\overline{\mathbf{Y}}_8$ and $\overline{\mathbf{Y}}_9$ generators, we find the solution given in equations \eqref{t=2GenDW1} to \eqref{t=2GenDW5} together with the solutions for $\varsigma_3$ and $\varsigma_4$ of the form
\begin{equation}
\varsigma_3=C_5\sqrt{1+4\varsigma_1^2} \qquad\text{ and }\qquad\varsigma_4=C_6\sqrt{1+4\varsigma_1^2}\, .
\end{equation}
\indent More interesting solutions are obtained by including the scalars corresponding to $\overline{\mathbf{Y}}_6$ and $\overline{\mathbf{Y}}_7$ generators. With the coset representative 
\begin{equation}\label{t=2cosetSpe}
V=e^{\varphi\overline{\mathbf{Y}}_{\boldsymbol{d}}+\phi_1\overline{\mathbf{Y}}_1+\phi_2\overline{\mathbf{Y}}_2
+\phi_3\overline{\mathbf{Y}}_6+\phi_4\overline{\mathbf{Y}}_7+\varsigma_1\overline{\mathbf{Y}}_3+\varsigma_2\overline{\mathbf{Y}}_4},
\end{equation}
we find that the scalar potential vanishes as in the previous case. There are also two superpotentials. One of them vanishes identically while the non-trivial one is given by
\begin{equation}
\mathcal{W}=\frac{g}{\sqrt{2}}e^{\varphi-12\phi_1}\sqrt{\cosh^2{2\phi_3}\cosh^2{2\phi_4}+\left(\varsigma_1-\varsigma_2+\cosh{2\phi_3}\cosh{2\phi_4}(\varsigma_1+\varsigma_2)\right)^2}\, .\label{W_Yrank2}
\end{equation}
Unlike the previous case, the Minkowski vacuum in this case is half-supersymmetric with the unbroken supersymmetry corresponding to the vanishing superpotential. This is very similar to $CSO(2,0,2)$ gauged supergravity in seven dimensions \cite{7D_Max_Gauging}.
\\
\indent Only the supersymmetry corresponding to the superpotential \eqref{W_Yrank2} is preserved by the domain wall. This again amounts to imposing a $\gamma^3$ projector of the form \eqref{t=2Gam3Proj}. Furthermore, consistency of the BPS equations from $\delta\chi_\pm$ requires $\varsigma_1=\varsigma_2=\varsigma$. It is useful to note the explicit form of the T-tensor for $\varsigma_1=\varsigma_2=\varsigma$ which is given by
\begin{equation}\label{t=2T-tensorSpe}
T^{\alpha\dot{\beta}}=\frac{e^{\varphi-12\phi_1}}{\sqrt{2}}\cosh{2\phi_3}\cosh{2\phi_4}\left[(\gamma^3)^{\alpha\beta}+ \Omega^{\alpha\beta}+2\varsigma\left((\gamma^{45})^{\alpha\beta}-(\gamma^{12})^{\alpha\beta}\right)\right]\delta^{\dot{\beta}}_\beta\, .
\end{equation}
\indent Using the Killing spinors \eqref{t=2Killing} subject to the projectors in \eqref{t=3GenDWProj} and the first projector in \eqref{t=2Gam3Proj}, we can derive the following BPS equations
\begin{eqnarray}\label{YZt=3BPS}
A'&=&\frac{g}{\sqrt{2}}e^{\varphi-12\phi_1}\cosh{2\phi_3}\cosh{2\phi_4}\sqrt{1+4\varsigma^2},\\
B'&=&\frac{2ge^{\varphi-12\phi_1}\cosh{2\phi_3}\cosh{2\phi_4}\varsigma}{\sqrt{2+8\varsigma^2}},\\
\varphi'&=&-\frac{ge^{\varphi-12\phi_1}\cosh{2\phi_3}\cosh{2\phi_4}(1+20\varsigma^2)}{5\sqrt{2+8\varsigma^2}},\\
\phi_1'&=&\frac{ge^{\varphi-12\phi_1}(\cosh^2{2\phi_3}\cosh^2{2\phi_4}+5)\,\text{sech}\,{2\phi_3}\,\text{sech}\,{2\phi_4}}{15\sqrt{2+8\varsigma^2}},\\
\phi_2'&=&\frac{ge^{\varphi-12\phi_1}(\cosh^2{2\phi_3}\cosh^2{2\phi_4}-1)\,\text{sech}\,{2\phi_3}\,\text{sech}\,{2\phi_4}}{3\sqrt{2+8\varsigma^2}},\\
\phi_3'&=&-\frac{\sqrt{2}ge^{\varphi-12\phi_1}\sinh{2\phi_3}\,\text{sech}\,{2\phi_4}}{\sqrt{1+4\varsigma^2}},
\end{eqnarray}
\begin{eqnarray}
\phi_4'&=&-\frac{\sqrt{2}ge^{\varphi-12\phi_1}\cosh{2\phi_3}\sinh{2\phi_4}}{\sqrt{1+4\varsigma^2}},\\
\varsigma'&=&-2ge^{\varphi-12\phi_1}\cosh{2\phi_3}\cosh{2\phi_4}\varsigma\sqrt{2+8\varsigma^2}\, .
\end{eqnarray}
Introducing a new radial coordinate $\rho$ via $\frac{d\rho}{dr}=\frac{e^{\varphi-12\phi_1}}{\sqrt{1+4\varsigma^2}}$, we eventually find a domain wall solution
\begin{eqnarray}
\phi_1&=&C_1+\frac{\phi_2}{5}-\frac{1}{10}\ln(e^{4\phi_3}-1)+\frac{1}{10}\ln(e^{4\phi_3}+1),\\
\phi_2&=&C_2-\frac{1}{12}\ln\left(e^{4\phi_3}+1\right)+\frac{1}{24}\ln\left[e^{2C_4}(1-2e^{4\phi_3}+e^{8\phi_3})-e^{4\phi_3}\right],\\
\phi_3&=&\frac{1}{4}\ln\left[\frac{1+2e^{2\sqrt{2}g\rho}+e^{4\sqrt{2}g\rho}+4e^{2C_4}}{1-2e^{2\sqrt{2}g \rho}+e^{4\sqrt{2}g \rho}+4e^{2C_4}}\right],\\
\phi_4&=&\frac{1}{4}\ln\left[\frac{e^{2\phi_3}-e^{C_4}+e^{C_4+4\phi_3}}{e^{2\phi_3}+e^{C_3}-e^{C_3+4\phi_3}}\right],\\
\varphi&=&C+\frac{1}{20}\ln(e^{4\phi_3}-1)+\frac{1}{10}\ln\left[e^{2C_4}(1-2e^{4\phi_3}+e^{8\phi_3})-e^{4\phi_3}\right]\\&&-\frac{1}{8}\ln\left[e^{4\phi_3}-e^{2C_4}(1-2e^{4\phi_3}+e^{8\phi_3})+4e^{2C_5}(1-2e^{4\phi_3}+e^{8\phi_3})\right],\quad\\
A&=&\frac{1}{8}\ln\left[\frac{e^{4\phi_3}-e^{2C_4}(1-2e^{4\phi_3}+e^{8\phi_3})+4e^{2C_5}(1-2e^{4\phi_3}+e^{8\phi_3})}{(e^{4\phi_3}-1)^2}\right],\\
\varsigma&=&\frac{e^{C_5}(e^{4\phi_3}-1)}{\sqrt{e^{2C_4}(1-2e^{4\phi_3}+e^{8\phi_3})-4e^{2C_5}(1-2e^{4\phi_3}+e^{8\phi_3})-e^{4\phi_3}}}.
\end{eqnarray}
\indent We end this section by noting that a domain wall solution with $\varsigma=0$ can similarly be obtained with the coordinate $\rho$ defined by $\frac{d\rho}{dr}=e^{\varphi-12\phi_1}$. In this case, the solutions for the dilaton and warped factor are given by
\begin{eqnarray}
\varphi&=&C+\frac{1}{20}\ln(e^{4\phi_3}-1)-\frac{1}{40}\ln\left[e^{4\phi_3}-e^{2C_4}(1-2e^{4\phi_3}+e^{8\phi_3})\right],\quad\\
A&=&-5\varphi
\end{eqnarray}
while solutions for the remaining scalars are the same as given above. 

\section{Conclusions and discussions}\label{Discuss}
We have constructed the embedding tensors of six-dimensional maximal $N=(2,2)$ gauged supergravity for various gauge groups with known seven-dimensional origins via an $S^1$ reduction. These gaugings are triggered by the embedding tensor in $\mathbf{15}^{-1}$ and $\overline{\mathbf{40}}^{-1}$ representations of $GL(5)\subset SO(5,5)$ duality symmetry. In $\mathbf{15}^{-1}$ representation, the corresponding gauge group is $CSO(p,q,5-p-q)$ which is the same as its seven-dimensional counterpart. On the other hand, for gaugings in $\overline{\mathbf{40}}^{-1}$ representation, additional translation groups $\mathbb{R}^n_{\boldsymbol{s}}$ associated with the shift symmetries on the scalar fields appear in the gaugings resulting in $CSO(p,q,4-p-q)\ltimes\mathbb{R}^4_{\boldsymbol{s}}$ gauge group. This is also the case for gaugings in $(\mathbf{15}+\overline{\mathbf{40}})^{-1}$ representation with gauge groups $SO(2,1)\ltimes\left(\mathbb{R}^4\times\mathbb{R}^2_{\boldsymbol{s}}\right)$, $SO(2)\ltimes\left(\mathbb{R}^6\times\mathbb{R}^2_{\boldsymbol{s}}\right)$, and $CSO(2,0,2)\ltimes\mathbb{R}^2_{\boldsymbol{s}}$. 

We have also studied supersymmetric domain wall solutions and found a large number of half-supersymmetric domain walls from purely magnetic and purely electric gaugings in $\mathbf{15}^{-1}$ and $\overline{\mathbf{40}}^{-1}$ representations, respectively. In addition, we have given $\frac{1}{4}$-supersymmetric domain walls for dyonic gaugings involving the embedding tensor in both $\mathbf{15}^{-1}$ and $\overline{\mathbf{40}}^{-1}$ representations. These are similar to the seven-dimensional solutions and in agreement with the general classification of supersymmetric domain walls in \cite{Eric_SUSY_DW} in which the existence of $\frac{1}{4}$-BPS domain walls has been pointed out. 

Apart from solutions with seven-dimensional analogues, we have also found solutions that are not uplifted to seven-dimensional domain walls due to the presence of axionic scalars leading to non-vanishing vector fields in seven dimensions. This can be explicitly seen from the truncation ansatz collected in appendix \ref{TrunAnsz}. Although this ansatz has originally been given only for $SO(5)$ gauge group, a similar ansatz with possibly suitable modifications in the tensor field content is also applicable for other gauge groups. In particular, the fact that a truncation of seven-dimensional vectors leads to axionic scalars in six dimensions is still true. Therefore, domain wall solutions with non-vanishing axionic scalars obtained in this work cannot be obtained from an $S^1$ reduction of any domain wall solutions in seven dimensions. Accordingly, these solutions are genuine six-dimensional domain walls without seven-dimensional analogues. As a final comment, we note that there is no $SO(5)$ symmetric domain wall in seven dimensions since there is no $SO(5)$ singlet scalar in $SL(5)/SO(5)$ coset. The six-dimensional $SO(5)$ symmetric domain wall, on the other hand, arises form an $S^1$ reduction of the supersymmetric $AdS_7$ vacuum by the general result of \cite{AdS_DW_Pope}. 

The seven-dimensional origin of all the gaugings considered in this work can also be embedded in ten or eleven dimensions, so the six-dimensional domain wall solutions can be embedded in string/M-theory via the corresponding seven-dimensional truncations. Accordingly, the solutions given here are hopefully useful in the study of DW$_6$/QFT$_5$ duality for maximal supersymmetric Yang-Mills theory in five dimensions from both six-dimensional framework and string/M-theory context. It is interesting to explicitly uplift the domain wall solutions to seven dimensions and subsequently to ten or eleven dimensions using the truncation ansatze given in \cite{11D_to_7D_Townsend,11D_to_7D_Nastase1,11D_to_7D_Nastase2,S3_S4_typeIIA,Henning_Emanuel}. 

Constructing truncation ansatze of string/M-theory to six dimensions using $SO(5,5)$ exceptional field theory given in \cite{SO5_5_EFT} is also of particular interest. This would allow uplifting the six-dimensional solutions directly to ten or eleven dimensions. In this paper, we have considered only gaugings with the embedding tensor in $\mathbf{15}^{-1}$ and $\overline{\mathbf{40}}^{-1}$ representations. It is natural to extend this study by performing a similar analysis for the embedding tensors in other $GL(5)$ representations as well as finding supersymmetric domain walls. Unlike the solutions obtained in this paper, these solutions will not have seven-dimensional counterparts via an $S^1$ reduction.  

It is also interesting to construct the embedding tensors for various gaugings under $SO(4,4)\subset SO(5,5)$. These gaugings can be truncated to gaugings in half-maximal $N=(1,1)$ supergravity coupled to four vector multiplets in which supersymmetric $AdS_6$ vacua are known to exist in the presence of both conventional gaugings and massive deformations \cite{D4D8,F4_flow,AdS6_Jan}. Finding supersymmetric solutions from these gauge groups could be useful in the study of AdS$_6$/CFT$_5$ correspondence. Finally, finding supersymmetric curved domain walls with non-vanishing vector and tensor fields as in seven-dimensional maximal gauged supergravity in \cite{7D_Defect, 7D_twist} is worth considering. This type of solutions can describe conformal defects or holographic RG flows from five-dimensional $N=4$ super Yang-Mills theories to lower dimensions. Along this line, examples of solutions dual to surface defects from $N=(1,1)$ gauged supergravity have appeared recently in \cite{6D_surface_defect}. 
\vspace{0.5cm}\\
{\large{\textbf{Acknowledgement}}} \\
This work is supported by the Second Century Fund (C2F), Chulalongkorn University. P. K. is supported by The Thailand Research Fund (TRF) under grant RSA6280022.
\appendix
\section{$GL(5)$ Branching rules}\label{AppA}
In this appendix, we collect all of the $SO(5,5)\rightarrow GL(5)$ branching rules used throughout the paper. Relevant decompositions have already been given in \cite{6D_Max_Gauging}, but in order to construct the embedding tensor, we need a concrete realization. Therefore, we will determine the decompositions for various representations of $SO(5,5)$ in terms of $GL(5)$ representations using explicit matrix forms.

\subsection{Vector}\label{VecBranc}
A vector or fundamental representation of $SO(5,5)$ decomposes under $GL(5)\subset SO(5,5)$ as $\mathbf{5}$ and $\overline{\mathbf{5}}$, i.e., $V_M=(V_m,V^m)$. The $SO(5,5)$ vector index $M=1,..,10$ can be raised and lowered through the following $SO(5,5)$ invariant metric in the light cone or off-diagonal basis
\begin{equation}\label{off-diag-eta}
\eta_{MN}\ =\ \eta^{MN}\ =\ \begin{pmatrix} 	0 & \mathds{1}_5 \\
							\mathds{1}_5 & 0    \end{pmatrix}
\end{equation}
in which $\mathds{1}_n$ is an ($n\times n$) identity matrix. For example, $V^M=\eta^{MN}V_N=(V^m,V_m)$. 

In vector representation, the $SO(5,5)$ algebra 
\begin{equation}\label{SO(5,5)algebra}
\left[\boldsymbol{t}_{MN},\boldsymbol{t}_{PQ}\right]\ =\ 4(\eta_{M[P}\boldsymbol{t}_{Q]N}-\eta_{N[P}\boldsymbol{t}_{Q]M})
\end{equation}
is realized by $SO(5,5)$ generators , $\boldsymbol{t}_{MN}=\boldsymbol{t}_{[MN]}$, of the form
\begin{equation}
{(\boldsymbol{t}_{MN})_P}^Q\ =\ 4\eta_{P[M}\delta^Q_{N]}
\end{equation}
where $\delta^M_N=\mathds{1}_{10}$. Defining an $\mathbb{R}^+\sim SO(1,1)\subset GL(5)\sim \mathbb{R}^+\times SL(5)$ generator by
\begin{equation}\label{DefDOp}
\boldsymbol{d}\ =\ {\boldsymbol{t}^m}_m\ =\ {\boldsymbol{t}^1}_1+{\boldsymbol{t}^2}_2+{\boldsymbol{t}^3}_3+{\boldsymbol{t}^4}_4+{\boldsymbol{t}^5}_5,
\end{equation}
we find the explicit form of the $\mathbb{R}^+$ generator in vector representation given by
\begin{equation}
\boldsymbol{d}\ =\ 2\begin{pmatrix} 	\mathds{1}_5 & 0 \\
						0 & -\mathds{1}_5    \end{pmatrix}.
\end{equation}
With an $SO(5,5)$ vector decomposing as $V_M=(V_m,V^m)$, we obtain the commutation relations
\begin{equation}
\left[\boldsymbol{d},V_m\right]\ =\ +2V_m \qquad \text{and} \qquad \left[\boldsymbol{d},V^m\right]\ =\ -2V^m\, .
\end{equation}
These imply that we can assign the $\mathbb{R}^+$ weights $\pm2$  to the $\mathbf{5}$ and $\overline{\mathbf{5}}$ representations of $SL(5)\subset GL(5)$. Therefore, the branching rule for a vector representation reads
\begin{equation}\label{VecDec}
\underbrace{\mathbf{10}}_{V_M}\ \rightarrow\ \underbrace{\mathbf{5}^{+2}}_{V_m}\,\oplus\,\underbrace{\overline{\mathbf{5}}^{-2}}_{V^m}\, .
\end{equation}

\subsection{Adjoint}\label{AdjDecompos}
The decomposition of adjoint representation follows from the branching rule of vector representations. Using \eqref{VecDec}, we can decompose the $SO(5,5)$ generators as 
\begin{equation}
\boldsymbol{t}_{MN}\ \rightarrow\ (\boldsymbol{t}_{mn},{\boldsymbol{t}^m}_n,\boldsymbol{t}^{mn})
\end{equation}
with ${\boldsymbol{t}_m}^n=-{\boldsymbol{t}^n}_m$. The 25 generators ${\boldsymbol{t}^m}_n$ of $GL(5)$ consist of the $\mathbb{R}^+$ generator defined in \eqref{DefDOp} and the $SL(5)$ generators given by
\begin{equation}\label{DefSL5Op}
{\boldsymbol{\tau}^m}_n\ =\ {\boldsymbol{t}^m}_n-\frac{1}{5}\,\boldsymbol{d}\,\delta^m_n
\end{equation}
with ${\boldsymbol{\tau}^m}_m=0$. 

We denote the shift and hidden generators by $\boldsymbol{s}_{mn}=\boldsymbol{t}_{mn}$ and $\boldsymbol{h}^{mn}=\boldsymbol{t}^{mn}$, respectively. In vector representation, the $SO(5,5)$ generators can be written as
\begin{equation}
{(\boldsymbol{t}_{MN})_P}^Q=\begin{pmatrix} {\boldsymbol{t}^m}_n & \boldsymbol{h}^{mn} \\
								\boldsymbol{s}_{mn} & -{\boldsymbol{t}^m}_n \end{pmatrix}.
\end{equation}
From the $SO(5,5)$ algebra, we can derive the following commutation relations 
\begin{eqnarray}
\left[\boldsymbol{d},\boldsymbol{d}\right]&=&0,\qquad\qquad\quad \left[\boldsymbol{d},{\boldsymbol{\tau}^m}_n\right]\ =\ 0,\label{R+ComRel1}\\
\left[\boldsymbol{d},\boldsymbol{s}_{mn}\right]&=&-4\boldsymbol{s}_{mn},\quad\quad\,\,\, \left[\boldsymbol{d},\boldsymbol{h}^{mn}\right]\ =\ +4\boldsymbol{h}^{mn},\label{R+ComRel2}\\
\left[\boldsymbol{s}_{mn},\boldsymbol{s}_{pq}\right]&=&0,\qquad\,\qquad \left[\boldsymbol{h}^{mn},\boldsymbol{h}^{pq}\right]\ =\ 0,\\
\left[{\boldsymbol{\tau}^m}_n,{\boldsymbol{\tau}^p}_q\right]&=&2(\delta_n^p{\boldsymbol{\tau}^m}_q-\delta_q^m{\boldsymbol{\tau}^p}_n),\label{SL5Algebra}\\
\left[{\boldsymbol{\tau}^m}_n,\boldsymbol{s}_{pq}\right]&=&2(\delta^m_q\boldsymbol{s}_{np}-\delta^m_p\boldsymbol{s}_{nq}+\frac{2}{5}\delta^m_n\boldsymbol{s}_{pq}),\\
\left[{\boldsymbol{\tau}^m}_n,\boldsymbol{h}^{pq}\right]&=&2(\delta^p_n\boldsymbol{h}^{mq}-\delta^q_n\boldsymbol{h}^{mp}-\frac{2}{5}\delta^m_n\boldsymbol{h}^{pq}),\\
\left[\boldsymbol{s}_{mn},\boldsymbol{h}^{pq}\right]&=&2(\delta^p_m{\boldsymbol{\tau}^q}_n-\delta^q_m{\boldsymbol{\tau}^p}_n-\delta^p_n{\boldsymbol{\tau}^q}_m+\delta^q_n{\boldsymbol{\tau}^p}_m)-\frac{2}{5}\,\boldsymbol{d}\,\delta^{[p}_m\delta^{q]}_n\nonumber\\
&=&2(\delta^p_m{\boldsymbol{t}^q}_n-\delta^q_m{\boldsymbol{t}^p}_n
-\delta^p_n{\boldsymbol{t}^q}_m+\delta^q_n{\boldsymbol{t}^p}_m)\label{shComRel}
\end{eqnarray}
in which $\delta^{[p}_m\delta^{q]}_n=\frac{1}{2}(\delta^{p}_m\delta^{q}_n-\delta^{q}_m\delta^{p}_n)$. In the second line of \eqref{shComRel}, we have used \eqref{DefSL5Op} to rewrite the commutation relation in terms of the $GL(5)$ generators. Note also that \eqref{SL5Algebra} is the $SL(5)$ algebra. It follows that the $GL(5)$ branching rule for adjoint representation is given by
\begin{equation}\label{SO55GenDec}
\underbrace{\mathbf{45}}_{\boldsymbol{t}_{MN}}\ \rightarrow\ \underbrace{\mathbf{1}^{0}}_{\boldsymbol{d}}\,\oplus\,\underbrace{\mathbf{24}^{0}}_{{\boldsymbol{\tau}^m}_n}\,\oplus\,\underbrace{\mathbf{10}^{-4}}_{\boldsymbol{s}_{mn}}\,\oplus\,\underbrace{\overline{\mathbf{10}}^{+4}}_{\boldsymbol{h}^{mn}}
\end{equation}
where the $\mathbb{R}^+$ weights are determined from the relations \eqref{R+ComRel1} and \eqref{R+ComRel2}. 

\subsection{Spinor}\label{SpinorBranc}
Unlike the vector, decomposition of $SO(5,5)$ spinor representation under $GL(5)$ is not straightforward. To describe this branching rule, we begin with the following two sets of $USp(4)\sim SO(5)$ gamma matrices satisfying
\begin{eqnarray}
\left\{\gamma_a,\gamma_b\right\}&=&2 \delta_{ab}\mathds{1}_4, \qquad \delta_{ab}\ =\ \text{diag}(+,+,+,+,+),\\
\left\{\gamma_{\dot{a}},\gamma_{\dot{b}}\right\}&=&2 \delta_{\dot{a}\dot{b}}\mathds{1}_4, \qquad \delta_{\dot{a}\dot{b}}\ =\ \text{diag}(+,+,+,+,+)
\end{eqnarray}
where $a,b,\ldots=1,\ldots,5$ and $\dot{a},\dot{b},\ldots=\dot{1},\ldots,\dot{5}$ are two sets of $SO(5)$ vector indices raised and lowered by $\delta_{ab}$ and $\delta_{\dot{a}\dot{b}}$, respectively. For both sets of $SO(5)$ gamma matrices, we will use the following explicit representation
\begin{eqnarray}
\gamma_1&=&-\sigma_2\otimes\sigma_2,\quad\, \gamma_2\ =\ \mathds{1}_2\otimes\sigma_1,\quad
\gamma_3\ =\ \mathds{1}_2\otimes\sigma_3,\nonumber\\ 
\gamma_4&=&\sigma_1\otimes\sigma_2,\qquad \gamma_5\ =\ \sigma_3\otimes\sigma_2
\end{eqnarray}
where $\{\sigma_1,\sigma_2,\sigma_3\}$ are the usual Pauli matrices given by
\begin{equation}
\sigma_1\ = \ \begin{pmatrix} 0 & 1\\ 1 & 0 \end{pmatrix},\qquad \sigma_2\ = \ \begin{pmatrix} 0 & -i\\ i & 0 \end{pmatrix},\qquad\sigma_3\ = \ \begin{pmatrix} 1 & 0\\ 0 & -1 \end{pmatrix}.
\end{equation}
\indent Each gamma matrix is a $4\times4$ matrix with the index structure ${(\gamma_a)_\alpha}^\beta$ and ${(\gamma_{\dot{a}})_{\dot{\alpha}}}^{\dot{\beta}}$. Indices $\alpha,\beta,\ldots=1,\ldots,4$ and $\dot{\alpha},\dot{\beta},\ldots=\dot{1},\ldots,\dot{4}$ are two sets of $USp(4)$ fundamental or $SO(5)$ spinor indices raised and lowered through two identical $USp(4)$ symplectic forms
\begin{equation}\label{USp(4)Omegas}
\Omega_{\alpha\beta}\ =\ \Omega_{\dot{\alpha}\dot{\beta}}\ =\ \mathds{1}_2\otimes i\sigma_2
\end{equation}
satisfying 
\begin{equation}
\Omega_{\beta\alpha}\ =\ -\Omega_{\alpha\beta},\quad \Omega^{\alpha\beta}\ =\ (\Omega_{\alpha\beta})^*,\quad \Omega_{\alpha\beta}\Omega^{\beta\gamma}\ =\ -\delta^\gamma_\alpha,
\end{equation}
and similarly for $\Omega_{\dot{\alpha}\dot{\beta}}$. Therefore, the matrices $(\gamma_a)_{\alpha\beta}=\Omega_{\beta\gamma}{(\gamma_a)_\alpha}^\gamma$ satisfy 
\begin{equation}
(\gamma_a)_{\beta\alpha}\ =\ -(\gamma_a)_{\alpha\beta},\quad \Omega^{\alpha\beta}(\gamma_a)_{\alpha\beta}\ =\ 0,\quad ((\gamma_a)_{\alpha\beta})^*\ =\ \Omega^{\alpha\gamma}\Omega^{\beta\delta}(\gamma_a)_{\gamma\delta},
\end{equation}
and similarly for $(\gamma_{\dot{a}})_{\dot{\alpha}\dot{\beta}}=\Omega_{\dot{\beta}\dot{\gamma}}{(\gamma_{\dot{a}})_{\dot{\alpha}}}^{\dot{\gamma}}$.

The 32-dimensional $SO(5,5)$ gamma matrices, $\tilde{\mathbf{\Gamma}}_{\underline{A}}=(\tilde{\mathbf{\Gamma}}_a,\tilde{\mathbf{\Gamma}}_{\dot{a}})$ with $\underline{A}=1,...,10$, satisfying the Clifford algebra
\begin{equation}
\left\{\tilde{\mathbf{\Gamma}}_{\underline{A}},\tilde{\mathbf{\Gamma}}_{\underline{B}}\right\}\ =\ 2 \eta_{\underline{A}\underline{B}}\mathds{1}_{32}
\end{equation}
with $\eta_{\underline{A}\underline{B}} = \textrm{diag}(+,+,+,+,+,-,-,-,-,-)$ can be constructed from the $SO(5)$ gamma matrices as 
\begin{equation}\label{DiagSO55Def}
\tilde{\mathbf{\Gamma}}_a\ =\ (\sigma_1\otimes\mathds{1}_4\otimes \gamma_a)\qquad\text{and}\qquad
\tilde{\mathbf{\Gamma}}_{\dot{a}}\ =\ (i\sigma_2\otimes \gamma_{\dot{a}}\otimes\mathds{1}_4).
\end{equation}
The matrices $\mathds{A}$, $\mathds{B}$, and $\mathds{C}$, which respectively realize Dirac, complex,
and charge conjugation, have the following defining properties
\begin{equation}
(\tilde{\mathbf{\Gamma}}_{\underline{A}})^\dagger\ = \ -\mathds{A}\tilde{\mathbf{\Gamma}}_{\underline{A}}\mathds{A}^{-1},\qquad(\tilde{\mathbf{\Gamma}}_{\underline{A}})^*\ = \ -\mathds{B}\tilde{\mathbf{\Gamma}}_{\underline{A}}\mathds{B}^{-1},\qquad(\tilde{\mathbf{\Gamma}}_{\underline{A}})^T\ = \ -\mathds{C}\tilde{\mathbf{\Gamma}}_{\underline{A}}\mathds{C}^{-1}.
\end{equation}
In our explicit representation, the matrices $\mathds{A}$ and $\mathds{B}$ are given by
\begin{equation}
\mathds{A}\ =\ \tilde{\mathbf{\Gamma}}_{6}\tilde{\mathbf{\Gamma}}_{7}\tilde{\mathbf{\Gamma}}_{8}\tilde{\mathbf{\Gamma}}_{9}
\tilde{\mathbf{\Gamma}}_{10}\qquad\text{and}\qquad \mathds{B}\ =\ \mathds{1}_2\otimes\Omega\otimes\Omega\, .
\end{equation}
The charge conjugation matrix $\mathds{C}$ can be obtained from $\mathds{A}$ and $\mathds{B}$ through the relation
\begin{equation}
\mathds{C}\ =\ \mathds{B}^{T}\mathds{A}\, .
\end{equation}
The $SO(5,5)$ chirality matrix takes the following diagonal form
\begin{equation}
\tilde{\mathbf{\Gamma}}_*\ =\ \tilde{\mathbf{\Gamma}}_{1}...\tilde{\mathbf{\Gamma}}_{10}\ =\ \text{diag}(\mathds{1}_{16},-\mathds{1}_{16}).
\end{equation}
Therefore, as seen from the definition \eqref{DiagSO55Def}, $SO(5,5)$ gamma matrices are chirally decomposed as
\begin{equation}
\tilde{\mathbf{\Gamma}}_a\ =\ \begin{pmatrix}  0 & \mathds{1}_4\otimes\gamma_a \\ \mathds{1}_4\otimes\gamma_a & 0 \end{pmatrix}\qquad \text{and} \qquad \tilde{\mathbf{\Gamma}}_{\dot{a}}\ =\ \begin{pmatrix} 0  & \gamma_{\dot{a}}\otimes\mathds{1}_4 \\ -\gamma_{\dot{a}}\otimes\mathds{1}_4 & 0 \end{pmatrix}.
\end{equation}
Elements of the $16\times 16$ $SO(5)$ gamma matrices $\mathds{1}_4\otimes\gamma_a$ and $\gamma_{\dot{a}}\otimes\mathds{1}_4$ are denoted by the following index structure
\begin{equation}
\mathds{1}_4\otimes\gamma_a\ =\ {(\gamma_a)_{\alpha\dot{\alpha}}}^{\beta\dot{\beta}}\ =\ {(\gamma_a)_{\alpha}}^{\beta}\delta_{\dot{\alpha}}^{\dot{\beta}} \quad \text{and}\quad \gamma_{\dot{a}}\otimes\mathds{1}_4\ =\ {(\gamma_{\dot{a}})_{\alpha\dot{\alpha}}}^{\beta\dot{\beta}}\ =\ \delta_{\alpha}^{\beta}{(\gamma_{\dot{a}})_{\dot{\alpha}}}^{\dot{\beta}}\, .
\end{equation}
\indent On the other hand, we can split a $32$-dimendional $SO(5,5)$ spinor index into $\mathcal{A}=(A,A')$ for $A,B,...=1,..,16$ and $A',B',...=17,...,32$ so that
\begin{equation}
{(\tilde{\mathbf{\Gamma}}_{\underline{A}})_{\mathcal{A}}}^{\mathcal{B}}\ =\ \begin{pmatrix}  0 & {(\Gamma_{\underline{A}})_{A}}^{B'} \\ {(\Gamma_{\underline{A}})_{A'}}^{B} & 0 \end{pmatrix}.
\end{equation}
We can then relate these two decompositions of $SO(5,5)$ spinor indices into $A,A'$ and a pair of $USp(4)$ indices $(\alpha\dot{\alpha})$ by using the following transformation matrices 
\begin{eqnarray}\label{thepmatrix}
\boldsymbol{p}_A^{\alpha\dot{\alpha}}&=&\delta_A^\alpha\delta_1^{\dot{\alpha}}+\delta_A^{\alpha+4}\delta_2^{\dot{\alpha}}+\delta_A^{\alpha+8}\delta_3^{\dot{\alpha}}+\delta_A^{\alpha+12}\delta_4^{\dot{\alpha}},\nonumber\\
\boldsymbol{p}^A_{\alpha\dot{\alpha}}&=&\delta^A_\alpha\delta^1_{\dot{\alpha}}+\delta^A_{\alpha+4}\delta^2_{\dot{\alpha}}
+\delta^A_{\alpha+8}\delta^3_{\dot{\alpha}}+\delta^A_{\alpha+12}\delta^4_{\dot{\alpha}}\, .
\end{eqnarray}
These matrices satisfy the relations 
\begin{equation}
\boldsymbol{p}_A^{\alpha\dot{\alpha}}\boldsymbol{p}^B_{\alpha\dot{\alpha}}\ =\ \delta_A^B\qquad\text{and}\qquad\boldsymbol{p}_A^{\alpha\dot{\alpha}}\boldsymbol{p}^A_{\beta\dot{\beta}}\ =\ \delta^{\alpha}_{\beta}\delta^{\dot{\alpha}}_{\dot{\beta}}\, .
\end{equation}
We can now write chiral $SO(5,5)$ gamma matrices in terms of the $SO(5)$ ones as
\begin{eqnarray}\label{ExplicitGamma1}
{(\Gamma_a)_{A}}^{B'}&=& \boldsymbol{p}_A^{\alpha\dot{\alpha}}{(\gamma_a)_{\alpha\dot{\alpha}}}^{\beta\dot{\beta}}\boldsymbol{p}^{B'}_{\beta\dot{\beta}},\qquad
{(\Gamma_{\dot{a}})_{A}}^{B'}\ =\ \boldsymbol{p}_A^{\alpha\dot{\alpha}}{(\gamma_{\dot{a}})_{\alpha\dot{\alpha}}}^{\beta\dot{\beta}}\boldsymbol{p}^{B'}_{\beta\dot{\beta}},\\
{(\Gamma_a)_{A'}}^{B}&=& \boldsymbol{p}_{A'}^{\alpha\dot{\alpha}}{(\gamma_a)_{\alpha\dot{\alpha}}}^{\beta\dot{\beta}}\boldsymbol{p}^B_{\beta\dot{\beta}},\qquad
{(\Gamma_{\dot{a}})_{A'}}^{B}\ =\ -\boldsymbol{p}_{A'}^{\alpha\dot{\alpha}}{(\gamma_{\dot{a}})_{\alpha\dot{\alpha}}}^{\beta\dot{\beta}}\boldsymbol{p}^B_{\beta\dot{\beta}}.
\end{eqnarray}
To raise and lower the spinor indices $A$ and $A'$, we use the charge conjugation matrix which in this basis takes the form of
\begin{equation}\label{CCDef}
\mathds{C}\ =\ \begin{pmatrix}  0 & \Omega\otimes\Omega \\ -\Omega\otimes\Omega & 0 \end{pmatrix}.
\end{equation}
Its elements can be explicitly expressed as 
\begin{equation}\label{CCDef2}
\mathds{C}_{\mathcal{A}\mathcal{B}}\ =\ \begin{pmatrix}  0 & \boldsymbol{c}_{AB'} \\ \boldsymbol{c}_{A'B} & 0 \end{pmatrix}\qquad\text{and}\qquad\mathds{C}^{\mathcal{A}\mathcal{B}}\ =\ \begin{pmatrix} 0  & \boldsymbol{c}^{AB'} \\ \boldsymbol{c}^{A'B} & 0 \end{pmatrix}.
\end{equation}
The $16\times 16$ matrix $\boldsymbol{c}_{A'B}$ is antisymmetric, $\boldsymbol{c}_{A'B}=-\boldsymbol{c}_{BA'}$. Similarly, the matrix $\boldsymbol{c}^{AB'}$ satisfying the relations 
\begin{equation}
\boldsymbol{c}_{AC'}\boldsymbol{c}^{C'B}=-\delta_A^B \qquad \textrm{and}\qquad  \boldsymbol{c}_{A'C}\boldsymbol{c}^{CB'}=-\delta_{A'}^{B'}
\end{equation}
is also antisymmetric $\boldsymbol{c}^{A'B}=-\boldsymbol{c}^{BA'}$.
\\
\indent By raising and lowering the $SO(5,5)$ spinor index, we can define gamma matrices with all upper or lower indices
\begin{eqnarray}
(\tilde{\mathbf{\Gamma}}_{\underline{A}})^{\mathcal{A}\mathcal{B}}& =& \mathds{C}^{\mathcal{A}\mathcal{C}}{(\tilde{\mathbf{\Gamma}}_{\underline{A}})_{\mathcal{C}}}^{\mathcal{B}}\ =\ \begin{pmatrix}  (\Gamma_{\underline{A}})^{AB} & 0 \\ 0 & (\Gamma_{\underline{A}})^{A'B'} \end{pmatrix},\\
(\tilde{\mathbf{\Gamma}}_{\underline{A}})_{\mathcal{A}\mathcal{B}}& =& {(\tilde{\mathbf{\Gamma}}_{\underline{A}})_{\mathcal{A}}}^{\mathcal{C}}\mathds{C}_{\mathcal{C}\mathcal{B}}\ =\ \begin{pmatrix}  (\Gamma_{\underline{A}})_{AB} & 0 \\ 0 & (\Gamma_{\underline{A}})_{A'B'} \end{pmatrix}
\end{eqnarray}
in which 
\begin{eqnarray}\label{RaiseLowerGamma1}
(\Gamma_{\underline{A}})^{AB}&=&\boldsymbol{c}^{AC'}{(\Gamma_{\underline{A}})_{C'}}^{B},\qquad(\Gamma_{\underline{A}})^{A'B'}\ =\ \boldsymbol{c}^{A'C}{(\Gamma_{\underline{A}})_{C}}^{B'},\nonumber\\ 
(\Gamma_{\underline{A}})_{AB}&=&{(\Gamma_{\underline{A}})_{A}}^{C'}\boldsymbol{c}_{C'B},\qquad(\Gamma_{\underline{A}})_{A'B'}\ =\ {(\Gamma_{\underline{A}})_{A'}}^{C}\boldsymbol{c}_{CB'}. 
\end{eqnarray}
In terms of the $USp(4)$ symplectic forms $\Omega_{\alpha\beta}$ and $\Omega_{\dot{\alpha}\dot{\beta}}$ which can be used to raise or lower $USp(4)$ fundamental indices, we can write the matrices $\boldsymbol{c}_{A'B}$ and $\boldsymbol{c}^{A'B}$ as
\begin{equation}
\boldsymbol{c}_{A'B}\ =\ \boldsymbol{p}_{A'}^{\alpha\dot{\alpha}}\boldsymbol{p}_B^{\beta\dot{\beta}}\Omega_{\alpha\beta}\Omega_{\dot{\alpha}\dot{\beta}}
\qquad\text{and}\qquad\boldsymbol{c}^{A'B}\ =\ \boldsymbol{p}^{A'}_{\alpha\dot{\alpha}}\boldsymbol{p}^B_{\beta\dot{\beta}}\Omega^{\alpha\beta}\Omega^{\dot{\alpha}\dot{\beta}}.
\end{equation}
With all these, we can eventually find the following relations
\begin{eqnarray}\label{RaiseLowerGamma2}
(\Gamma_a)^{AB}&=&\boldsymbol{p}^A_{\alpha\dot{\alpha}}\boldsymbol{p}^{B}_{\beta\dot{\beta}}(\gamma_a)^{\alpha\dot{\alpha},\beta\dot{\beta}},\qquad(\Gamma_{\dot{a}})^{AB}=-\boldsymbol{p}^A_{\alpha\dot{\alpha}}\boldsymbol{p}^{B}_{\beta\dot{\beta}}(\gamma_{\dot{a}})^{\alpha\dot{\alpha},\beta\dot{\beta}},\nonumber\\
(\Gamma_a)^{A'B'}&=&\boldsymbol{p}^{A'}_{\alpha\dot{\alpha}}\boldsymbol{p}^{B'}_{\beta\dot{\beta}}(\gamma_a)^{\alpha\dot{\alpha},\beta\dot{\beta}},\qquad(\Gamma_{\dot{a}})^{A'B'}\ =\ \boldsymbol{p}^{A'}_{\alpha\dot{\alpha}}\boldsymbol{p}^{B'}_{\beta\dot{\beta}}(\gamma_{\dot{a}})^{\alpha\dot{\alpha},\beta\dot{\beta}},\nonumber\\
(\Gamma_a)_{AB}&=&\boldsymbol{p}_{A}^{\alpha\dot{\alpha}}\boldsymbol{p}_{B}^{\beta\dot{\beta}}(\gamma_a)_{\alpha\dot{\alpha},\beta\dot{\beta}},\qquad(\Gamma_{\dot{a}})_{AB} =\ \boldsymbol{p}_{A}^{\alpha\dot{\alpha}}\boldsymbol{p}_{B}^{\beta\dot{\beta}}(\gamma_{\dot{a}})_{\alpha\dot{\alpha},\beta\dot{\beta}},\nonumber\\
(\Gamma_a)_{A'B'}&=&\boldsymbol{p}_{A'}^{\alpha\dot{\alpha}}\boldsymbol{p}_{B'}^{\beta\dot{\beta}}(\gamma_a)_{\alpha\dot{\alpha},\beta\dot{\beta}},\qquad(\Gamma_{\dot{a}})_{A'B'}\ =\ -\boldsymbol{p}_{A'}^{\alpha\dot{\alpha}}\boldsymbol{p}_{B'}^{\beta\dot{\beta}}(\gamma_{\dot{a}})_{\alpha\dot{\alpha},\beta\dot{\beta}}
\end{eqnarray}
with
\begin{eqnarray}
(\gamma_a)^{\alpha\dot{\alpha},\beta\dot{\beta}}&=&\Omega^{\alpha\delta}\Omega^{\dot{\alpha}\dot{\delta}}{(\gamma_a)_{\delta\dot{\delta}}}^{\beta\dot{\beta}}\ = \ (\gamma_a)^{\alpha\beta}\Omega^{\dot{\alpha}\dot{\beta}},\nonumber\\
(\gamma_{\dot{a}})^{\alpha\dot{\alpha},\beta\dot{\beta}}&=&\Omega^{\alpha\delta}\Omega^{\dot{\alpha}\dot{\delta}}{(\gamma_{\dot{a}})_{\delta\dot{\delta}}}^{\beta\dot{\beta}}\ = \ \Omega^{\alpha\beta}(\gamma_{\dot{a}})^{\dot{\alpha}\dot{\beta}},\nonumber\\
(\gamma_a)_{\alpha\dot{\alpha},\beta\dot{\beta}}&=&\Omega_{\beta\delta}\Omega_{\dot{\beta}\dot{\delta}}{(\gamma_a)_{\alpha\dot{\alpha}}}^{\delta\dot{\delta}}\ = \ (\gamma_a)_{\alpha\beta}\Omega_{\dot{\alpha}\dot{\beta}},\nonumber\\
(\gamma_{\dot{a}})_{\alpha\dot{\alpha},\beta\dot{\beta}}&=&\Omega_{\beta\delta}\Omega_{\dot{\beta}\dot{\delta}}{(\gamma_{\dot{a}})_{\alpha\dot{\alpha}}}^{\delta\dot{\delta}}\ = \ \Omega_{\alpha\beta}(\gamma_{\dot{a}})_{\dot{\alpha}\dot{\beta}}.
\end{eqnarray}
\indent We now transform all these results to the basis with off-diagonal $\eta_{MN}$ given in \eqref{off-diag-eta}. Denoting $SO(5,5)$ gamma matrices in this basis by $\tilde{\mathbf{\Gamma}}_M$, we can write the corresponding Clifford algebra as
\begin{equation}
\left\{\tilde{\mathbf{\Gamma}}_M,\tilde{\mathbf{\Gamma}}_N\right\}\ =\ 2 \eta_{MN}\mathds{1}_{32}\, .
\end{equation}
From \cite{6D_Max_Gauging}, the relation between diagonal and off-diagonal $\eta$ is given by 
\begin{equation}
\eta_{MN}\ =\ {\mathbb{M}_M}^{\underline{A}}\,{\mathbb{M}_N}^{\underline{B}}\,\eta_{\underline{A}\underline{B}}
\end{equation}
with
\begin{equation}\label{offDiagTrans}
\mathbb{M}\ =\ \frac{1}{\sqrt{2}}\begin{pmatrix} \mathds{1}_5 & \mathds{1}_5 \\ \mathds{1}_5 & -\mathds{1}_5 \end{pmatrix}.
\end{equation}
We can then find the following relation between these two sets of gamma matrices  
\begin{equation}
\tilde{\mathbf{\Gamma}}_M\ =\ {\mathbb{M}_M}^{\underline{A}}\tilde{\mathbf{\Gamma}}_{\underline{A}}
\end{equation}
with the same chiral decomposition of the form
\begin{equation}
{(\tilde{\mathbf{\Gamma}}_M)_{\mathcal{A}}}^{\mathcal{B}}\ =\ \begin{pmatrix}  0 & {(\Gamma_M)_{A}}^{B'} \\ {(\Gamma_M)_{A'}}^{B} & 0 \end{pmatrix}.
\end{equation}
\indent Moreover, we can still raise and lower the chirally decomposed spinor indices with the charge conjugation matrix given in \eqref{CCDef} such that
\begin{equation}
(\tilde{\mathbf{\Gamma}}_M)_{\mathcal{A}\mathcal{B}}\ =\ {(\tilde{\mathbf{\Gamma}}_M)_{\mathcal{A}}}^{\mathcal{C}}\mathds{C}_{\mathcal{C}\mathcal{B}}\ =\ \begin{pmatrix}  (\Gamma_M)_{AB} & 0 \\ 0 & (\Gamma_M)_{A'B'} \end{pmatrix}
\end{equation}
with $(\Gamma_M)_{AB}={(\Gamma_M)_{A}}^{C'}\boldsymbol{c}_{C'B}$ and $(\Gamma_M)_{A'B'}={(\Gamma_M)_{A'}}^{C}\boldsymbol{c}_{CB'}$. We will see in the following analysis that $(\Gamma_M)_{AB}$ play an important role in determining specific forms of the embedding tensor.

In spinor representation, the $SO(5,5)$ generators satisfying \eqref{SO(5,5)algebra} are given by
\begin{equation}\label{SO55GenSpin}
{(\boldsymbol{t}_{MN})_A}^B\ =\ {(\Gamma_{MN})_A}^B.
\end{equation}
In $32\times32$ representation, we can write
\begin{eqnarray}
{(\tilde{\mathbf{\Gamma}}_{MN})_{\mathcal{A}}}^{\mathcal{B}}&=&\ \frac{1}{2}\left({(\tilde{\mathbf{\Gamma}}_M)_{\mathcal{A}}}^{\mathcal{C}}{(\tilde{\mathbf{\Gamma}}_N)_{\mathcal{C}}}^{\mathcal{B}}-{(\tilde{\mathbf{\Gamma}}_N)_{\mathcal{A}}}^{\mathcal{C}}{(\tilde{\mathbf{\Gamma}}_M)_{\mathcal{C}}}^{\mathcal{B}}\right)\\
&=&\ \begin{pmatrix} {(\Gamma_{MN})_A}^B  & 0 \\ 0 & {(\Gamma_{MN})_{A'}}^{B'} \end{pmatrix}\nonumber
\end{eqnarray}
with
\begin{eqnarray}
{(\Gamma_{MN})_A}^B&=&\frac{1}{2}\left[{(\Gamma_M)_A}^{C'}{(\Gamma_N)_{C'}}^B-{(\Gamma_N)_A}^{C'}{(\Gamma_M)_{C'}}^B\right],\\
{(\Gamma_{MN})_{A'}}^{B'}&=&\frac{1}{2}\left[{(\Gamma_M)_{A'}}^C{(\Gamma_N)_C}^{B'}-{(\Gamma_N)_{A'}}^C{(\Gamma_M)_C}^{B'}\right].
\end{eqnarray}
It should be noted that the $SO(5,5)$ generators in spinor representation given in \eqref{SO55GenSpin} also decompose according to \eqref{SO55GenDec} and satisfy the same algebra given in \eqref{R+ComRel1} to \eqref{shComRel} for vector representation. 

As pointed out in \cite{6D_Max_Gauging}, the branching rules for spinor and conjugate spinor representations of $SO(5,5)$ are respectively given by
\begin{equation}
\mathbf{16}_s\ \rightarrow\ \overline{\mathbf{5}}^{+3}\,\oplus\,\mathbf{10}^{-1}\,\oplus\,\mathbf{1}^{-5}\qquad\text{and}\qquad\mathbf{16}_c\ \rightarrow\ \mathbf{5}^{-3}\,\oplus\,\overline{\mathbf{10}}^{+1}\,\oplus\,\mathbf{1}^{+5}.
\end{equation}
To find the corresponding decompositions of spinor indices, we define the following transformation matrices
\begin{eqnarray}
\mathbb{T}_{Am}&=&\frac{1}{2\sqrt{2}}(\Gamma_m)_{AB}\boldsymbol{p}^B_{\alpha\beta}\Omega^{\alpha\beta},\label{TranMatTDef1}\\
\mathbb{T}_{A}^{mn}&=&\frac{1}{4\sqrt{2}}{(\Gamma^{mn})_A}^B\boldsymbol{p}_B^{\alpha\beta}\Omega_{\alpha\beta},\label{TranMatTDef2}\\
\mathbb{T}_{A\ast}&=&\frac{1}{10}{({\Gamma^m}_m)_A}^B\boldsymbol{p}_B^{\alpha\beta}\Omega_{\alpha\beta}\, .\label{TranMatTDef3}
\end{eqnarray}
In these equations, $\boldsymbol{p}^A_{\alpha\beta}$ matrices are defined in the same way as $\boldsymbol{p}^A_{\alpha\dot{\beta}}$ in \eqref{thepmatrix}. 

We can now decompose an $SO(5,5)$ spinor in $\mathbf{16}_s$ representation as
\begin{equation}
\Psi_A\ =\ \mathbb{T}_{Am}\Psi^m+\mathbb{T}_{A}^{mn}\Psi_{mn}+\mathbb{T}_{A\ast}\Psi_\ast
\end{equation}
with $\Psi_{mn}=\Psi_{[mn]}$. The commutation relations between these components and the $\mathbb{R}^+$ generator are given by
\begin{eqnarray}
\left[\boldsymbol{d},\mathbb{T}_{Am}\Psi^m\right]&=&+3\mathbb{T}_{Am}\Psi^m,\\
\left[\boldsymbol{d},\mathbb{T}_{A}^{mn}\Psi_{mn}\right]&=&-\mathbb{T}_{A}^{mn}\Psi_{mn},\\
\left[\boldsymbol{d},\mathbb{T}_{A\ast}\Psi_\ast\right]&=&-5\mathbb{T}_{A\ast}\Psi_\ast
\end{eqnarray}
in accord with the branching rule 
\begin{equation}
\underbrace{\mathbf{16}_s}_{\Psi_A}\ \rightarrow\ \underbrace{\overline{\mathbf{5}}^{+3}}_{\Psi^m}\,\oplus\,\underbrace{\mathbf{10}^{-1}}_{\Psi_{[mn]}}\,\oplus\,\underbrace{\mathbf{1}^{-5}}_{\Psi_\ast}\, .
\end{equation}
\indent The inverse matrices of $\mathbb{T}_A$ are simply given by their complex conjugation $\mathbb{T}^A=(\mathbb{T}_A)^{-1}=(\mathbb{T}_A)^*$ satisfying
\begin{eqnarray}
\mathbb{T}^{Am}\mathbb{T}_{An}\ =\ \delta^m_n,\qquad\quad\mathbb{T}^{A}_{mn}\mathbb{T}_{A}^{pq}& =& \delta^{[p}_{m}\delta^{q]}_{n},\qquad\mathbb{T}^{A}_\ast\mathbb{T}_{A\ast}\ =\ 1,\nonumber\\
\mathbb{T}^{Am}\mathbb{T}_{Anp}\ =\ 0,\qquad\quad\,\mathbb{T}^{Am}\mathbb{T}_{A\ast}& =& 0,\qquad\quad\mathbb{T}^{A}_{mn}\mathbb{T}_{A\ast}\ =\ 0
\end{eqnarray}
together with
\begin{equation}
\mathbb{T}^{Am}\mathbb{T}_{Bm}+\mathbb{T}^{A}_{mn}\mathbb{T}_{B}^{mn}+\mathbb{T}^{A}_\ast\mathbb{T}_{B\ast}=\delta^A_B\, .
\end{equation}
In addition, we also note that a complex conjugation of the $SO(5,5)$ gamma matrices is related to raising the indices $((\Gamma_M)_{AB})^*=(\Gamma^M)^{AB}$. We can then similarly decompose a conjugate spinor of $SO(5,5)$ transforming in $\mathbf{16}_c$ as follows
\begin{equation}
\Psi^A\ =\ \mathbb{T}^{Am}\Psi_m+\mathbb{T}^{A}_{mn}\Psi^{mn}+\mathbb{T}^{A}_{\ast}\Psi_\ast\, .
\end{equation}
The following commutation relations 
\begin{eqnarray}
\left[\boldsymbol{d},\mathbb{T}^{Am}\Psi_m\right]&=&-3\mathbb{T}^{Am}\Psi_m,\\
\left[\boldsymbol{d},\mathbb{T}^{A}_{mn}\Psi^{mn}\right]&=&+\mathbb{T}^{A}_{mn}\Psi^{mn},\\
\left[\boldsymbol{d},\mathbb{T}^{A}_{\ast}\Psi_\ast\right]&=&+5\mathbb{T}^{A}_{\ast}\Psi_\ast
\end{eqnarray}
imply the branching rule
\begin{equation}\label{16cDec}
\underbrace{\mathbf{16}_c}_{\Psi^A}\ \rightarrow\ \underbrace{\mathbf{5}^{-3}}_{\Psi_m}\,\oplus\,\underbrace{\overline{\mathbf{10}}^{+1}}_{\Psi^{[mn]}}\,\oplus\,\underbrace{\mathbf{1}^{+5}}_{\Psi_\ast}.
\end{equation}

\subsection{Vector-spinor}\label{Apptheta}
The vector-spinor of $SO(5,5)$ we are interested in is given by $\theta^{AM}\in\mathbf{144}_c$, which parameterizes the embedding tensor. It transforms according to
\begin{equation}\label{thetaTrans}
\left[\boldsymbol{t}_{MN},\theta^{AP}\right]\ =\ -{(\boldsymbol{t}_{MN})_Q}^P\theta^{AQ}-{(\boldsymbol{t}_{MN})_B}^A\theta^{BP}.
\end{equation}
Here, $\theta^{AM}$ is a ($16\times10$) matrix subject to 
\begin{equation}\label{LC}
(\Gamma_M)_{AB}\,\theta^{BM}\ =\ 0
\end{equation}
which is the linear constraint required by supersymmetry, reducing $160$ components of the $\theta^{AM}$ to $144$ in $\mathbf{144}_c$ representation. 

To determine the decomposition of the vector-spinor representation under $GL(5)$, we first split the $SO(5,5)$ vector index $M$ as $\theta^{AM}=(\theta^{Am},\theta^{A}_m)$. Then, with the inverse of the transformation matrices $\mathbb{T}_A$ given in \eqref{TranMatTDef1} to \eqref{TranMatTDef3}, $\theta^{Am}$ and $\theta^{A}_m$ can be further decomposed into the following six components
\begin{eqnarray}
\theta^{Am}&=&\mathbb{T}^{An}{(\vartheta_1)_n}^{m}+\mathbb{T}^{A}_{np}(\vartheta_3)^{np,m}+\mathbb{T}^{A}_\ast(\vartheta_5)^{m},\label{SixBlockTheta1}\\
 \theta^{A}_m&=&\mathbb{T}^{An}(\vartheta_2)_{nm}+\mathbb{T}^{A}_{np} (\vartheta_4)^{np}_{m}+\mathbb{T}^{A}_\ast(\vartheta_6)_m\, .\label{SixBlockTheta2}
\end{eqnarray}
It is straightforward to show that their commutation relations with the $\mathbb{R}^+$ generators in both vector and spinor representations are given by
\begin{eqnarray}
\left[\boldsymbol{d},\mathbb{T}^{An}{(\vartheta_1)_n}^{m}\right]&=&-5\mathbb{T}^{An}{(\vartheta_1)_n}^{m},\qquad \left[\boldsymbol{d},\mathbb{T}^{An}(\vartheta_2)_{nm}\right]\ =\ -\mathbb{T}^{An}(\vartheta_2)_{nm},\nonumber\\\left[\boldsymbol{d},\mathbb{T}^{A}_{np}(\vartheta_3)^{np,m}\right]&=&-\mathbb{T}^{A}_{np}(\vartheta_3)^{np,m},\qquad\, \left[\boldsymbol{d},\mathbb{T}^{A}_{np} (\vartheta_4)^{np}_{m}\right]\ =\ +3\mathbb{T}^{A}_{np} (\vartheta_4)^{np}_{m},\nonumber\\\left[\boldsymbol{d},\mathbb{T}^{A}_\ast(\vartheta_5)^{m}\right]&=&+3\mathbb{T}^{A}_\ast(\vartheta_5)^{m},\qquad\quad \left[\boldsymbol{d},\mathbb{T}^{A}_\ast(\vartheta_6)_m\right]\ =\ +7\mathbb{T}^{A}_\ast(\vartheta_6)_m.\nonumber\\
\end{eqnarray}
The branching rule for $\mathbf{144}_c$ representation is then given by 
\begin{equation}\label{thetaDec}
\underbrace{\mathbf{144}_c}_{\theta^{AM}}\ \rightarrow\ \underbrace{\overline{\mathbf{5}}^{+3}}_{J^m}\,\oplus\,\underbrace{\mathbf{5}^{+7}}_{K_m}\,\oplus\,\underbrace{\mathbf{10}^{-1}}_{Z_{mn}}\,\oplus\,\underbrace{\mathbf{15}^{-1}}_{Y_{mn}}\,\oplus\,\underbrace{\mathbf{24}^{-5}}_{{S_m}^n}\,\oplus\,\underbrace{\overline{\mathbf{40}}^{-1}}_{U^{mn,p}}\,\oplus\,\underbrace{\overline{\mathbf{45}}^{+3}}_{W^{np}_m}
\end{equation}
in agreement with that given in \cite{6D_Max_Gauging}. We now explicitly construct a number of possible embedding tensors arising from various components of the above decomposition. 

\subsubsection{$\mathbf{24}^{-5}$ representation}
With only $\vartheta_1\neq 0$, we have $\theta^{A}_m=0$, and $\theta^{Am}$ is parametrized by a $5\times5$ matrix ${(\vartheta_1)_n}^{m}$. Therefore, the embedding tensor is given by 
\begin{equation}
\theta^{AM}=\left(\,\mathbb{T}^{An}{(\vartheta_1)_n}^{m}\,,\, 0\,\right).
\end{equation}
The linear constraint \eqref{LC} implies that ${(\vartheta_1)_n}^{m}$ is traceless or
\begin{equation}
{(\vartheta_1)_n}^{m}={S_n}^m
\end{equation}
for ${S_m}^m=0$. This leads to an embedding tensor in $\mathbf{24}^{-5}$ representation of $GL(5)$ given by
\begin{equation}
\theta_{\mathbf{24}^{-5}}^{AM}=\left(\,\mathbb{T}^{An}{S_n}^{m}\,,\, 0\,\right).
\end{equation}

\subsubsection{$\mathbf{15}^{-1}$ representation}
For $\theta^{Am}=0$, $\theta^{A}_m$ is parametrized by a $5\times5$ matrix $(\vartheta_2)_{mn}$ which can further be decomposed in terms of symmetric and antisymmetric parts, $Y_{mn}=Y_{(mn)}$ and $Z_{mn}=Z_{[mn]}$, as
\begin{equation}
\theta^{A}_m=\mathbb{T}^{An}(\vartheta_2)_{nm}=\mathbb{T}^{An}(Y_{nm} + Z_{nm}).
\end{equation}
The linear constraint \eqref{LC} requires $Z_{mn}=0$, so the embedding tensor is given by
\begin{equation}
\theta_{\mathbf{15}^{-1}}^{AM}=\left(\,0\,,\,\mathbb{T}^{An}Y_{nm}\,\right)
\end{equation}
in $\mathbf{15}^{-1}$ representation of $GL(5)$.

\subsubsection{$\overline{\mathbf{40}}^{-1}$ representation}
With only $(\vartheta_3)^{np,m}$ non-vanishing, we have $\theta^{A}_m=0$ and $\theta^{Am}$ given by
\begin{equation}
\theta^{Am}=\mathbb{T}^{A}_{np}(\vartheta_3)^{np,m}\, .
\end{equation}
The tensor $(\vartheta_3)^{np,m}$ can in turn be parametrized as 
\begin{equation}
(\vartheta_3)^{np,m}\ = \ U^{np,m} + \frac{1}{2}\varepsilon^{mnpqr} \zeta_{qr}\, .
\end{equation}
$U^{mn,p}=U^{[mn],p}$ with $U^{[mn,p]}=0$ and $\zeta_{mn}=\zeta_{[mn]}$ correspond to $\overline{\mathbf{40}}^{-1}$ and $\mathbf{10}^{-1}$ representations, respectively. The condition \eqref{LC} requires $\zeta_{qr}=0$ resulting in the embedding tensor in $\overline{\mathbf{40}}^{-1}$ representation of the form
\begin{equation}\label{fulltheta3}
\theta_{\overline{\mathbf{40}}^{-1}}^{AM}=\left(\,\mathbb{T}^{A}_{np}U^{np,m}\,,\,0\,\right).
\end{equation}

\subsubsection{$\mathbf{10}^{-1}$ representation}
Turning on $\mathbf{10}^{-1}$ irreducible part of both $(\vartheta_2)_{nm}$ and $(\vartheta_3)^{np,m}$ by setting $Y_{mn}=0$ and $U^{mn,p}=0$, we find the embedding tensor of the form
\begin{equation}\label{fulltheta3}
\theta_{\mathbf{10}^{-1}}^{AM}=\left(\,\frac{1}{2}\mathbb{T}^{A}_{np}\varepsilon^{mnpqr} \zeta_{qr}\,,\,\mathbb{T}^{An}Z_{nm}\,\right).
\end{equation}
The condition \eqref{LC} is satisfied for
\begin{equation}
\zeta_{mn}=\frac{\sqrt{2}}{3}Z_{mn}\, .
\end{equation}
Therefore, the embedding tensor in $\mathbf{10}^{-1}$ representation is given by
\begin{equation}
\theta_{\mathbf{10}^{-1}}^{AM}=\left(\frac{1}{3\sqrt{2}}\mathbb{T}^{A}_{np}\varepsilon^{mnpqr} Z_{qr}\,,\,\mathbb{T}^{An}Z_{nm}\,\right).
\end{equation}

\subsubsection{${\overline{\mathbf{45}}}^{+3}$ representation}
In this case, we consider non-vanishing $(\vartheta_4)^{np}_{m}$ which can be decomposed into $\overline{\mathbf{45}}^{+3}$ and $\overline{\mathbf{5}}^{+3}$ irreducible representations of the form
\begin{equation}
(\vartheta_4)^{np}_{m}\ =\ W^{np}_{m}+J^{[n}\delta^{p]}_m
\end{equation}
with $W^{np}_m=W^{[np]}_m$ satisfying $W^{nm}_m=0$. The linear constraint \eqref{LC} requires $J^m=0$ leading to the embedding tensor in $\overline{\mathbf{45}}^{+3}$ representation given by
\begin{equation}\label{firsttheta4}
\theta_{{\overline{\mathbf{45}}}^{+3}}^{AM}=\left(\,0\,,\,\mathbb{T}^{A}_{np} W^{np}_{m}\,\right).
\end{equation}

\subsubsection{$\overline{\mathbf{5}}^{+3}$ representation}
We now consider non-vanishing $\overline{\mathbf{5}}^{+3}$ components from both $(\vartheta_5)^{m}$ and $(\vartheta_4)^{m}$ in terms of which the embedding tensor is given by
\begin{equation}
\theta_{\overline{\mathbf{5}}^{+3}}^{AM}=\left(\,\mathbb{T}^{A}_{\ast}\xi^m\,,\,\mathbb{T}^{A}_{np} J^{[n}\delta^{p]}_m\,\right).
\end{equation}
This satisfies the linear constraint \eqref{LC} for
\begin{equation}
\xi^m=-\frac{2\sqrt{2}}{5}J^m.
\end{equation}
Therefore, we find the embedding tensor in $\overline{\mathbf{5}}^{+3}$ representation given by 
\begin{equation}
\theta_{\overline{\mathbf{5}}^{+3}}^{AM}=\left(\,-\frac{2\sqrt{2}}{5}\mathbb{T}^{A}_{\ast}J^m\,,\,\mathbb{T}^{A}_{np} J^{[n}\delta^{p]}_m\,\right).
\end{equation}

\subsubsection{$\mathbf{5}^{+7}$ representation}
Finally, we consider $(\vartheta_6)_{m}$ corresponding to $\mathbf{5}^{+7}$ representation. This can be parameterized by an arbitrary $GL(5)$ vector of the form $(\vartheta_6)_{m}=K_m$. The corresponding embedding tensor is also in $\mathbf{5}^{+7}$ representation and takes the form
\begin{equation}
\theta_{\mathbf{5}^{+7}}^{AM}=\left(\,0\,,\,\mathbb{T}^{A}_{\ast}K_m\,\right)
\end{equation}
which automatically satisfies the condition \eqref{LC}.
\\
\indent We end this appendix by giving the full parametrization of the embedding tensor $\theta^{AM}=(\theta^{Am},\theta^{A}_m)$ under $GL(5)$ 
\begin{eqnarray}
\theta^{Am}&=&\mathbb{T}^{An}{S_n}^m+\mathbb{T}^{A}_{np}\left(U^{np,m} + \frac{1}{3\sqrt{2}}\varepsilon^{mnpqr}Z_{qr}\right)-\frac{2\sqrt{2}}{5}\mathbb{T}^{A}_{\ast}J^m,\label{ExpliTheta1}\\
 \theta^{A}_m&=&\mathbb{T}^{An}(Y_{nm} + Z_{nm})+\mathbb{T}^{A}_{np}(W^{np}_{m}+J^{[n}\delta^{p]}_m)+\mathbb{T}^{A}_\ast K_m\, .\label{ExpliTheta2}
\end{eqnarray}

\section{Symplectic-Majorana-Weyl spinors in six dimensions}\label{SMWspinor}
In this appendix, we collect conventions and fundamental relations involving irreducible spinors in six-dimensional space-time used throughout this work. In six dimensions, there exist Dirac spinors with $16$ real components. The Dirac spinors are reducible and can be decomposed into two irreducible Weyl spinors of opposite chirality with $8$ real components each. The six-dimensional Clifford algebra is defined by the relation
\begin{equation}\label{6DClifford}
\{\hat{\gamma}_{\hat{\mu}}\hat{\gamma}_{\hat{\nu}}+\hat{\gamma}_{\hat{\nu}}\hat{\gamma}_{\hat{\mu}}\}=2\eta_{\hat{\mu}\hat{\nu}}\mathds{1}_8.
\end{equation}
Here, $\hat{\gamma}_{\hat{\mu}}$ are $8\times8$ Dirac matrices and $\eta_{\hat{\mu}\hat{\nu}}=\text{diag}(-1,+1,+1,+1,+1,+1)$ with $\hat{\mu},\hat{\nu},...=0,1,...,5$ being six-dimensional flat space-time indices. We will use the following explicit representation of the gamma matrices 
\begin{eqnarray}\label{6Dgamma}
& &\hat{\gamma}_{0}=\sigma_1\otimes i\sigma_2\otimes\mathds{1}_2,\qquad \hat{\gamma}_{1}=\sigma_2\otimes\mathds{1}_2\otimes \sigma_2,\qquad\hat{\gamma}_{2}=\sigma_1\otimes \sigma_1\otimes\mathds{1}_2,\nonumber\\
& &\hat{\gamma}_{3}=\sigma_2\otimes\mathds{1}_2\otimes \sigma_1,\qquad\hat{\gamma}_{4}=\sigma_2\otimes\mathds{1}_2\otimes \sigma_3,\qquad \hat{\gamma}_{5}=\sigma_1\otimes \sigma_3\otimes\mathds{1}_2\, .
\end{eqnarray}
In this representation, the Dirac, complex, and charge conjugation matrices are respectively given by
\begin{equation}\label{6DABCExpli}
\hat{\mathds{A}}=\hat{\gamma}_0,\qquad\hat{\mathds{B}}=-i\hat{\gamma}_3\hat{\gamma}_4,\quad\hat{\mathds{C}}=i\hat{\gamma}_0\hat{\gamma}_3\hat{\gamma}_4\, .
\end{equation}
They satisfy the following relations
\begin{equation}
(\hat{\gamma}_{\hat{\mu}})^\dagger\ = \ -\hat{\mathds{A}}\hat{\gamma}_{\hat{\mu}}\hat{\mathds{A}}^{-1},\qquad(\hat{\gamma}_{\hat{\mu}})^*\ = \ -\hat{\mathds{B}}\hat{\gamma}_{\hat{\mu}}\hat{\mathds{B}}^{-1},\qquad(\hat{\gamma}_{\hat{\mu}})^T\ = \ -\hat{\mathds{C}}\hat{\gamma}_{\hat{\mu}}\hat{\mathds{C}}^{-1}
\end{equation}
together with the identities
\begin{equation}\label{6DABCProp}
\hat{\mathds{B}}^T=\hat{\mathds{C}}\hat{\mathds{A}}^{-1},\qquad \hat{\mathds{B}}^*\hat{\mathds{B}}=-\mathds{1}_8,\qquad \hat{\mathds{C}}^T=-\hat{\mathds{C}}^{-1}=-\hat{\mathds{C}}^{\dagger}=\hat{\mathds{C}}\, .
\end{equation}
The chirality operator $\hat{\gamma}_\ast$ can be defined as
\begin{equation}
\hat{\gamma}_\ast=\hat{\gamma}_0\hat{\gamma}_1\hat{\gamma}_2\hat{\gamma}_3\hat{\gamma}_4\hat{\gamma}_5=\text{diag}(\mathds{1}_{4},-\mathds{1}_{4})\quad \text{with}\quad \hat{\gamma}_\ast\hat{\gamma}_\ast=\mathds{1}_8\, .
\end{equation}
The diagonal form of $\hat{\gamma}_\ast$ implies that a Dirac spinor $\Psi$ can be chirally decomposed as 
\begin{equation}
\Psi=\Psi_++\Psi_-\qquad\text{with}\qquad \mathbb{P}_{\pm}\Psi_{\pm}=\pm\Psi_{\pm}
\end{equation}
where the projection operators are given by
\begin{equation}
\mathbb{P}_{\pm}=\frac{1}{2}(\mathds{1}_8\pm\hat{\gamma}_\ast).
\end{equation}
Therefore, we can define two irreducible Weyl spinors $\psi_+$ and $\chi_-$ from the Dirac spinor $\Psi$ by
\begin{equation}
\Psi=\begin{pmatrix} \psi_+ \\ \chi_- \end{pmatrix},\qquad \Psi_+=\begin{pmatrix} \psi_+ \\ 0 \end{pmatrix},\qquad \Psi_-=\begin{pmatrix} 0 \\ \chi_- \end{pmatrix}.
\end{equation}
\indent Although the second property in \eqref{6DABCProp} implies that a reality condition cannot be imposed on the Dirac or Weyl spinors, we can define a symplectic-Majorana-Weyl spinor of the form
\begin{equation}
\Psi_{+\alpha}=\begin{pmatrix} \psi_{+\alpha} \\ 0 \end{pmatrix}\qquad\text{and}\qquad \Psi_{-\dot{\alpha}}=\begin{pmatrix} 0 \\ \chi_{-\dot{\alpha}} \end{pmatrix}
\end{equation}
with $\Psi_{+\alpha}$ and $\Psi_{-\dot{\alpha}}$ satisfying the pseudo-reality condition given by
\begin{equation}
\Psi_{+}^\alpha=(\Psi_{+\alpha})^*=\Omega^{\alpha\beta}\hat{\mathds{B}}\Psi_{+\beta}\qquad\text{and}\qquad
\Psi_{-}^{\dot{\alpha}}=(\Psi_{-\dot{\alpha}})^*=\Omega^{\dot{\alpha}\dot{\beta}}\hat{\mathds{B}}\Psi_{-\dot{\beta}}\, .
\end{equation}
$\Omega^{\alpha\beta}$ and $\Omega^{\dot{\alpha}\dot{\beta}}$ are symplectic forms of the two $USp(4)$ factors in the R-symmetry $USp(4)\times USp(4)$ under which $\Psi_{+\alpha}$ and $\Psi_{-\dot{\alpha}}$ transform separately.

\section{Truncation ansatze}\label{TrunAnsz}
In this appendix, we collect some useful formulae for a consistent truncation of seven-dimensional $SO(5)$ gauged supergravity on a circle ($S^1$), giving rise to $SO(5)$ gauged supergravity in six dimensions. This truncation has been constructed in \cite{6D_SO(5)}. 

The truncation ansatze for the seven-dimensional metric, scalar, vector, and tensor fields are respectively given by
\begin{eqnarray}
d\hat{s}_7^2&=&e^{\frac{\sigma}{\sqrt{10}}}ds_6^2+e^{-\frac{4\sigma}{\sqrt{10}}}(dz+A_{(1)})^2,\label{MetricAntz}\\
{\hat{\Pi}_I}^{\ i}(x^\mu,z)&=&{\Pi_I}^i(x^\mu),\label{ScalarAntz}\\
{\hat{B}_{(1)I}}^{\quad\ J}&=&{B_{(1)I}}^J+{B_{(0)I}}^J(dz+A_{(1)}),\label{VecAntz}\\
{\hat{S}_{(3)I}}&=&{S_{(3)I}}+{S_{(2)I}}(dz+A_{(1)})
\end{eqnarray}
in which hatted quantities refer to seven-dimensional fields. $x^\mu$ are six-dimensional space-time coordinates, and $z$ is the coordinate on $S^1$. Here, $I,J,...=1,...,5$ are vector indices of the gauge group $SO(5)$ while $i,j,...=1,...,5$ are vector indices for the local composite $SO(5)_c$. 
\\
\indent There are $(1+14+10)$ scalars denoted by $\{\sigma,\, {\Pi_I}^i,\, {B_{(0)I}}^J\}$ in the six-dimensional theory. These are given by the dilaton scalar field from the matric ansatz \eqref{MetricAntz}, fourteen scalars parametrizing $SL(5)/SO(5)_c$ coset, and ten axionic scalar fields from the truncation ansatz of vector fields \eqref{VecAntz}. There are also $(10+1)$ vectors $\{{B_{(1)I}}^J,\, A_{(1)}\}$ together with five two-form potentials $S_{(2)I}$. The five three-form potentials $S_{(3)I}$ do not contribute to the Lagrangian of the $SO(5)$ gauged theory in six dimensions since the seven-dimensional self-duality condition allows to eliminate them in favor of the two-form potentials.

For supersymmetric domain wall solutions considered in this work, we can set $A_{(1)}={B_{(1)I}}^J={S_{(2)I}}=0$. Using the domain wall ansatz for the matric from \eqref{DWmetric} together with $\varphi=\frac{\sigma}{2\sqrt{10}}$, we find that \eqref{MetricAntz} becomes
\begin{eqnarray}
e^{2\hat{A}}dx^2_{1,5}+d\hat{r}^2&=&e^{2A+2\varphi}dx^2_{1,4}+e^{-8\varphi}dz^2+e^{2\varphi}dr^2\nonumber\\
&=&e^{\frac{8A}{5}}(dx^2_{1,4}+dz^2)+e^{2\varphi}dr^2
\end{eqnarray}
In the second line, we have substituted $\varphi=-\frac{A}{5}$ from the domain wall solutions given here. This is also necessary for $dx_{1,4}^2$ and $dz^2$ to form a six-dimensional flat space-time matching $dx_{1,5}^2$ on the left hand side. We can also see the relations between the warped factors $\hat{A}=\frac{4A}{5}$ and the radial coordinates $d\hat{r}=e^\varphi dr$. 

The ansatz \eqref{ScalarAntz} implies that the scalars parametrizing $SL(5)/SO(5)_c$ coset in seven- and six-dimensional supersymmetric domain walls are the same since they are independent of $z$ and depend only on the corresponding radial coordinates
\begin{equation}
{\hat{\Pi}_I}^{\ i}(\hat{r})={\Pi_I}^i(r).
\end{equation}
For ten axionic scalars ${B_{(0)I}}^J$, which are called shift scalars in this work, we can see from \eqref{VecAntz} that they give rise to non-vanishing vector fields in seven dimensions
\begin{equation}\label{OURVecAntz}
{\hat{B}_{(1)I}}^{\quad\ J}={B_{(0)I}}^Jdz.
\end{equation}
Therefore, domain wall solutions with non-vanishing axionic scalars obtained in this work cannot be obtained from an $S^1$ reduction of any domain wall solutions in seven dimensions.


\end{document}